 \documentclass[11pt, oneside]{amsart}   	
\usepackage[letterpaper, margin=1in]{geometry}                	
\usepackage[parfill]{parskip} 
\usepackage[nopatch=eqnum]{microtype}
\usepackage{amssymb, bm, enumitem, bbm, mathtools}
\usepackage[obeyspaces,hyphens,spaces]{url}

\usepackage{soul}
\setstcolor{red}

\usepackage{graphicx,overpic}	
\usepackage[usenames,dvipsnames]{xcolor}

\allowdisplaybreaks
\numberwithin{equation}{section}

\usepackage{float}

\usepackage{tikz}
\usetikzlibrary{arrows, positioning, calc, intersections}
\usetikzlibrary{decorations.pathreplacing, decorations.markings}
\usepackage{relsize}					
\usepackage{exscale}	
\usepackage{pgfplots}
\pgfplotsset{compat=newest}
\tikzset{->-/.style={decoration={markings, mark=at position .7 with {\arrow[scale=0.8] {stealth}} },postaction={decorate}}}
\tikzset{-<-/.style={decoration={markings, mark=at position .7 with {\arrow[scale=0.8] {stealth reversed}} },postaction={decorate}}}

\pgfplotsset{
   axis line style = {black!50},
   every axis/.style={scale only axis}
}

\definecolor{shadecolor}{rgb}{0.95, 0.95, 0.86}

\makeatletter 
\@ifdefinable\@latex@chi{\let\@latex@chi\chi}
\renewcommand*\chi{{\@latex@chi\smash[t]{\mathstrut}}} 
\makeatletter

\newtheorem{theorem}{Theorem}[section]
\newtheorem{lemma}[theorem]{Lemma}

\newtheorem{remark}[theorem]{Remark}

\newtheorem{prop}[theorem]{Proposition}

\theoremstyle{definition}

\newtheorem{assume}[theorem]{Assumption}
\newtheorem{RHP}[theorem]{Riemann-Hilbert Problem}
\newtheorem{dbar.rhp}[theorem]{$\dbar$-Riemann-Hilbert Problem}

\newcommand{\R}{{\mathbb R}}

\newcommand{\N}{{\mathbb N}}
\newcommand{\C}{\mathbb{C}}

\newcommand{\dd}{\mathrm{d}}
\newcommand{\dbar}{\overline{\partial}}

\renewcommand{\Re}{\mathop{ \mathrm{Re}}\nolimits}
\renewcommand{\Im}{\mathop{ \mathrm{Im} }\nolimits}
\newcommand{\ii}{\mathrm{i}}
\newcommand{\eee}{\mathrm{e}}

\newcommand{\bigo}[2][]{\mathcal{O}_{#1}\!\left( #2 \right)}

\newcommand{\littleo}[1]{{o}\left( #1 \right) }

\newcommand{\lp}{\left(}
\newcommand{\rp}{\right)}

\newcommand{\restrict}[1]{{\raisebox{-0.75ex}{\Big\vert}}_{\mathrlap{#1}}\;}

\newcommand{\vect}[1]{\boldsymbol{\mathbf{#1}}}

\newcommand{\triu}[2][1]{\begin{bmatrix} #1 & #2 \\ 0 & #1 \end{bmatrix}}
\newcommand{\tril}[2][1]{\begin{bmatrix} #1 & 0 \\ #2 & #1 \end{bmatrix}}
\newcommand{\diag}[2]{\begin{bmatrix} #1 & 0 \\ 0 & #2 \end{bmatrix}}
\newcommand{\offdiag}[2]{\begin{bmatrix} 0 & #1 \\ #2 & 0 \end{bmatrix}}

\newcommand{\sigone}{\vect{\sigma}_1}
\newcommand{\sigtwo}{\vect{\sigma}_2}
\newcommand{\sigthree}{\vect{\sigma}_3}

\newcommand{\ie}{i.e.}
\newcommand{\eg}{e.g.}
\newcommand{\cf}{cf.}

\newcommand{\eps}{\epsilon}

\DeclareMathOperator{\Wr}{Wr}
\DeclareMathOperator*{\res}{Res}

\DeclareMathOperator{\dist}{dist}

\DeclareMathOperator{\inside}{int}

\DeclareMathOperator{\dn}{dn}

\newcommand{\Alow}{\mathcal{A}^{\mathrm{low}} }
\newcommand{\Aup}{\mathcal{A}^{\mathrm{up}} }

\usepackage{calligra}
\DeclareMathAlphabet{\mathcalligra}{T1}{calligra}{m}{n}
\DeclareFontShape{T1}{calligra}{m}{n}{<->s*[2.2]callig15}{}
\newcommand{\scripty}[1]{\ensuremath{\mathcalligra{#1}}}

\newcommand{\scalemath}[2]{\scalebox{#1}{\mbox{\ensuremath{\displaystyle #2 } } } }

\newcommand{\staritem}{\global\asterisktrue\item}
\newcommand{\perhapsasterisk}{%
  \ifasterisk*\global\asteriskfalse\fi
}
\newif\ifasterisk

\def\be{\begin{equation}}
\def\ee{\end{equation}}

\def\bi{\begin{itemize}}
	
\def\ei{\end{itemize}}

\def\bea{\begin{eqnarray}}
	
\def\eea{\end{eqnarray}}

\def\bl{\begin{lemma}}
\def\el{\end{lemma}}

\def\g{\gamma}

\def\G{\Gamma}

\def\l{\lambda}
\def\s{\sigma}
\def\o{\omega}
\def\\e{\left}
\def\ri{\right}

 \usepackage[backend=biber, maxbibnames=99, style=numeric, sortcites=true]{biblatex}
\begin{document}

\title[Thermodynamic limit of the fNLS hierarchy]{Approximation of the thermodynamic limit of finite-gap solutions to the focusing NLS hierarchy by multisoliton solutions}

\author{Robert Jenkins}
\author{Alexander Tovbis}
\address{Department of Mathematics, University of Central Florida, Orlando, Florida 32816}
\email{robert.jenkins@ucf.edu, alexander.tovbis@ucf.edu}

\date{\today}

\begin{abstract}
In this paper we approximate the thermodynamic limit of finite-gap solutions to any integrable equations in the focusing NLS hierarchy (NLS, mKdV,  \ldots) with an associated multisoliton solutions using the Riemann-Hilbert Problem approach. Moreover, we show that both the finite-gap and multisoliton solutions are approximated in the thermodynamic limit by a generalization of the primitive potentials  introduced by V. Zakharov and his collaborators in the KdV context. 
Under certain assumptions on the spectral data for the finite gap potentials, we provide error estimates for the approximation on compact subsets of the $(x,t)$-plane.
\end{abstract}

\maketitle

\tableofcontents

\section{Introduction}

Solitons---localized traveling wave solutions with particle like interactions---are essential to our understanding of integrable systems. 
Many integrable systems, including fundamental models such as the nonlinear Schr\"{o}dinger (NLS) and Korteweg-de Vries (KdV) equations, admit multi-soliton solutions. 
By carefully tuning the parameters in these multi-soliton solutions they can combine to form macroscopic coherent structures such as dispersive shock waves and modulated wave trains \cite{Maiden18}, \cite{ElHoefer16}. 
Recently, it has been proposed that these same multi-soliton solutions should be able to describe incoherent, random wave motion in integrable systems, sometimes referred to as integrable turbulence. 
Developing a sound mathematical theory for describing the evolution and interaction of these random, many-particle, (soliton) wave-fields, informally dubbed \emph{soliton gases}, is an active area of research with many open questions. 

The wave-fields soliton gas theory seeks to describe can be dense. 
In such case, a  description in terms of localized solutions like solitons seems to be not the best approach; one needs a more general family of solutions, like finite gap solutions, adapted to describing dense wave-fields.   
Like solitons, finite-gap solutions are another family of exact solutions possessed by many integrable PDEs. 
They represent multiphase, quasi-periodic waves with a particularly simple spectrum (compared to general quasi-periodic solutions). By properly tuning the parameters defining a finite-gap solution, one can approximate a soliton or a multi-soliton on compact subsets of the $(x,t)$ plane. 
Conversely, it has been shown that the finite-gap potentials belong to the closure of the set of multi-soliton solutions \cite{Gesztesy92}, \cite{Marchenko91} \cite{Zaitsev83}, and so are, in some sense, a natural framework for modeling soliton gases. 

In recent years the so called thermodynamic limit of finite-gap solutions to integrable equations has emerged as an important theoretical model for describing soliton gases. 
The thermodynamic limit of finite-gap solutions describes the spectral theory of soliton gases, see \cite{El2003, ET2020, Elrev}. 
The spectral theory of soliton gases is concerned with the some  observables of  a soliton gas, and not necessarily with the corresponding  wave-fields, i.e., with particular realizations of the soliton gas.  
Example of observables in a soliton gas include the effective speed of an element of the gas (a tracer), the average intensity of the soliton gas, the gas kurtosis, etc., see, for example, \cite{AgZak2015}.  
 A brief introduction to the spectral theory of fNLS gases could be found in Section \ref{sec-backgr}. 
 
Another important approach for investigating integrable soliton gases is to study the large $N$ (many-particle) limit  of $N$-soliton solutions to integrable equations (KdV, mKdV, NLS). 
Necessarily, this approach deals with realizations of soliton gases, since the object of study are exact solutions of the underlying integrable equation. 
To obtain large $N$ asymptotics of these solutions, one must make assumptions on the spectral data defining the sequence of $N$-soliton solutions. 
In a series of papers \cite{GGJM, GGJMM, GJMnew, BGO23}, it was proved under the assumptions that the discrete eigenvalues of the system accumulate on an allowable spectral domain with some given asymptotic density, while the norming constants are sampled from some sufficiently smooth function, the corresponding $N$-soliton solutions converge as $N \to \infty$ to soliton gas realizations described in terms of so-called primitive potentials. In particular, the results of \cite{GGJMM} demonstrated how some spectral observables, such as effective soliton velocities, emerge naturally in an actual (mKdV) soliton gas realization.  
The primitive potentials describing the soliton gas realizations were first proposed by V. Zakharov and his collaborators \cite{DZZ, DNZZ, NabZZ} as a model of soliton gases using a more formal approach. 

To date, neither the thermodynamic limit approach nor the $N$-soliton limit approach has satisfactorily integrated randomness into its theory. 
They are largely deterministic models. 
In the thermodynamic limit approach one works with averaged or derived quantities which are independent of the randomness in the initial phase distribution of particular realizations.   
Despite this, the predictions and results of the thermodynamic limit theory have been experimentally confirmed in both optical \cite{Suret23, Schwache} and hydrodynamic  \cite{Suret20, Redor19, Fache24} contexts.   
In the $N$-soliton approach, current methods require smoothness assumptions on data, which make the inclusion of random spectral data problematic. However, some initial  progress has been reported recently on the inclusion of randomness into both approaches to the theory \cite{GM24+, Bonnemain22}.

The present paper is the first study of realizations of soliton gases obtained as thermodynamic limits of finite-gap solutions, i.e., when the initial phases of finite-gap solutions are fixed in a certain way. It represents a step toward the direct analysis of the thermodynamic limit of finite-gap potentials, with the long-range goal of fully integrating randomness into the model. 
Roughly speaking, our results show that under certain restrictions on the initial phases in the finite-gap solutions, there exists a corresponding sequence of multi-soliton solutions that accurately approximates the finite-gap solutions on compact subsets of the $(x,t)$-plane in the thermodynamic limit. Moreover, under the aforementioned restriction on initial phases, both the finite-gap and multi-soliton solutions converge to a primitive potential solution of the underlying integrable PDE analogous to those introduced by \cite{DZZ} for the KdV equation. This converge is uniform on compact subsets of the $(x,t)$-plane as $N \to \infty$, where $N$ is the number of solitons in the multi-soliton solution.  However, \textit{a posteriori}, if certain mild assumptions are satisfied, the convergence is uniform on sets which grow slowly with $N$.  
As such, our results are similar in spirit to those of \cite{GGJM, GGJMM} in that we establish the convergence of realizations to a primitive potential limit, but our results extend this approach in several meaningful way. First, our main results, Theorems \ref{thm:fg-sol} and \ref{thm:fg-prim}, hold for the entirety of the focusing NLS hierarchy of integrable equations which includes the focusing NLS and mKdV equations as its first two nontrivial flows. 
Second, we consider a much larger class of primitive potentials generalizing the model introduced in \cite{DZZ}, which include the full family of hyper-elliptic finite-gap solutions of the focusing NLS hierarchy. To demonstrate the utility of our results, some concrete examples are provided in Section \ref{sec:examples}.

Our approach to this problem is algebraic in nature, and we believe it should extend in a straightforward way to other families of integrable PDEs. 
Our main tool of analysis is the Riemann-Hilbert Problem (RHP) approach to formulating the inverse scattering transformation (IST).

\subsection*{Organization of the rest of the paper}
In the remainder of the introduction we introduce the focusing NLS (AKNS) hierarchy of integrable PDEs, see Section 1.1. Section 1.2 describes  the Riemann-Hilbert problem formulation of the IST for different classes of solutions to the focusing NLS hierarchy that are needed to describe our results. Section 1.3 precisely characterizes the thermodynamic limit of finite-gap solutions and introduces the assumptions needed to state our main results, which are given in Section 1.4. 
Section 2 provides a brief introduction to the spectral theory for focusing NLS gases. Sections 3-5 encapsulate the argument proving our main results. Section 3 describes how one approximates finite-gap solutions by multi-soliton solutions in the thermodynamic limit; Section 4 introduces the approximation of the discrete large $N$ models by a continuous (primitive potential) model; finally in Section 5 the main theorems are proved by showing that, in the thermodynamic limit, the ratio between the finite-gap potentials and continuous primitive potential model satisfies a small-norm Riemann-Hilbert problem that can be explicitly estimated. The last section, Section 6, introduces some concrete examples applying the main results of the paper.

\subsection{The focusing NLS hierarchy}
We consider the evolution of finite-gap solutions under the time-flows of the focusing NLS hierarchy. By the focusing NLS hierarchy we mean the AKNS hierarchy with the non-self-adjoint symmetry reduction $q(x,\vect{t}) = - \overline{p(x,\vect{t})} = \psi(x,\vect{t})$. Here $x \in \R$ and for any $\mathcal{K} \in \N$, $\vect{t} = \begin{bmatrix} t_1 & \dots & t_\mathcal{K} \end{bmatrix} \in \R^{\mathcal{K}}$. 

The focusing NLS hierarchy \cite{FNR} is the family of integrable PDEs with Lax-Pair structure
\begin{subequations}\label{lax.pair}
\begin{align}
	\label{zs}
	&\partial_x\mathbf{w} = \mathbf{L} \mathbf{w}(x,\vect{t}) 
	&&\mathbf{L} := - \ii z \sigthree + \begin{bmatrix} 0 & \psi(x,\vect{t}) \\ - \overline{\psi(x,\vect{t})} & 0 \end{bmatrix}  \\
	\label{time.flow}
	&\partial_{t_k} \mathbf{w} = \mathbf{B}^{(k)} \mathbf{w}(x,\vect{t}),
	&&\mathbf{B}^{(k)} := \ii z^k \sigthree + \sum_{\ell=0}^{k-1} \vect{B}^{(k)}_\ell(x,\vect{t}) z^\ell
\end{align}
\end{subequations}
where the $2\times 2$ zero-trace matrix coefficients in each $\vect{B}^{(k)}$ are determined to ensure that the zero-curvature equation
\begin{equation}\label{zero.curvature}
	\partial_{t_k} \vect{L} - \partial_x \vect{B}^{(k)} + \left[ \vect{L} , \vect{B}^{(k)} \right] = 0
\end{equation}
is isospectral, \ie, the dependence on $z$ cancels out leaving only the $k$-th flow of the hierarchy
\begin{equation}
\label{nls.flows}
	\partial_{t_k} \begin{pmatrix} 0 & \psi(x,t) \\ -\overline{\psi(x,t)} & 0 \end{pmatrix} - \partial_x \vect{B}^{(k)}_0  + \left[ \begin{pmatrix} 0 & \psi(x,t) \\ -\overline{\psi(x,t)} & 0 \end{pmatrix} , \vect{B}^{(k)}_0 \right] = 0.
\end{equation}

The coefficients $\vect{B}^{(k)}_\ell$ are polynomial in $\psi, \bar\psi$, and their $x$ derivatives; in particular all the matrix coefficients can be uniquely determined if one insists that all integration constants are identically zero. More details are given in Appendix~\ref{sec:AKNS}.
The flows for the first few times $t_k$ are given by
\begin{subequations}\label{flows}
\begin{gather}
	\partial_{t_0} \psi(x,\vect{t})  = 0, \\
\label{convect}	\partial_{t_1}\psi(x,\vect{t}) + \partial_x \psi(x,\vect{t})  = 0, \\
\label{fnls}		\partial_{t_2} \psi(x,\vect{t}) + \frac{1}{2} \partial_x^2 \psi(x,\vect{t}) + | \psi(x,\vect{t})|^2 \psi(x,\vect{t}) = 0, \\
\label{mkdv}	\partial_{t_3} \psi(x,\vect{t}) + \frac{1}{4} \partial_x^3 \psi + \frac{3}{2} |\psi|^2 \psi_x.
\end{gather}
\end{subequations}
The zeroth and first flows are trivial (constant and advective flow respectively); we recognize in the flows for $t_2$ and $t_3$ the focusing nonlinear Schr\"odinger  and (complex-valued) modified Korteweg-de Vries equations respectively.

It's well known that the time-flows also commute, i.e.,
\begin{equation}
	\partial_{t_k} \vect{B}^{(j)} - \partial_{t_j} \vect{B}^{(k)} + \left[ \vect{B}^{(j)} , \vect{B}^{(k)} \right] = 0
\end{equation}
which implies that all members of the hierarchy can be solved simultaneously. \cite{AKNS, FNR}

Integrability of this system allows for the construction of solutions to \eqref{nls.flows} through the inverse scattering transform (IST). 
Depending on the class of initial data considered, one constructs a forward scattering map $\psi_0 \mapsto \mathcal{D}$ which computes spectral data $\mathcal{D}$ from the initial data $\psi_0 = \psi(\, \cdot\, , \vect{0})$. The type of scattering data needed depends on the class of initial data (\ie, rapid decay, step-like, periodic, etc. ). 
The inverse problem can then be expressed in terms of Riemann-Hilbert problems (RHP) defined in terms of the given scattering data $\mathcal{D}$. 

\subsection{A trio of Riemann-Hilbert problems}

In both the thermodynamic limit approach to modeling soliton gas \cite{Elrev}, and the primitive potential approach \cite{GGJM} 
the local behavior of soliton gas realizations is described by quasi-periodic multi-phase wave solutions of the underlying PDEs. In order to precisely describe these models we introduce below three different classes of Riemann-Hilbert problem, characterized by the geometric properties of their scattering data, whose solutions encode, respectively, the finite-gap, generalized soliton, and primitive potential solutions of the NLS hierarchy. The majority of the analysis done in the subsequent sections will be to show that, under appropriate assumptions on the scattering data, both the thermodynamic limit of high genus finite-gap solutions and the many soliton limit of generalized multi-soliton solutions converge to the same soliton gas model which can be described by a primitive potential type RHP.  

The family of finite-gap solutions of the NLS hierarchy are defined by quasi-periodic potentials. These may be characterized in terms of scattering data
\begin{equation}	
	\mathcal{D}_{\mathrm{fg}} := \left( \Gamma, \Lambda \right),
\end{equation}
where $\Gamma$ is a finite collection of nonintersecting oriented simple arcs (i.e. curves homeomorphic to a closed line segment)  in $\C$ that are anti-Schwarz symmetric (\ie, $\overline \Gamma = - \Gamma$ in the sense of path algebra), and $\Lambda:\Gamma \to \C$ is a Schwarz symmetric ($\Lambda(\bar z) = \overline{\Lambda(z)}$) function that is  constant on each connected component of $\Gamma$. The data $\mathcal{D}_{\mathrm{fg}}$ defines the following problem.

\begin{RHP}[\textbf{Finite-gap RHP}]\label{rhp:n-phase}
Given $\mathcal{D}_{\mathrm{fg}}=(\Gamma, \Lambda)$, find a matrix-valued function $\vect{M}$ such that for each $(x, 
	\vect{t}) \in \R \times \R^\mathcal{K} $ :
\begin{enumerate}[align=left, leftmargin=*]
	\item[ \textbf{Analyticity}:] $\vect{M}(z;x,\vect{t})$ is analytic for $z \in \C \setminus \Gamma$. 
	\item[ \textbf{Normalization}:] $\vect{M}(z;x,\vect{t}) = \vect{I} + \bigo{z^{-1}}$ as $z \to \infty$. 
	\item[ \textbf{Endpoint growth}:] At any endpoint $p$ of $\Gamma$ the matrix $\vect{M}$ admits $\tfrac{1}{4}$-root singularities:
	\[
		\vect{M}(z;x,\vect{t}) = \bigo{ |z-p|^{-1/4} }, \qquad \text{ as } z \to p \in \partial \Gamma.
	\]
	\item[ \textbf{Jump Condition}:] $\vect{M}$ has continuous non-tangential boundary values $\vect{M}_\pm(z;x,\vect{t})$ as $z \to \Gamma$ satisfying the jump relation
	\begin{gather}
		\label{fg.jump}
		\vect{M}_+(z;x,\vect{t}) =\vect{M}_-(z;x,\vect{t})  (-\ii \sigma_2) \eee^{2 \ii (\varphi(z;x,\vect{t} ) + \Lambda(z) )\sigthree}, \quad z \in \Gamma \\
		\label{dispersive.phase}
		\varphi(z;x,\vect{t}) := x z  + \sum_{k=0}^\mathcal{K} t_k z^k .
	\end{gather}
\end{enumerate}
\end{RHP}
If a solution of this problem can be found then the function 
\begin{equation}
\label{recovery}
	\psi(x,\vect{t}) = 2\ii \lim_{z  \to \infty} z \vect{M}_{12}(z;x,\vect{t}) 
\end{equation}
is a solution of the first $\mathcal{K}$ flows of \eqref{nls.flows} \cite{BC84}.

The existence, uniqueness, and methods for constructing explicit solutions to finite-gap problems like RHP~\ref{rhp:n-phase} are well known \cite{ItsBelo,DIZ} . Let $\mathfrak{X}= \mathfrak{X}_G$ denote the hyperelliptic Riemann surface of genus $G$ associated with the cut plane $\overline{\C} \setminus \Gamma$, where $G+1$ is the number of component arcs in $\Gamma$. Then the solution of RHP~\ref{rhp:n-phase} is expressed in terms of ratios of certain Riemann $\Theta$-functions composed with the Abel-Jacobi map for the surface $\mathfrak{X}$ \cite{DIZ}.

Soliton solutions (with vanishing boundary conditions) of the NLS hierarchy are spectrally characterized by having a purely discrete spectra. More generally, breather solutions of NLS are solitons on an asymptotically (nonzero) constant background. The spectral data for breathers is characterized by both discrete spectrum and a single fixed band $\Gamma_0$ encoding the constant background. Recently, there has been interest in adding discrete spectrum to general finite-gap spectral curves $\Gamma_G$, corresponds to potentials with solitons evolving on a finite-gap `background' potential. 
The evolution of such systems was recently studied in the context of the KdV equation \cite{BJT22}. 
We call these potentials generalized soliton solutions of the NLS hierarchy. 
Spectrally, they are described by adding discrete spectral data to the spectrum of the finite-gap problem. 

We denote the generalized soliton scattering data as follows.
Fix a genus $G$ finite-gap (background) potential by choosing scattering data $(\Gamma_G, \Lambda:\Gamma_G \to \C)$, as specified above.
For solitons on a vanishing background we take $G=-1$ and $\Gamma_G = \emptyset$. 
To the finite-gap scattering data, append a Schwarz-symmetric finite set $\mathcal{Z} \subset \C \setminus (\R \cup \Gamma_G)$. We partition $\mathcal{Z} = \mathcal{Z}_\mathrm{low} \cup \mathcal{Z}_\mathrm{up}$ such that $\overline{\mathcal{Z}_\mathrm{low}} = \mathcal{Z}_\mathrm{up}$ and $\mathcal{Z}_\mathrm{low} \cap \mathcal{Z}_\mathrm{up} = \emptyset$. Finally, on $\mathcal{Z}_\mathrm{low}$ define a non-vanishing function $c:\mathcal{Z}_{\mathrm{low}} \to \C \setminus \{0\}$.

Let
\begin{equation}\label{gensol.data}
	\mathcal{D}_\mathrm{sol} = \left( \Gamma_G, \Lambda; \mathcal{Z}_\mathrm{low}, c \right).
\end{equation}
This data defines the following problem:
\begin{RHP}[\textbf{generalized soliton RHP}]\label{rhp:fg-gen}
Given $\mathcal{D}_{\mathrm{sol}}=\left(\Gamma_G, \Lambda; \mathcal{Z}_\mathrm{low}, c \right)$, find a matrix-valued function $\vect{M}$ such that for each $(x, \vect{t}) \in \R \times \R^\mathcal{K} $ :
\begin{enumerate}[align=left, leftmargin=*]
	\item[ \textbf{Analyticity}:] 
	  $\vect{M}(z;x,\vect{t})$ is analytic for $z \in \C \setminus (\Gamma_G \cup \mathcal{Z}_\mathrm{low} \cup \overline{\mathcal{Z}_\mathrm{low} } )$.
	\item[ \textbf{Normalization}:] 
	  $\vect{M}(z;x,\vect{t}) = \vect{I} + \bigo{z^{-1}}$ as $z \to \infty$. 
	  \item[ \textbf{Endpoint growth}:] 
	At any endpoint $p$ of $\Gamma_G$ the matrix $\vect{M}$ admits $\tfrac{1}{4}$-root singularities:
	  \[
		\vect{M}(z;x,\vect{t}) = \bigo{ |z-p|^{-1/4} }, \qquad \text{ as } z \to p \in \partial \Gamma_G.
	  \]
	\item[ \textbf{Jump Condition}:] 
	  $\vect{M}$ has continuous non-tangential boundary values $\vect{M}_\pm(z;x,\vect{t})$ as $z \to \Gamma_G$ 
	  satisfying the jump relation
	  \begin{gather}\label{fg-gen.jump}
		\vect{M}_+(z;x,\vect{t}) =\vect{M}_-(z;x,\vect{t})  (-\ii \sigma_2) 
		\eee^{2 \ii (\varphi(z;x,\vect{t} ) + \Lambda(z) )\sigthree}, \quad z \in \Gamma_G 
	  \end{gather}
	\item[\textbf{Residues}:] At each point $\zeta \in \mathcal{Z}_\mathrm{low}$ and its complex conjugate 
	  $\vect{M}(z ; x,\vect{t})$ has simple poles satisfying the residue relations
		\begin{equation}\label{fg-gen.res}
			\begin{aligned}
			&\res_{z = \zeta} \vect{M}(z;x,\vect{t}) = \lim_{z \to \zeta}  \vect{M}(z;x,\vect{t}) 
			\tril[0]{c(\zeta) \eee^{\ii \varphi(z;x,\vect{t}\, )}} \\
			&\res_{z = \bar \zeta} \vect{M}(z;x,\vect{t}) = \lim_{z \to \bar \zeta}  \vect{M}(z;x,\vect{t}) 
			\triu[0]{-\overline{c(\bar \zeta)} \eee^{-\ii \varphi(z;x,\vect{t}\, )}}
			\end{aligned}
		\end{equation}
\end{enumerate}
\end{RHP}

\begin{remark}
In the literature, the standard setup for the RHP characterizing the inverse scattering transform confines $\mathcal{Z}_\mathrm{low}$ to either $\C^+$ or $\C^-$. 
Here, we consider a more general situation in which $\mathcal{Z}_\mathrm{low} \subset \C \setminus \R$. 
So, $\vect{M}(\,\cdot\,;x,\vect{t})$ may have residues of the lower triangular type in \eqref{fg-gen.res} in both $\C^+$ and $\C^-$. Symmetry still demands that $\vect{M}(z ; x,\vect{t})$ have upper triangular residues at each point $\bar \zeta \in \overline{\mathcal{Z}_\mathrm{low}}$. See Appendix~\ref{app:normalization} for more details.
\end{remark}

\begin{figure}[ht]
	\hspace*{\stretch{1}}
	\includegraphics[width=.3\textwidth]{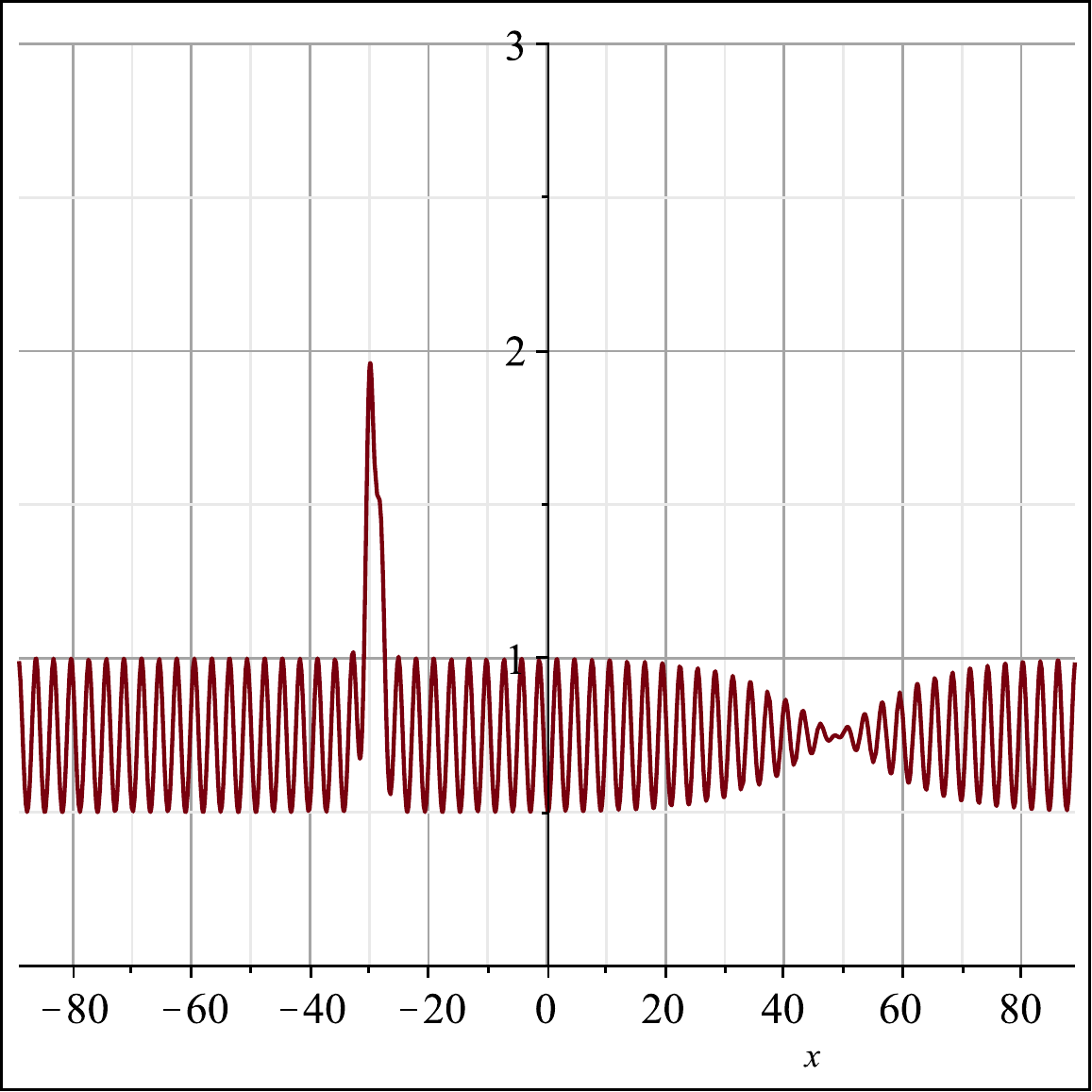}
	\hspace*{\stretch{2}}
	\includegraphics[width=.3\textwidth]{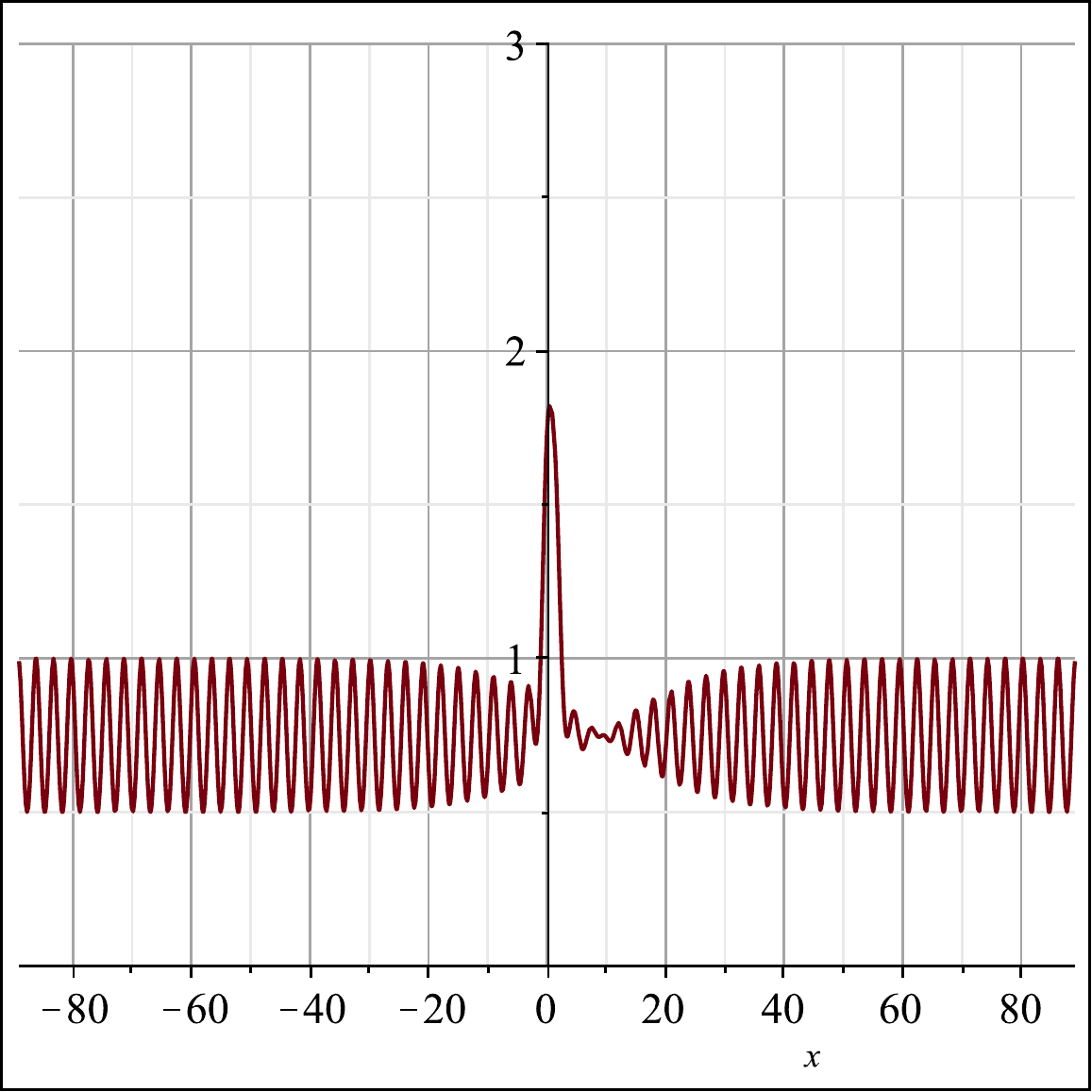}
	\hspace*{\stretch{2}}
	\includegraphics[width=.3\textwidth]{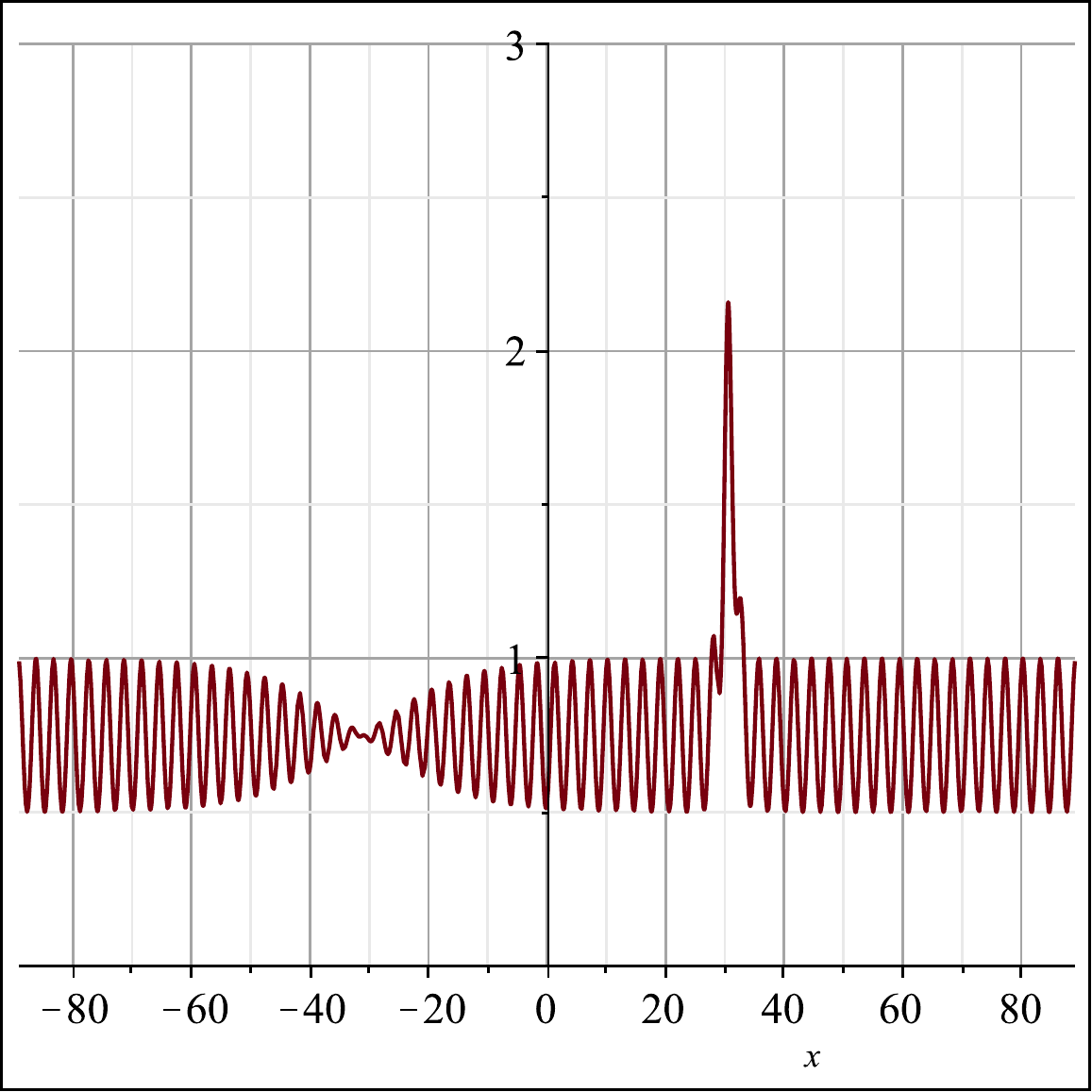}
	\hspace*{\stretch{1}}
\caption{An example of a generalized soliton in the context of the KdV equation. Frames show the evolution of generalized two-soliton solution of the KdV equation on an elliptic background. In this example, the `bright' soliton peak has positive velocity, while the 'dark' soliton depression wave has negative velocity. Reprinted from \cite{BJT22} with permission. 
}
\label{fig:kdv-solitons}
\end{figure}

The Riemann-Hilbert problem with data $\mathcal{D}_\mathrm{sol}$ encodes a solution of the NLS hierarchy which can be thought of as a generalized soliton solution where the `background' solution is not zero (as it would be if $\Gamma = \emptyset$) but instead the finite-gap potential with spectral data $(\Gamma, \Lambda)$, see Figure~\ref{fig:kdv-solitons}. In the simplest non-trivial case, when $\Gamma$ consists of a single arc ($G=0$) crossing the real axis,
RHP~\ref{rhp:fg-gen} describes the evolution of multi-breather solutions (that is solitons on a plane-wave background) of the NLS hierarchy.  More complicated cases of solitons on the hyperelliptic ($G\geq 1$) backgrounds were studied in the recent paper \cite{BJT22} in the context of KdV.

The solution of RHP~\ref{rhp:fg-gen} exists and  is unique.  For given data $\mathcal{D}_\mathrm{sol}$ it can be constructed explicitly. One approach is to first construct the solution of the finite-gap RHP~\ref{rhp:n-phase} with data $(\Gamma, \Lambda)$ without discrete spectrum in terms of Riemann $\Theta$-functions. Using this as a `seed solution', one can then introduce a series of Darboux transformations which insert a pair of poles at $\zeta, \bar \zeta$ for each $\zeta \in \mathcal{Z}_\mathrm{low}$ such that \eqref{fg-gen.res} is satisfied without effecting the jump relation \eqref{fg-gen.jump}. This approach was recently used in \cite{GGJMM} to construct solutions of mKdV where a single large soliton interacts with a soliton gas. An alternative other approach used in \cite{BJT22} is based on calculating limits of  Riemann theta functions on the degenerated Riemann surfaces.

Roughly speaking, a soliton gas realization is a solution of the NLS hierarchy that can be understood in the limit as an infinite collection of solitons.
There are different approaches for giving some rigorous meaning to this understanding. 
One of them is the thermodynamic limit approach, where one models macroscopic characteristics of the soliton gas using finite-gap potentials, whose bands grow in number as their individual band lengths shrink to zero while accumulating in some compact set $\mathcal{A}$. 
This is the approach taken by G. El  (\cite{El2003})  in the context of the KdV, and later by G. El and one of the authors \cite{ET2020} in the context of focusing NLS (fNLS).
It has been very successful in calculating various observables for soliton gases, such as the effective speed of a tracer soliton (an element of soliton gas) \cite{ET2020} and the average kurtosis \cite{Congy24}.
Many of these calculations have been verified in both optical \cite{Suret23, Schwache} and hydrodynamic \cite{Suret20, Redor, Redor19} experiments. 
Thus far, in this approach, actual soliton gas realizations have not been studied as it is not clear how to handle randomly chosen initial phases in the sequence of finite-gap potentials (RHP \ref{rhp:n-phase}) considered in the thermodynamic limit.
However, see \cite{TW2022} for the idea of periodic fNLS soliton/breather gases that develop from semiclassical periodic potentials after long-time evolution.
An alternative approach directly models the soliton gas as a limit of infinitely many solitons that spectrally belong to a compact set. 
One still has the problem of how to choose the norming constants $c_k$ in RHP~\ref{rhp:fg-gen} as one passes to the infinite soliton limit. 
In both approaches, the exact solutions described above become analytically intractable as the number of bands/discrete spectral points becomes arbitrarily large. 
To describe the limiting soliton gas we want a different class of solutions---with its own corresponding scattering data and Riemann-Hilbert problem---which we can describe analytically and for which we can control the convergence of the finite-gap/soliton solutions to an element of the limiting class of solutions.

A candidate for the limiting model for soliton gases is the class of primitive potentials first described  by Zakarhov et. al. \cite{DZZ} via IST in the context of the KdV equation. A formal argument for viewing the family of primitive potentials was given in \cite{DZZ}. In \cite{GGJM} the convergence of multi-solitons to (a subset of the) primitive potential solutions was rigorously proven and the far-field asymptotics of the resulting primitive potentials was described. Subsequently, this approach was used to precisely describe the interaction of a large solitons with a smaller amplitude soliton gas realization in the context of mKdV \cite{GGJMM} and to describe novel soliton gas realizations of the NLS equation whose discrete spectrum accumulates on a two-dimensional set \cite{BGO23}. More recently, it has been shown that the full family of primitive potentials described by Zakharov can be realized as the infinite limit of multi-soliton solutions \cite{GJMnew}.

\begin{figure}[ht]
	\begin{overpic}[height =.32\textwidth]{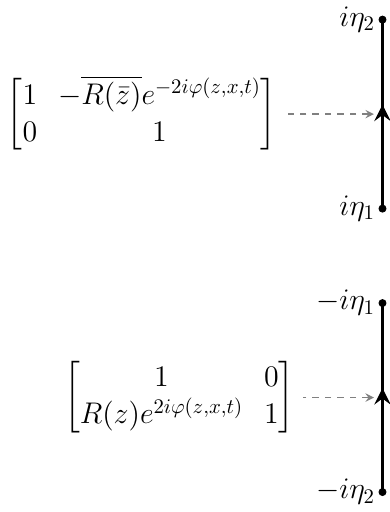}
	\put(30,-7){$(a)$}
	\end{overpic}
	\hspace*{\stretch{2}}
	\begin{overpic}[width=.32\textwidth]{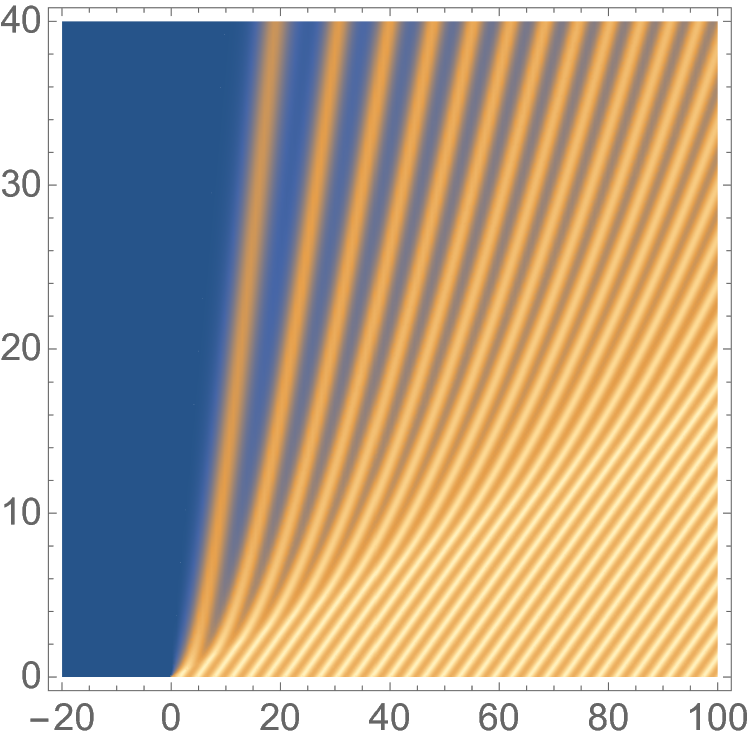}
	\put(50,-7.5){$(b)$}
	\end{overpic}
	\hspace*{\stretch{1}}
	\raisebox{0.25\height}{
	\begin{overpic}[width=.32\textwidth]{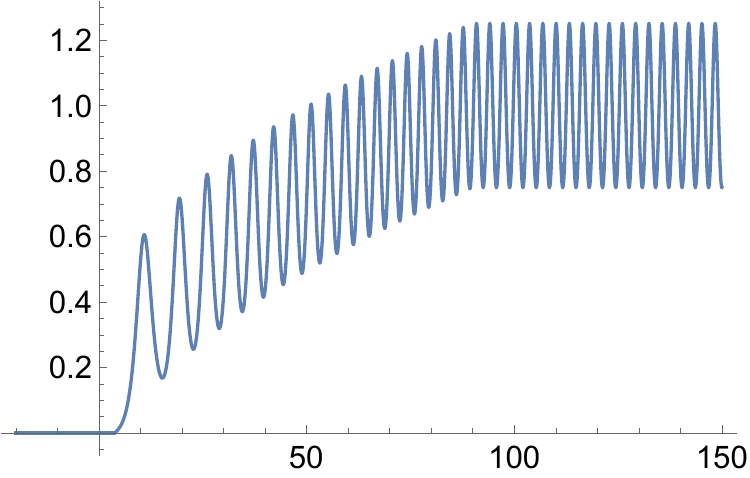}
	\put(53,-23){$(c)$}
	\end{overpic} 
	}
	\vspace*{10pt}
	\caption{An example of the primitive potential described by RHP~\ref{rhp:accumulated} on a vanishing bachground, i.e., $\Gamma_G = \emptyset$; 
	(a) the spectral data is one-dimensional with $\mathcal{A}^\mathrm{low} = [-\ii \eta_2, -\ii \eta_1]$;
	(b) Density plot of the (real) solution $\psi(x,t)$ of mKdV (so in \eqref{accumulated.jump}  $\varphi=  xz+4t z^3$) corresponding to spectral data in panel (a) with $\eta_1 = 0.25$, $\eta_2 = 1$, and $R(z) = i$;  
	(c) plot of the solution from panel (b) at $t=20$. 
	}
	\label{fig:primitive-example}
\end{figure}

Here, we introduce a further generalization of the primitive potential model by considering the accumulation of generalized soliton solutions. That is, we consider a fixed finite-gap potential with scattering data $(\Gamma_G, \Lambda)$. To this finite-gap background we append a (Schwarz-symmetric) compact set $\mathcal{A} = \Alow \cup \Aup$ with $\Aup = \overline{\Alow}$, $\Alow \cap \Aup = \emptyset$ which should be thought of as the accumulation domain of the generalized solitons and can be either one or two dimensional. The final piece of scattering data for our primitive potentials is a function $R:\Alow \to \C$ measuring the ``non-analyticity" of the solution $\mathbf{M}$ on $\Alow$. This collection of scattering data 
\begin{equation}
	\mathcal{D}_{\mathrm{prim}} = \left( \Gamma_G, \Lambda; \Alow, R \right).
\end{equation}
defines a Riemann-Hilbert, or more generally a $\dbar$-problem when the set $\mathcal{A}$ is 2-dimensional, encoding the primitive potential solution of the NLS hierarchy defined by this scattering data. See Figure~\ref{fig:primitive-example}.

\begin{dbar.rhp}\label{rhp:accumulated} 
Given scattering data $\mathcal{D}_{\mathrm{prim}} = \left( \Gamma_G, \Lambda; \Alow, R \right)$
let $\Alow = \Alow_1 \cup \Alow_2$, 
where $\Alow_k$, $k=1,2$, denotes the $k$-dimensional components of $\Alow$. Similarly, let $\mathcal{A}_k=\Alow_k\cup \Aup_k$, 
where $\Aup_k = \overline{\Alow_k}$, $k=1,2$.
Find a $2\times2$ matrix-valued function $\vect{M}(z ;x,\vect{t})$ with the following properties for each $(x, \vect{t})\in \R \times \R^\mathcal{K} $:

\begin{enumerate}[align=left, leftmargin=*]
	\item[\textbf{Continuity}:] $\vect{M}$ is a continuous function of $z$ in $\C \setminus (\Gamma_G \cup \mathcal{A}_1)$.
	\item[\textbf{Normalization}:] $\vect{M}(z;x,\vect{t}) = \vect{I} + \bigo{z^{-1}}$ as $z \to \infty$. 
	\item[\textbf{Analyticity}:]  $\vect{M}(\,\cdot\, ; x, t)$ is analytic in $\C \setminus (\Gamma_G \cup \mathcal{A})$.
	\item[\textbf{Endpoint growth}:] At any endpoint $p$ of $\Gamma_G$ the matrix $\vect{M}$ admits $\tfrac{1}{4}$-root singularities:
	\[
		\vect{M}(z;x,t) = \bigo{ |z-p|^{-1/4} }, \qquad \text{ as } z \to p \in \partial \Gamma_G.
	\]
	\item[\textbf{Jump Conditions}:] $\vect{M}$ has continuous non-tangential boundary values $\vect{M}_\pm(z;x,\vect{t})$ as $z \to   \Gamma_G \cup \mathcal{A}_1$ satisfying the jump relation
	\begin{equation}\label{accumulated.jump}
	\begin{gathered}
		\vect{M}_+(z;x,\vect{t}) = \vect{M}_-(z;x,\vect{t}) \vect{V}_M(z;x,\vect{t}) \\  
		\vect{V}_M(z;x,\vect{t}) = \begin{dcases}
			(-\ii \sigtwo) \eee^{2\ii (\varphi(z;x,\vect{t}) + \Lambda(z)) \sigthree} & z \in \Gamma_G \\
			\tril{ R(z) \eee^{2\ii \varphi(z;x,\vect{t})}} & z \in \Alow_1\\
			\triu{ -\overline{R(\bar z)}  \eee^{-2\ii \varphi(z;x,\vect{t})}} & z \in \Aup_1
		\end{dcases}
	\end{gathered}
	\end{equation}
	\item[\textbf{Non-analyticity:}] $\vect{M}$ is not analytic for $z \in \mathcal{A}_2$. It is locally a (weak) solution of the partial differential equation 
	\begin{equation}
	\begin{gathered}
		\dbar \vect{M}(z;x,\vect{t}) = \vect{M}(z;x,\vect{t}) \vect{W}(z;x,\vect{t}) \\
		\vect{W}(z;x,\vect{t}) = \begin{dcases}
			\tril[0]{ R(z) \eee^{2\ii \varphi(z;x,\vect{t})}} & z \in \Alow_2,\\
			\triu[0]{ -\overline{R(\bar z)} \eee^{-2\ii \varphi(z;x,\vect{t})}} & z \in \Aup_2
		\end{dcases}
	\end{gathered}
	\end{equation}
\end{enumerate}
\end{dbar.rhp}

\begin{remark}
The existence and uniqueness of the solution of general $\dbar$-Riemann-Hilbert problems is a non-trivial question. 
However, $\dbar$-RHP~\ref{rhp:accumulated} is equivalent by an explicit invertible transformation to a  RHP problem without $\dbar$ derivatives for which the existence and uniqueness of solutions is easily established. 
See transformation \eqref{P.infty.def} and Proposition~\ref{prop:Qinfty.exists} below for more details. 
\end{remark}

\subsection{The thermodynamic limit setup}
\label{sec-therm}

In the thermodynamic limit, we are interested in the behavior of finite-gap solutions of the NLS hierarchy as the number of non-intersecting component arcs in $\Gamma$ (see RHP~\ref{rhp:n-phase}) approaches infinity, while their arc lengths simultaneously shrink in such a way that they accumulate in some compact subset of the complex plane. We allow for the possibility that a finite number of arcs are fixed and, for simplicity,  do not change in this process. Fix a collection, possibly empty, of $G+1$ simple, non-intersecting, anti-Schwarz symmetric arcs $\Gamma_G$. See Figure~\ref{fig:fg-accumulation}.

Here, we gather all of the technical assumptions we make in order to prove the results in this paper. We will postpone their motivation until the need for each assumption arises in our subsequent argument, and will refer back to this section for the necessary formulas. The reader should feel free to briefly skim this section on a first pass as it sets much of the notation which will appear in what follows. Two of these assumptions which are essential to our approach are emphasized by asterisks.
\begin{figure}[ht]
\centering
\begin{overpic}[width=.75\textwidth]{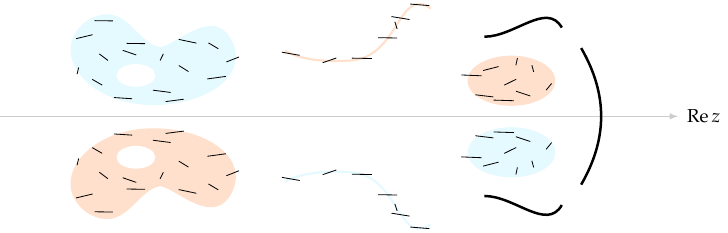}
\put(5,25){\color{CornflowerBlue}{\scalemath{.8}{\Alow_2}}}
\put(5,5){\color{RedOrange}{\scalemath{.8}{\Aup_2}}}
\put(59, 11){\color{CornflowerBlue}{\scalemath{.8}{\Alow_2}}}
\put(59, 21){\color{RedOrange}{\scalemath{.8}{\Aup_2}}}
\put(47,2){\color{CornflowerBlue}{\scalemath{.8}{\Alow_1}}}
\put(47,27){\color{RedOrange}{\scalemath{.8}{\Aup_1}}}
\end{overpic}
\caption{An example spectral curve $\Gamma_G^{(n)} = \Gamma_G \cup \gamma^{(n)}$ defining a finite-gap potential in a thermodynamic limit sequence 
(shown here with $G=2$ and $n=64$). 
The three ($G+1$) thick contours represent the fixed path $\Gamma_G$ defining a genus $G$ ``background" Riemann-surface. 
The set of small thin contours represent the sequence of shrinking contours $\gamma^{(n)}$ which accumulate as $n \to \infty$ on a set $\mathcal{A}$ (the shaded set in the figure). 
}
\label{fig:fg-accumulation}
\end{figure}
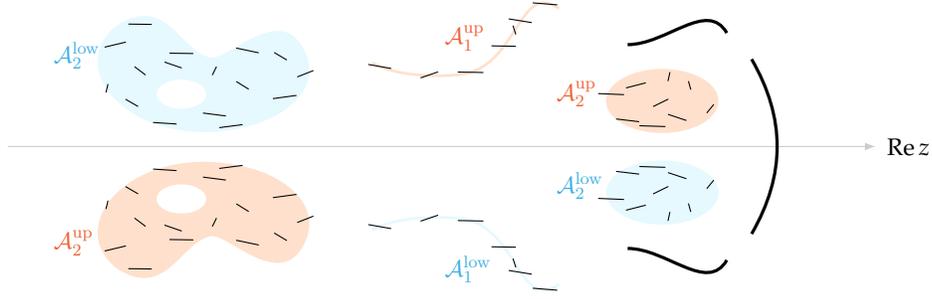
\begin{assume}\label{data.assumptions} 
In everything that follows we consider sequences of finite-gap potentials described by RHP~\ref{rhp:n-phase} with data $\mathcal{D}^{(n)}_{\mathrm{fg}} = \left( \Gamma_G^{(n)}, \Lambda^{(n)} \right)$ for each $n \in 2\N$. 
The data $\mathcal{D}^{(n)}_{\mathrm{fg}}$ must satisfy each of the following: 
\begin{enumerate}[label={\protect\perhapsasterisk\arabic*.}, ref=\arabic*]
\item The contour 
	\begin{equation}
		\Gamma_{G}^{(n)} = \Gamma_G \cup \gamma^{(n)}
	\end{equation}
	is a collection of nonintersecting, anti-Schwarz symmetric, simple arcs. Here, $\Gamma_G$ consists of $G+1$ simple arcs for some $G \geq -1$, which are fixed and independent of $n$, while $\gamma^{(n)}$ is a collection of $n$ simple arcs whose lengths all shrink to zero as $n \to \infty$. We label the component arcs of $\gamma^{(n)}$ by $\gamma_k^{(n)}$:
	\[
		\gamma^{(n)} =\bigcup_{k=1}^{n} \gamma^{(n)}_k . 
	\]
	For each component $\gamma_{k}^{(n)}$ of $\gamma^{(n)}$, if $\alpha^{(n)}_{2k-1}$ and $\alpha^{(n)}_{2k}$ denote its initial and terminal points respectively, let
	\begin{equation}
		\zeta^{(n)}_k := \frac{ \alpha^{(n)}_{2k} + \alpha^{(n)}_{2k-1}}{2}, 
		\qquad
		\delta^{(n)}_k := \frac{ \alpha^{(n)}_{2k} - \alpha^{(n)}_{2k-1}}{2},
	\end{equation}
	denote the midpoint and endpoint difference of $\gamma_k^{(n)}$. 
	To avoid pathological contours, we require that there exists a $C \geq 2$, independent of $k$ and $n$, such that
	$ | \gamma_k^{(n)} | \leq C  |\delta_k^{(n)}|$, $1 \leq k \leq n$, where $| \gamma_k^{(n)} | $ denotes the arclength of $ \gamma_k^{(n)} $.
	
	\item The  arcs in $\gamma^{(n)}$ are smooth whereas the arcs in $\Gamma_G $ are piece-wise smooth and $\C\setminus \Gamma_G^{(n)} $ is a connected set.

\item As $n \to \infty$, let
	\begin{equation}
		r_{n,k} =  \min_{ j \neq k} |  \zeta^{(n)}_{k} - \zeta_j^{(n)} |, 
	\end{equation}
	denote the center-to-center distance between any two component arcs of  $\gamma^{(n)}$. Define
	\begin{equation}\label{r def}
		r_n = \min_{1 \leq k \leq n} r_{n,k}, \qquad  \scripty{r}_n = \max_{1 \leq k \leq n} r_{n,k}.
	\end{equation}
	We assume that both $r_n , \scripty{r}_n  \to 0$ as $n \to \infty$ and for some $C_0>0$ 
	\begin{equation}\label{gamma separation}
		\dist( \gamma^{(n)} ,  \R \cup \Gamma_G)  > C_0 \scripty{r}_n.
	\end{equation}
	
\item\label{assume.measure} As $n \to \infty$, the centers $\zeta_k^{(n)}$ of the shrinking arcs accumulate on a compact (Schwarz-symmetric) set $\mathcal{A}$ in the sense that  $\lim_{n \to \infty} \zeta_{k}^{(n)} = \zeta^{(\infty)}_k  \in \mathcal{A}$. The set $\mathcal{A}$ consists of a union of finitely many disjoint, connected, compact components with piecewise smooth boundary. We assume $\inside(\mathcal{A}) \cap (\R \cup \Gamma_G) = \emptyset$ and that intersections of the boundary $\partial \mathcal{A}$ with $\R \cup \Gamma_G$ are finite in number and occur non-tangentially. 
The points $\zeta_k^{(n)}$ accumulate on $\mathcal{A}$ such that for some probability measure $\mu$ on $\mathcal{A}$, absolutely continuous with respect to Lebesgue measure, there exists a partition of $\mathcal{A} = \bigcup_{k=1}^n \mathfrak{a}_k^{(n)}$ with the properties that for some fixed $p>0$:
\begin{equation}\label{point.measure.limit}
	\begin{aligned}
	&n \mu(\mathfrak{a}_k^{(n)} ) = 1 + \bigo{n^{-p}}, &&n \to \infty \\
	&\sup_{\xi \in \mathfrak{a}_k^{(n)}} | \zeta^{(n)}_k - \xi| = \bigo{ \scripty{r}_n }.
	\end{aligned}
 \end{equation}  

\item\label{assume.densities}  Each connected component of $\mathcal{A}$ can be either one or two dimensional. Let $\chi_{\mathcal{A}_1}$ and $\chi_{\mathcal{A}_2}$ be the characteristic functions of the one and two dimensional components of $\mathcal{A}$. Then we will write the probability measure $\dd \mu(z)$ in the form
\begin{subequations}\label{eig.measure.density}
\begin{equation}\label{eig.measure.density.prob}
	\dd \mu(z)  =  \chi_{\mathcal{A}_1}(z) \rho_1(z) |dz|  + \chi_{\mathcal{A}_2}(z) \rho_2(z,\bar z) \frac{\ii (\dd z \wedge \dd \bar z)}{2};
\end{equation}
when we want to emphasize that it is a probability measure, but we will also use the following, equivalent\footnote{
The relation between the positive and complex densities are: 
$\varrho_1(z) = 2\pi \ii \rho_1(z) \frac{|\dd z|}{\dd z}$ and $\varrho_2(z,\bar z) = - \frac{1}{2\pi} \rho_2(z,\bar z)$.
}, 
notation which will be more convenient when working with complex integration
\begin{equation}\label{eig.measure.density.complex}
	\dd \mu(z)  =  \chi_{\mathcal{A}_1}(z)  \varrho_1(z) \frac{dz}{2\pi \ii} +  \chi_{\mathcal{A}_2}(z) \varrho_2(z,\bar z)  \frac{\dd z \wedge \dd \bar z}{2\pi \ii}.
\end{equation}
\end{subequations}

\item\label{assume.touching} At any point $p \in \partial\mathcal{A} \cap ( \Gamma_G \cup \R)$---if the intersection is non-empty---there are a finite number of disjoint closed sectors centered at $p$ such that for all sufficiently small disks $\mathcal{U}$ centered at $p$, $\mathcal{A} \cap \mathcal{U}$ consists of a fixed number of connected components each enclosed in a separate sector.  We further assume that for all sufficiently large $n$, any band centers $\zeta_k^{(n)} \in \mathcal{U}$ are contained in these sectors as well. Finally,  the measure $\mu$ vanishes linearly at each point of contact, i.e., the densities $\rho_1(z)$ and $\rho_2(z,z)$ in \eqref{eig.measure.density} satisfy
\begin{equation}\label{measure.vanish}
	\rho_1(z) = \bigo{ (z-p_0)}, \text{ and }\rho_2(z,\bar z) = \bigo{ (z-p_0)}, \qquad z \to p_0, \quad p_0 \in \partial{\mathcal{A}} \cap ( \R \cup \Gamma_G).
\end{equation}

\staritem\label{assume.shrinkrate} The arc length of each shrinking band in $\gamma^{(n)}$ decays much faster than the distance between bands. 
	Let 
	\begin{gather}\label{delta def} 
		\delta_n = \max_{1 \leq k \leq n} |\delta^{(n)}_{k} |
	\end{gather}
	denote the the maximal half chord length of any shrinking arc.
	Define	
	\begin{equation}\label{band.gap.ratio}
		\varsigma(n) :=  \frac{|\delta_n|}{r_n}  .
	\end{equation}
	In order for our main estimates to work we require that
	\begin{equation}\label{band.gap.ratio.rate}
		\lim_{n \to \infty}  \varsigma(n)^2 \left( \frac{n}{r_n} + \frac{1}{n r_n^3} \right) = 0.
	\end{equation}

\staritem\label{assume.phases} 
The piecewise constant function $\Lambda^{(n)}: \Gamma_G^{(n)} \to \C$, when restricted to the fixed bands in $\Gamma_G$ is independent of $n$
\begin{equation}\label{Lambda.G.limit}
	\Lambda^{(n)}(z) \restrict{z \in \Gamma_G} = \Lambda^{(\infty)}(z)   
\end{equation}
Let $\Lambda^{(n)}_k := \Lambda^{(n)}(z) \restrict{z \in \gamma^{(n)}_k}$ 
denote the constant complex phases on the shrinking arcs accumulating on the set $\mathcal{A}$. 
We make the following assumption on the behavior of the phases $\Lambda_k^{(n)}$ as $n \to \infty$.
Split $\mathcal{A}$ into two disjoint compact subsets  $\mathcal{A} = \Alow \cup \Aup$, $\Aup = \overline\Alow$.  
We require that for some non-vanishing  $\alpha$-H\"older continuous function $\beta$, $0< \alpha \leq 1$,  defined on an open neighborhood of $\Alow$, that the following estimate holds
\begin{subequations}\label{Lambda.cond}
\begin{equation}\label{A+}
	\left| \frac{\ii n}{2} \delta^{(n)}_k \eee^{\ii \Lambda^{(n)}_k} - \beta \left( \zeta_k^{(n)} \right) \right|= \bigo{n^{-q}}, \quad 
	\quad \text{whenever} \quad
	\lim_{n \to \infty} \zeta_k^{(n)}  \in \Alow,
\end{equation}
for some $q>0$ and any $k$, $1 \leq k \leq n$.
Due to the Schwarz symmetry of $\Lambda^{(n)}(z)$ and anti-Schwarz symmetry of $\gamma^{(n)}$, for any $k$, $1 \leq k \leq n$,  there exists a $j \neq k$, $1 \leq j \leq n$, such that
$\gamma_j^{(n)} = - \overline{\gamma}_k^{(n)}$ (so  $\delta^{(n)}_j = - \bar \delta_k^{(n)}$ and $ \zeta^{(n)}_j = \bar \zeta_k^{(n)}$) and $\Lambda^{(n)}_j = \overline{\Lambda}^{(n)}_k$. It follows that
\begin{equation}\label{A-}
	\left| \frac{\ii n}{2} \delta^{(n)}_j \eee^{-\ii \Lambda^{(n)}_j} - \overline{\beta \left( \bar \zeta_j^{(n)} \right)} \right|= \bigo{n^{-q}},
	\quad \text{whenever} \quad
	\lim_{n \to \infty} \zeta_j^{(n)}  \in \Aup.
\end{equation} 
\end{subequations}

\end{enumerate}
\end{assume}
	
\subsection{Main Results}

Consider a sequence of finite-gap potentials with scattering data $\mathcal{D}_{\mathrm{fg}}^{(n)} = \left(\Gamma_{G}\cup\gamma^{(n)}, \Lambda^{(n)}\right)$, which satisfy the conditions in Assumption~\ref{data.assumptions}, in the (thermodynamic) limit as $n \to \infty$.
The main results of our work is to show that 
one can construct explicit scattering data \eqref{gensol.data} for a sequence of generalized soliton potentials solving RHP~\ref{rhp:fg-gen}
that have the same limiting behavior as the sequence of finite-gap solutions as
$n\to\infty$.
Moreover, their joint limiting potential is well-defined and can be expressed in terms of a solution of the $\dbar$-Riemann-Hilbert Problem~\ref{rhp:accumulated}
with data $\mathcal{D}_\mathrm{prim} = \left(\Gamma_G,  \Lambda^{(\infty)}; \Alow, R \right)$, which is explicitly computable from the sequence of finite-gap data $\mathcal{D}_{\mathrm{fg}}^{(n)}$. In particular, the region of non-analyticity is the accumulation set $\mathcal{A}= \Alow \cup \overline{\Alow}$ on which the sequence of finite-gap spectral curves accumulate in the thermodynamic limit.

In order to precisely state our results, we first need to define some related sets of scattering data. Suppose that for each $n\in 2\N$
\[
	\mathcal{D}_{\mathrm{fg}}^{(n)} = \left(\Gamma_{G}\cup\gamma^{(n)}, \Lambda^{(n)}\right)
\]
is a sequence of finite-gap scattering data for RHP~\ref{rhp:n-phase} (see also Theorem~\ref{thm:fg-sol}| below) satisfying the conditions of Assumption~\ref{data.assumptions}. Let 
$\psi(x;\vect{t} \,\vert\, \mathcal{D}^{(n)}_{\mathrm{fg}})$ denote the corresponding finite-gap solutions of the focusing NLS hierarchy. 

For each $n \in 2\N$, define discrete (soliton) scattering data in terms of the given finite-gap data:
\begin{gather}\label{soliton.ensemble.data}
	\mathcal{D}_{\mathrm{sol}}^{(n)} = \left( \Gamma_G,  \Lambda_G^{(n)}; \mathcal{Z}_\mathrm{low}^{(n)}, c^{(n)} \right); \\
	\begin{gathered}\label{Zn.def}
	\Lambda_G^{(n)} = \Lambda^{(n)} \, \big{\vert}_{z \in \Gamma_G}, \\
	\mathcal{I}^{(n)}_\mathrm{low} = \left\{ k\,:\, 1 \leq k \leq n, \ \ \lim_{m \to \infty} \zeta_k^{(n)} \in \mathcal{A}^\mathrm{low} \right\}, \qquad
	\mathcal{Z}_\mathrm{low}^{(n)} = \left\{ \zeta_k^{(n)} \, :\, k \in \mathcal{I}^{(n)}_\mathrm{low}. \right\}, \\
	c^{(n)} : \mathcal{Z}_\mathrm{low}^{(n)} \to \C \setminus \{ 0 \}, \quad c^{(n)}(\zeta_k^{(n)}) = \frac{ \ii}{2} \delta^{(n)}_k \eee^{\ii \Lambda^{(n)}_k},
	\end{gathered}
\end{gather}
and let $\psi(x;\vect{t}\,\vert\, \mathcal{D}^{(n)}_{\mathrm{sol}})$ denote the solution of the focusing NLS hierarchy described by RHP~\ref{rhp:fg-gen}. 

Both the sequence of finite-gap potentials $\psi \left(x,\vect{t} \,\vert\, D^{(n)}_{\mathrm{fg}}\right)$ and of multi-soliton potentials $\psi \left(x,\vect{t} \,\vert\, \mathcal{D}^{(n)}_{\mathrm{sol}}\right)$, with data satisfying Assumption~\ref{data.assumptions}, are models of soliton gas realization on a genus $G$ background as $n \to \infty$. 
The key hypotheses of Assumption~\ref{data.assumptions} are that: $i.)$ the sequence of point measures $\dd\mu^{(n)}(z) := \frac{1}{n} \sum_{k=1}^n \delta(z-\zeta^{(n)}_k)$ converge to a measure $\dd\mu(z)$ supported on a smooth finitely connected set $\Alow$ with density $\varrho(z)$ with respect to the natural complex measures on each component of $\Alow$ (\cf \eqref{eig.measure.density}); $ii.)$ there exists a non-vanishing $\alpha$-H\"older continuous function $\beta$ on an open neighborhood of $\Alow$ such that
\begin{equation}\label{beta.thm}
	n c^{(n)}(\zeta_k^{(n)}) = \beta(\zeta^{(n)}_k) + \bigo{n^{-q}};
\end{equation}
$iii.)$ when restricted to the fixed contour $\Gamma_G$, $\Lambda^{(n)}$ has a well defined limit
\begin{equation}
	\Lambda^{\infty}: \Gamma_G \to \C, \qquad \Lambda^{(\infty)}(z) = \lim_{n \to \infty} \Lambda^{(n)}(z).
\end{equation}
Let $R:\Alow \to \C$ be defined by  
\begin{equation}\label{R.thm}
	R(z) = \varrho(z) \beta(z),
\end{equation}
where $\varrho(z) = \varrho_1(z)$ for $z \in \mathcal{A}_1$ and $\varrho(z) = \varrho_2(z,\bar z)$ for $z \in \mathcal{A}_2$.

These limiting quantities define a primitive potential $\psi(x;\vect{t}\,\vert\, \mathcal{D}^{(n)}_{\mathrm{prim}})$ solution of the focusing NLS hierarchy described by the $\dbar$-RHP~\ref{rhp:Pinfty} with scattering data 
\begin{equation}\label{prim.limit.data}
	\mathcal{D}_\mathrm{prim} = \left(\Gamma_G,  \Lambda^{(\infty)}; \Alow, R \right).
\end{equation}

The following three theorems are the main results of this paper.

\begin{theorem}\label{thm:fg-sol}
Fix $\mathcal{K} \in \N$. For each $n \in 2\N$, let $\psi \left(x,\vect{t} \,\vert\, D^{(n)}_{\mathrm{fg}}\right)$ be the sequence of finite-gap solutions of the first $\mathcal{K}$ flows of the focusing NLS hierarchy described by RHP~\ref{rhp:n-phase} with scattering data $\mathcal{D}_{\mathrm{fg}}^{(n)} = \left(\Gamma_{G}\cup\gamma^{(n)}, \Lambda^{(n)}\right)$ satisfying Assumption~\ref{data.assumptions}  in the thermodynamic limit. 
Then the corresponding sequence, $\psi \left(x, \vect{t}\, \vert \mathcal{D}_\mathrm{sol}^{(n)}\right)$, of generalized $n/2$-soliton solutions on a genus $G$ background, of the same flows of the focusing NLS hierarchy, described by RHP~\ref{rhp:fg-gen} with scattering data $\mathcal{D}_{\mathrm{sol}}^{(n)} = \left( \Gamma_G,  \Lambda_G^{(n)} ; \mathcal{Z}^{(n)}, c^{(n)} \right)$ is such that 
\[
	 	\left| \psi \left(x,\vect{t} \,\vert\, \mathcal{D}^{(n)}_{\mathrm{fg}}\right) - \psi \left(x, \vect{t} \vert \mathcal{D}_\mathrm{sol}^{(n)}\right) \right|
		= \bigo{ \varsigma(n)^2 \left(  \frac{n}{r_n} + \frac{1}{n r_n^3} \right)} \quad \text{as~} n\to \infty
\]
uniformly for $(x,\vect{t})$ in compact subsets of $\R\times\R^\mathcal{K}$.
\end{theorem}

The above theorem makes precise the informal statement that given a soliton gas realization modeled by the   thermodynamic limit of finite-gap solutions we can construct a  sequence of soliton solutions which has the same behavior in the large $n$ limit. The next theorem describes what that limiting behavior actually is:

\begin{theorem}\label{thm:fg-prim}
Let $\psi \left(x,\vect{t} \,\vert\, \mathcal{D}^{(n)}_{\mathrm{fg}}\right)$ and $\psi \left(x, \vect{t} \vert \mathcal{D}_\mathrm{sol}^{(n)}\right)$ be the sequences of finite-gap and generalized soliton solutions of the focusing NLS hierarchy described in Theorem \ref{thm:fg-sol}. Then, as $n\to\infty$, both sequences converge to the primitive potential solution of the focusing NLS hierarchy $\psi(x,\vect{t}\,\vert\, \mathcal{D}_{\mathrm{prim}})$ described by the $\dbar$-RHP~\ref{rhp:accumulated} with scattering data $\mathcal{D}_{\mathrm{prim}}$ given by \eqref{prim.limit.data}
\begin{align}
&\begin{gathered}
	\left| \psi \left(x,\vect{t} \,\vert\, \mathcal{D}^{(n)}_{\mathrm{fg}}\right) - \psi \left(x, \vect{t} \, \vert \, \mathcal{D}_\mathrm{prim}\right) \right|
		= \bigo{ f(n) }, \\
	\left| \psi \left(x,\vect{t} \,\vert\, \mathcal{D}^{(n)}_{\mathrm{sol}}\right) - \psi \left(x, \vect{t} \, \vert \, \mathcal{D}_\mathrm{prim}\right) \right|
		= \bigo{ f(n) }
\end{gathered}
&& n \to \infty
\end{align}
uniformly for $(x,\vect{t})$ in compact subsets of $\R\times\R^\mathcal{K}$, where 
\[
	f(n) := \begin{dcases}
	\max \{ n^{-q}, n^{-p}, \scripty{r}_n^{\, \alpha} \} & \mathcal{A} \cap (\Gamma_G \cup \R) = \emptyset \\
	\max \{ n^{-q}, n^{-p}, \scripty{r}_n^{\, \alpha},  \scripty{r}_n^{\frac{d}{1+d}} \} & \mathcal{A} \cap (\Gamma_G \cup \R) \neq \emptyset .\\
	\end{dcases}
\]
Here, $\scripty{r}_n$, defined by \eqref{r def}, measures the rate at which shrinking bands accumulate; $p>0$ is given in \eqref{point.measure.limit}; $q>0$ is given in \eqref{Lambda.cond}; $\alpha \in (0,1]$ is the H\"{o}lder exponent of function $\beta$; 
and $d \in \{1,2\}$ is the smallest dimension of any connected component of $\mathcal{A}$ which intersects $\R \cap \Gamma_G$. 

\end{theorem}

\begin{remark}
We observe here that if $d=1$ in the above theorem, meaning that the accumulation set $\mathcal{A}$ is entirely one-dimensional, then the primitive potential problem $\dbar$-RHP~\ref{rhp:accumulated} has no $\dbar$-component and the problem reduces to a standard Riemann-Hilbert problem. For readers unfamiliar with $\dbar$ problems, given any  scattering data $\mathcal{D}_{\mathrm{prim}}$ of the form in \eqref{prim.limit.data}, the primitive potential can always be expressed as the solution of a RHP, albeit on a somewhat arbitrary contour. This is described by RHP~\ref{rhp:Qinfty} below. 
\end{remark}

Theorems \ref{thm:fg-sol} and \ref{thm:fg-prim} show that while both sequences $\psi \left(x,\vect{t} \,\vert\, D^{(n)}_{\mathrm{fg}}\right)$ and $\psi \left(x, \vect{t} \vert D_\mathrm{sol}^{(n)}\right)$ converge to $\psi(x,\vect{t}\,\vert\, \mathcal{D}_{\mathrm{prim}})$, the rate of convergence between the $n^{th}$ terms of these sequences could be much faster than their individual rates of convergence to the common limit $\psi(x,\vect{t}\,\vert\, \mathcal{D}_{\mathrm{prim}})$.

We now would like to know if the validity of Theorems \ref{thm:fg-sol} and \ref{thm:fg-prim} can be extended to unbounded subsets of $(x, \vect{t}) \in \mathbb{R} \times \mathbb{R}^\mathcal{K}$.  The following Theorem \ref{thm-unbound} shows that this is possible subject to some technical condition expressed through the solution to the RHP~\ref{rhp:Pinfty}.

\begin{theorem}\label{thm-unbound}
	If the solution $\vect{P}^{(\infty)}(z;x,\vect{t})$ of RHP~\ref{rhp:Pinfty} is of exponential order in both $x$ and $\vect{t}$ for $z$ in any compact set containing the accumulation set $\mathcal{A}$ in its interior, then there exists a constant $H>0$ such that for any $s$, $0 < s < 1$ the results of Theorem~\ref{thm:fg-sol} and Theorem~\ref{thm:fg-prim} remain true for $(x, \vect{t}) \in \mathbb{R} \times \mathbb{R}^\mathcal{K}$ in unbounded sets in $n$ such that
	\[
		\sqrt{x^2+ \sum_{k=1}^\mathcal{K} t_k^2} \leq H s \log \left( \frac{1}{f(n)} \right)
	\]
	except that the error estimates in Theorem~\ref{thm:fg-prim} are replaced by the slower decay rate $\bigo{ f(n)^{1-s} }$. 
\end{theorem}
	
\subsection{Notational conventions}\label{sec:notation}

Throughout the paper matrices and matrix-valued functions are all denoted by bold-face symbols, \eg, $\vect{A}, \vect{B}$, etc. In particular 
\begin{equation}
	\sigone = \offdiag{1}{1}, \qquad 
	\sigtwo=\begin{bmatrix*}[r] 0 & -\ii \\ \ii & 0 \end{bmatrix*}, \qquad 
	\sigthree = \diag{1}{-1},
\end{equation}
denote the three Pauli matricies. 
For any complex number $z$, $\bar z$ denotes its complex conjugate; similarly, for $\mathcal{A} \subset \C$, $\overline{\mathcal{A}}$ denotes the complex conjugate set. 
For complex functions $f(z)$, $f^*(z) := \overline{f(\bar z)}$ denotes the Schwarz reflection of $f$. 
For matrices $\mathbf{M}$, $\mathbf{M}^\dagger$ denotes the hermitian conjugate. 
When discussing analyticity of  functions, given an oriented contour $\gamma$ and a function $f$, we write $f_+(z)$ and $f_-(z)$ to denote the non-tangential limits of $f$ as the argument approaches $z \in \gamma$ from the left and right side of the contour with respect to the orientation of $\gamma$. We will also on occasion write $f(z) = f(z, \bar z)$ when we want to stress that a $f$ is non-analytic. 

We use standard asymptotic analysis notation. Given two functions $f$ and $g$ and a point $p \in \C \cup \{ \infty \}$ we say a function 
\begin{align}
	&f(z) = \bigo{g(z)} \text{ as } z \to p && \text{if} \quad  |f(z)| \leq K |g(z)| & z \in \mathcal{U}(p) 
\end{align} 
for some constants $K > 0$ and $\mathcal{U}(p)$ a neighborhood of $p$.

\section{Background: spectral theory of fNLS soliton gases and their realizations}\label{sec-backgr}

In this section we give some brief background information on the spectral theory of soliton gases and their realizations.

\subsection{Brief introduction to the spectral theory of fNLS soliton gases}

It is well known that the focusing NLS equation is integrable \cite{ZS}; 
the Cauchy (initial value) problem for the focusing NLS can be solved using the inverse scattering transform (IST) method for various classes of initial data (potentials).
The scattering transform connects a given potential with its scattering data expressed in terms of the spectral variable $z\in\C$.
In particular, scattering data consisting of only one pair of spectral points $z=a\pm i b$, $b>0$, and a (norming) constant $c\in \C$,
defines the famous soliton solution 
\begin{equation}\label{nls_soliton}
\psi_S (x,t)= 2ib \, \hbox{sech}[2b(x+4at-x_0)]e^{-2i(ax + 2(a^2-b^2)t)+i\phi_0},
\end{equation}
to the focusing NLS equation (normalized, in this section only, as $i\psi_t+\psi_{xx}+2|\psi|^2\psi=0$). 
It is characterized by two independent parameters: the discrete spectral value $z = a +\pm i b$ determines the soliton amplitude $2b$ and its (free) velocity $s=-4 a$; the norming constant $c$ determines the initial position $x_0$ of its center and the initial phase $\phi_0$.
Scattering data that consists of several points $z_j \in \C^+ $ (and their complex conjugates $\bar z_j$), $j\in \N$, together with their norming constants corresponds to multi-soliton solutions. 
Assuming that at $t=0$ the centers of individual solitons are far from each other, 
we can represent the focusing NLS time evolution of a multi-soliton solution as propagation and interaction of the 
individual solitons.

It is well known that the interaction of solitons reduces to only two-soliton elastic collisions, where the faster soliton (let us say corresponding to the spectral parameter $z_m$) gets a forward shift (\cite{ZS})
 $\frac{\Delta_{mj}}{\Im z_m}$, where
\be\label{Delta}
\Delta_{mj}= \log\left|\frac{z_m-\overline{z}_j}{z_m-z_j}
\right|,
\ee
and the slower ``$z_j$-soliton'' is shifted backwards by  $-\frac{\Delta_{mj}}{\Im z_j}$.

Suppose now we have a large ensemble  (a ``gas'') of solitons \eqref{nls_soliton} whose  spectral parameters $z$ are distributed over a compact set $\G^+\subset \C^+$ with some probability density $\rho(z)$.
Assume also that the locations (centers) of these solitons are distributed uniformly on $\R$ and that the density of states $u(z)$ is small, i.e, the gas 
is diluted. 
Let us consider the speed of a $z$-soliton in the gas. 
Since it undergoes rare but sustained collisions with other solitons, the speed $s_0(z)=-4\Re z$ of a free solution must be modified as
\begin{equation} \label{mod-speed}
s(z)=s_0(z) + \frac{1}{\Im z} \int_{\G^+}\log \left|\frac{w-\bar z}{w-z}\right|[s_0(z)-s_0(w)]u(w)d\l(w),
\end{equation}
where $\l$ is a reference measure on $\G^+$ (arc length measure if $\G^+$ is a curve or area measure if $\G^+$ is a 2D region.
Similar modified speed formula was first obtain by V. Zakharov \cite{Za71} in the context of the Korteweg-de Vries
(KdV) equation. Without the ``diluted gas'' assumption, i.e, with $u(z)=O(1)$, the equation \eqref{mod-speed} for $ s(z)$ turns into
into the integral equation
\begin{equation} \label{eq-state}
s(z)=s_0(z) + \frac{1}{\Im z} \int_{\Gamma^+}\log \left|\frac{w-\bar z}{w-z}\right|[s(z)-s(w)]u(w)d\l(w),
\end{equation}
known as the equation of state for the fNLS soliton gas. This equation was obtained 
in \cite{ElKamch} using purely physical reasoning. In \eqref{eq-state},  $s(z)$ has the meaning of the effective speed of the ``element of the gas'' (also called a tracer) associated with the spectral parameter $z$ (note that when $u(z)=O(1)$ we cannot distinguish individual solitons).

If we now assume some dependence of $s$ and $u$ on the space time parameters $(x,t)$ that  
occurs on very large spatiotemporal scales, then we complement the equation of state 
\eqref{eq-state} by the continuity equation for the density of states
\begin{equation} \label{kinet}
\partial_t u+\partial_x (su)=0, 
\end{equation}
Equations \eqref{eq-state},
\eqref{kinet} form the kinetic equation for  a dynamic (non-equilibrium) focusing NLS soliton gas \cite{ElKamch}. 
The kinetic equation for the KdV soliton gas was originally derived in  \cite{El2003}.
It is remarkable that recently a kinetic equation having similar structure was derived in the framework of “generalized hydrodynamics”
for quantum many-body integrable systems, see, for example, \cite{DYC, DSY, VY}. 

\begin{figure}[ht]
\begin{center}
\begin{tikzpicture}[scale=.85]
\draw[black,thick,postaction = {decorate, decoration = {markings, mark = at position 1 with {\arrow[black,thick]{>}}}}] (-3,0) -- (6,0);
\node[below] at (6,0) {$\Re z$};
\draw[blue,thick,postaction = decorate, decoration = {markings, mark = at position .5 with {\arrow[blue,thick]{>}}}] (4,0) arc [start angle=0, end angle=50, x radius=15pt, y radius=30pt];
\draw[blue,thick] (4,0) arc [start angle=0, end angle=-50, x radius=15pt, y radius=30pt]; 
 
\draw[blue,thick,,postaction = decorate, decoration = {markings, mark = at position .5 with {\arrow[blue,thick]{>}}}] (3.3,1.4) -- (2.95,1.65);
\draw[blue,thick,,postaction = decorate, decoration = {markings, mark = at position .5 with {\arrow[blue,thick]{>}}}] (2.4,2) -- (1.95,2.2);
\draw[blue,thick,postaction = decorate, decoration = {markings, mark = at position .25 with {\arrow[blue,thick]{>}}}] (1.4,2.4) -- (0.9,2.5);
\draw[blue,thick,postaction = decorate, decoration = {markings, mark = at position .5 with {\arrow[blue,thick]{>}}}] (0.35,2.56) -- (-0.25,2.6);
\draw[blue,thick,postaction = decorate, decoration = {markings, mark = at position .5 with {\arrow[blue,thick]{>}}}] (-0.8,2.6) -- (-1.4,2.6);
 
\draw[blue,thick,postaction = decorate, decoration = {markings, mark = at position .5 with {\arrow[blue,thick]{<}}}] (3.3,-1.4) -- (2.95,-1.65);
\draw[blue,thick,postaction = decorate, decoration = {markings, mark = at position .5 with {\arrow[blue,thick]{<}}}] (2.4,-2) -- (1.95,-2.2);
\draw[blue,thick,postaction = decorate, decoration = {markings, mark = at position .25 with {\arrow[blue,thick]{<}}}] (1.4,-2.4) -- (0.9,-2.5);
\draw[blue,thick,postaction = decorate, decoration = {markings, mark = at position .5 with {\arrow[blue,thick]{<}}}] (0.35,-2.56) -- (-0.25,-2.6);
\draw[blue,thick,postaction = decorate, decoration = {markings, mark = at position .5 with {\arrow[blue,thick]{<}}}] (-0.8,-2.6) -- (-1.4,-2.6);
  
\draw[black,postaction = decorate, decoration = {markings, mark = at position .0 with {\arrow[black,thick]{>}}}] (1.15,2.45) circle [radius=15pt];
\node[below] at (1.15, 3.6) {$\mathrm{A}_j$};
\node at (0.97,2.67) {$\textcolor{red}{\g_j}$};
 
\node at (4.3,-0.3) {$\textcolor{red}{\g_0}$};
  
\draw[black,postaction = decorate, decoration = {markings, mark = at position .74 with {\arrow[black,thick]{>}}}] (3.9,0.6) .. controls (3.5,0.3) and (1.0,1.8) .. (1.15,2.45);
\draw[black,dashed,postaction = decorate, decoration = {markings, mark = at position .5 with {\arrow[black,thick]{<}}}] (3.9,0.6) .. controls (4.9,1) and (1.5,3.75) .. (1.15,2.45);
\node[below] at (2.4,1) {$\mathrm{B}_j$};
\node at (3,1) {$\textcolor{red}{}$};
 
\draw[black,postaction = decorate, decoration = {markings, mark = at position .74 with {\arrow[black,thick]{<}}}] (3.9,-0.6) .. controls (3.5,-0.3) and (1.0,-1.8) .. (1.15,-2.45);
\draw[black,dashed,postaction = decorate, decoration = {markings, mark = at position .5 with {\arrow[black,thick]{>}}}] (3.9,-0.6) .. controls (4.9,-1) and (1.5,-3.75) .. (1.15,-2.45);
\node[below] at (2.4,-0.5) {$\mathrm{B}_{-j}$};
\node at (3,-1.1) {$\textcolor{red}{}$};
 
\draw[black,postaction = decorate, decoration = {markings, mark = at position .0 with {\arrow[black,thick]{<}}}] (1.15,-2.45) circle [radius=15pt];
\node[above] at (1.3,-3.7) {$\mathrm{A}_{{}-j}$};
\node at (1,-2.7) {$\textcolor{red}{\g_{-j}}$};

\node at (-1.1,2.85) {$\textcolor{red}{\g_1}$};
 
\node at (4.5, 0.5) {$\textcolor{black}{ \G}$};
 
\node at (-1.1,-2.85) {$\textcolor{red}{\g_{-1}}$};
\end{tikzpicture}
\end{center}
\vskip -0.3cm
\vspace{-5pt}
\caption{The spectral bands $\g_{\pm j}$ and the cycles $\mathrm{A}_{\pm j}, \mathrm{B}_{\pm j}.$
 The 1D Schwarz symmetrical curve $\G$ consists of the bands $\g_{\pm j}$, $j=0,\dots,N$, and the gaps $c_{\pm j}$ between the bands (the gaps are  not shown on this figure).}
\label{Fig:Cont}
\vspace{-5pt}
\end{figure}
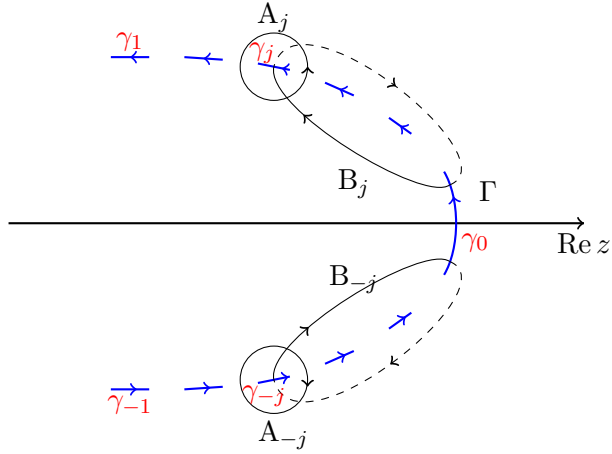

A mathematical approach for deriving  \eqref{eq-state} was presented in \cite{ET2020}. 
This derivation is based on the idea of the thermodynamic limit for a family of finite-gap solutions of the focusing NLS equation, which was originally developed for the KdV equation in \cite{El2003}. 
Finite-gap solutions are quasi-periodic functions in $x,t$ that can be spectrally represented by a finite
number of Schwarz symmetrical arcs (bands) on the complex $z$-plane. Here Schwarz symmetry means that 
a band $\g$ either coincides with its Schwarz symmetrical image, i.e. $\g = \bar \g$,  or  $\g$ and $\bar\g$ are distinct disjoint spectral bands of the finite-gap solution. Let  $\mathfrak R_N$ denotes a hyperelliptic Riemann surface based on $2N+1$ Schwarz symmetric bands $\g_n$, $|n|=0,1,\dots,N$, with only one of them,  $\g_0$, satisfying $\g_0\cap\R\not=\emptyset$.

We define the quasimomentum $dp_N$ and quasienergy $dq_N$ differentials on $\mathfrak R_{N}$ as the unique real normalized (all the periods of $dp_N,dq_N$ are real) meromorphic differentials of the second kind with poles only at $z=\infty$ on both sheets which admit local expansions
\begin{equation}\label{pq}
	dp_N \sim [\pm 1 + \mathcal{O}(z^{-2})]dz , \qquad dq_N \sim [\mp 4bbz+\mathcal{O}(z^{-2})]dz 
\end{equation}
near $z=\infty$ on the main and second sheet respectively (see e.g. \cite{ForLee, Krichever, BT15}).
As was shown in \cite{ET2020}, the wave numbers $k_j$, $\tilde k_j$ and the frequencies $w_j$, $\tilde w_j$ of a quasi-periodic
finite-gap solution determined by $\mathfrak R_{N}$ are given by
\begin{align}
	k_j &= 
	\oint_{\mathrm{A}_j}dp_N, \quad w_j = 
	\oint_{\mathrm{A}_j}dq_N, \quad |j|=1, \dots, N, \label{kdp}\\
	\tilde k_j &= \oint_{\mathrm{B}_j}dp_N, \quad \tilde w_j = \oint_{\mathrm{B}_j}dq_N, \quad |j| = 1, \dots, N, \label{omdq}
\end{align}
where the cycles $A_j,B_j$ are shown on Figure \ref{Fig:Cont}.  Due to symmetry, all the periods of the cycles $A_{\pm j}$,  $B_{\pm j}$ are equal. 

The Riemann bilinear relation
\be\label{RBR}
\sum_j\left[\oint_{A_j}\o_m\oint_{B_j}dp_N-\oint_{A_j}dp_N\oint_{B_j}\o_m\right]=2\pi i \sum {\rm Res}(\int \o_m dp_N),  ~~|m|=1,\dots,N,
\ee
where $\o_j$ are the normalized holomorphic differentials on  $\mathfrak{R}_N$, defines the first (discrete) system comprising the generalized (complex) Nonlinear Dispersion Relations  (NDR).  
Replacing $dp_N$ by $dq_N$ in \eqref{RBR}  provides the second system of generalized NDR.

Taking the real and imaginary parts of \eqref{RBR}, one gets
\bea\label{RBR-decouple}
\sum_j k_j \Im \oint_{B_j}\o_m=-2\pi \Re\left(  \sum {\rm Res}(\int \o_m dp_N)   \ri) \cr
\tilde k_m  -  \sum_j k_j \Re \oint_{B_j}\o_m=-2\pi \Im\left(  \sum {\rm Res}(\int \o_m dp_N)   \ri).
\eea
The system \eqref{RBR-decouple} effectively decouples the ``tilded" (carrier)  and untilded (solitonic) wavenumbers and frequencies.
Since the matrix $\Im \oint_{B_j}w_m$ is positive definite, equations \eqref{RBR} always have a unique solution.  
Equations connecting the wavenumbers and frequencies via the underlying Riemann surface $\mathfrak{R}_N$ explain the name NDR.
Note that if a certain  band $\g_m$ is shrinking, then  $ \Im \oint_{B_j}\o_m$ must grow and, thus, the corresponding $k_m$ must go to zero. Therefore, $k_m,\o_m$ are called solitonic wavenumbers/frequencies. 
The remaining  $\tilde k_m,\tilde\o_m$ are called carrier wavenumbers/frequencies. 

In the thermodynamic limit, where the number of bands is growing and the size of each band is shrinking much faster than $(1/N)$, see \cite{ET2020}, \cite{TW2022} for details, the equation \eqref{RBR} becomes  
\be \label{NDRg1}
\tilde u(z)-\int_{\G^+}g(z,w)u(w)d\l(w) + i\s(z)u(z)=z,\quad z\in\G^+,
\ee
where: $\l(w)$ is the reference measure on $\G^+$; 
the centers $z_j$  of shrinking bands $\g_j$, $j=1,\dots,N$, accumulate on a  compact  $\G^+\subset \C^+$ with the probability density $\rho(z)>0$;
 $g(z,w)$, $z,w\in \G^+$,  is the thermodynamic limit of the off-diagonal entries of the Riemann period matrix $\tau=\oint_{B_j}w_m$; 
 $2\tilde u(z)$ interpolates the values of $\tilde k_j$ on $ \G^+$;  
 $2\pi u(z)$  interpolates the values of $Nk_j\rho(z_j)$ on $ \G^+$; $\s(z)= \frac{2\nu(z)}{\rho(z)}$, and; $z$ is the thermodynamic limit of the rhs of \eqref{RBR}. Here we assume that  the bands $|\g_j|=e^{-N\nu_j (1+o(1))}$ and 
$\nu(z)$ interpolates $\nu_j \geq 0$ on $\G^+$.

The unknown functions $u$ and $\tilde u$ in \eqref{NDRg1}  are called solitonic and carrier densities of states (DOS) for the soliton gas, associated with a sequence $\mathfrak{R}_N$ and the 
$dp_N$, $dq_N$ on $\mathfrak{R}_N$. 
Similar to \eqref{NDRg1}, one derives equation for the densities of fluxes (DOF) $\tilde v(z), v(z)$ given by 
\be \label{NDRg2}
\tilde v(z)-\int_{\G^+}g(z,w)v(w)d\l(w)+ i\s(z)v(z)=-2z^2,  \quad z\in\G^+.
\ee 
Note that $u$ must satisfy $u(z)\geq 0$ whereas $v(z) $ is real on $\G^+$. The latter should also be true for $\tilde u(z),\tilde v(z)$.  One can now obtain equations (NDR) for the solitonic components of DOS and DOF 
\bea \label{NDR-sol}
-\int_{\G^+}\Im g(z,w)u(w)d\l(w) + \s(z)u(z)=\Im z,  \quad z\in\G^+, \cr
- \int_{\G^+}\Im g(z,w)v(w)d\l(w)+ \s(z)v(z)=-2\Im z^2,  \quad z\in\G^+,
\eea
by taking imaginary parts of the general NDR \eqref{NDRg1}-\eqref{NDRg2}.
According to \cite{ET2020}, \cite{TW2022},
 $\Im g(z,w)=- \Delta(z,w)$, see \eqref{Delta}. We want to emphasize that the DOS $u(z)$ represents the scaled  thermodynamic limit of the $A$-periods of $dp_N$ and  the DOF $v(z)$ the same for $dq_N$.
In the case of s focusing NLS soliton gas,  existence and uniqueness of solutions to \eqref{NDR-sol}, as well as the fact that $u(z)\geq 0$, was proven in \cite{KT21}, subject to mild restrictions on $\G^+$ and $\s$.

It is easy to observe that \eqref{eq-state} is a direct consequence of
\eqref{NDR-sol}, where $s(z)= \frac{v(z)}{u(z)}$.
Indeed, after multiplying   the first equation \eqref{NDR-sol} by $s(z)$,
substituting $v(z)=s(z)u(z)$ into  the second equation \eqref{NDR-sol},  subtracting the second equation from the first one
and dividing both parts by $\Im z$ we obtain exactly \eqref{eq-state}.

\section{Approximating $n$ shrinking phases by solitons}

A sequence of finite-gap spectral data $\mathcal{D}_{\mathrm{fg}}^{(n)} = (\Gamma_{G}^{(n)}, \Lambda^{(n)})$ satisfying Assumption~\ref{data.assumptions} for each $n \in 2 \N$, defines a sequence of finite-gap potentials $\psi\left(x,t\, \vert\, \mathcal{D}_{\mathrm{fg}}^{(n)} \right)$ via their associated RHP, RHP~\ref{rhp:n-phase}. 
For each $n \in 2\N$, let $\vect{M}^{(n)}(z;x,t)$ denote the solution of RHP~\ref{rhp:n-phase} with data $\mathcal{D}_{\mathrm{fg}}^{(n)}$. 
In this section, we describe a series of transformation which lead to an approximation of $\vect{M}^{(n)}(z;x,t)$ by the solution of a generalized soliton RHP---RHP~\ref{rhp:fg-gen}---with an explicit mapping from the given finite-gap spectral data $\mathcal{D}_{\mathrm{fg}}^{(n)}$ to soliton scattering data $\mathcal{D}_{\mathrm{sol}}^{(n)}$. 
The accuracy of this approximation depends on having estimates for the large $n$ behavior of the sequence of solutions to the soliton RHPs. We take this issue up in the following sections. 

\subsection{The one-cut model function: from micro-arcs to micro-loops}
The motivation for our approach is that at any particular $z$ which is far from the shrinking arcs---measured relative to the length of the shrinking arcs---the vanishing spectral curve $\gamma^{(n)}$ looks more and more like a set of discrete spectral points. However, locally, on the scale of their arclength, the shrinking arcs remain relevant. Following an idea similar in spirit to \cite{BG15}, we proceed by first deforming the jump contours away the arcs $\gamma^{(n)}$ onto loops that enclose each component arc $\gamma^{(n)}_{k}$ by introducing local model functions inside each loop, which solve the jump condition along $\gamma^{(n)}_k$, but no others.  

To define our local models, consider an arbitrary oriented simple arc $\gamma$ in the complex plane with initial and terminal endpoints $a$ and $b$ respectively. Define 
\begin{equation}
	\eta(z;\gamma) = \left( \frac{ z- b}{z-a} \right)^{1/4}, 
\end{equation}
to be analytic in $\C \setminus \gamma$ such that $\eta(z;\gamma) = 1 + \bigo{z^{-1}}$ as $z \to \infty$ and normalized so that $\eta_+(z;\gamma) = \ii \eta_-(z;\gamma)$ for $z \in \gamma$.  
\begin{prop}
Given a simple oriented arc $\gamma$ and a constant $\Lambda \in \C$, the function
\begin{equation}\label{one.cut}
	\vect{\psi}(z;\gamma,\Lambda) := \eee^{-\frac{\ii}{2} \Lambda \sigthree} \eta(z;\gamma)^{-\sigtwo} \eee^{\frac{\ii}{2} \Lambda \sigthree} 
	=\begin{bmatrix}
	\frac{\eta(z;\gamma)+\eta(z;\gamma)^{-1}}{2} & - \eee^{-\ii \Lambda} \frac{\eta(z;\gamma)-\eta(z;\gamma)^{-1}}{2\ii} \\
	\eee^{\ii \Lambda} \frac{\eta(z;\gamma)-\eta(z;\gamma)^{-1}}{2\ii} & \frac{\eta(z;\gamma)+\eta(z;\gamma)^{-1}}{2} 
	\end{bmatrix} 
\end{equation}
is the unique function analytic in $\C \setminus \gamma$ with the following properties:
\begin{enumerate}[label = \roman*.]
	\item $\vect{\psi}(z;\gamma,\Lambda) = I + \bigo{z^{-1}}$ as $z \to \infty$.
	\item $\vect{\psi}(z;\gamma,\Lambda)$ admits at worst ${1/4}$-root singularities at the endpoints of $\gamma$. 
	\item For $z \in \gamma$, the boundary values $\vect{\psi}_\pm(z;\gamma, \Lambda)$ satisfy the jump relation
	\begin{equation}
		\vect{\psi}_+(z;\gamma,\Lambda) = \vect{\psi}_-(z;\gamma,\Lambda) (-\ii \sigtwo) e^{i \Lambda \sigthree}, \qquad z \in \gamma.
	\end{equation}	
\end{enumerate}
Defining $z_0 = \tfrac{1}{2}(b+a)$ and $\delta = \tfrac{1}{2}(b-a)$, for $z$ satisfying $|z-z_0| > \max_{w \in \gamma} |w-z_0|$ we have
	\begin{equation}\label{one.cut.expand}
		\vect{\psi}(z;\gamma,\Lambda) = \left( \sum_{\ell=0}^\infty \eta_{2\ell} \frac{\delta^{2\ell}}{(z-z_0)^{2\ell} } \right) \vect{I} 
		- \left( \sum_{\ell=0}^\infty \eta_{2\ell+1} \frac{\delta^{2\ell+1}}{(z-z_0)^{2\ell+1} } \right)    \sigtwo \eee^{\ii \Lambda \sigthree}
	\end{equation}
	where $\eta_j$ are the Maclaurin series coefficients 
	\[
		\left( \frac{1-x}{1+x} \right)^{1/4} = \sum_{\ell=0}^\infty \eta_\ell \, x^\ell = 1 - \frac{1}{2} x + \bigo{x^2} 
	\]
\end{prop}

\begin{proof}
	That \eqref{one.cut} satisfies property $i.$ follows from inserting the large $z$ expansion of $\eta(z;\gamma)$ into \eqref{one.cut}.
	Property $ii.$ is obvious. That \eqref{one.cut} satisfies the jump condition $iii$. follows from inserting the jump relation $\eta_+(z;\gamma) = \ii \eta_-(z;\gamma)$ into \eqref{one.cut}.  
	Uniqueness follows from a standard Liouville-type argument which relies on the proscribed growth bounds at each endpoint of $\gamma$. 
	Finally, \eqref{one.cut.expand} follows from writing $\eta(z;\gamma) = \left( \tfrac{ z-z_0 - \delta}{z-z_0+\delta}\right)^{1/4}$ and using the Laurent expansion about $z_0$.
\end{proof}

To define our first transformation we patch together many of these local models, one around each of the shrinking bands in $\gamma^{(n)}$. 
Since the arcs shrink faster than they accumulate, we introduce
\begin{equation}
	\hat \gamma^{(n)} := \bigcup_{k=1}^{n} \hat \gamma_{k}^{(n)}, 
\end{equation}
a collection of $n$ nonintersecting simple closed arcs such that each $\hat \gamma_{k}^{(n)}$ encloses its respective shrinking arc $\gamma_{k}^{(n)}$. 
Condition \eqref{band.gap.ratio.rate} in Assumption~\ref{data.assumptions} guarantees that we can always choose 
the loops $\hat \gamma^{(n)}$ such that there exists a constant $c>0$ independent of $k,n$, such that
\begin{equation}\label{loop.dist}
	c \, r_n \leq \min_{1\leq k \leq n} \dist( \hat \gamma_{k}^{(n)}, \gamma^{(n)} ) \leq c^{-1} r_n
\end{equation}
We orient each $\hat \gamma^{(n)}_{k}$ positively (counterclockwise). 

Our sequence of finite-gap RHPs $\vect{M}^{(n)}$ have a jump on each shrinking arc $\gamma^{(n)}_k$ given by \eqref{fg.jump} parameterized by the complex constant
$\Lambda^{(n)}_{k} := \Lambda^{(n)}(z) \big\vert_{z \in \gamma_k^{(n)}}$. Define
\begin{gather}
	\vect{\psi}^{(n)}_{k}(z;x,\vect{t}) := \eee^{-\ii \varphi(z;x,\vect{t}) \sigthree}\, \vect{\psi}(z; \gamma^{(n)}_{k}, \Lambda^{(n)}_{k}) \,\eee^{\ii \varphi(z;x,\vect{t}) \sigthree}, \qquad k=1,\dots,n
\end{gather}
where $\varphi(z;x,\vect{t})$ is given by \eqref{dispersive.phase}.
Using this collection of functions and the loop contours $\hat \gamma^{(n)}_{k}$, define the change of variables
\begin{equation}\label{O.def}
	\begin{gathered}
	\vect{O}^{(n)}(z;x,\vect{t}) := \vect{M}^{(n)}(z;x,\vect{t}) \vect{\Psi}^{(n)}(z;x,\vect{t})^{-1},
	\\
	\vect{\Psi}^{(n)}(z;x,\vect{t}) := 
		\begin{cases}
		\vect{\psi}_{k}^{(n)}(z;x,\vect{t}) & z \in \inside(\hat \gamma^{(n)}_{k}), \qquad  k=1,\dots,n\\
		\vect{I} &\text{elsewhere}
	\end{cases}
	\end{gathered}
\end{equation}

\begin{figure}[ht]
	\includegraphics[width=.8\textwidth]{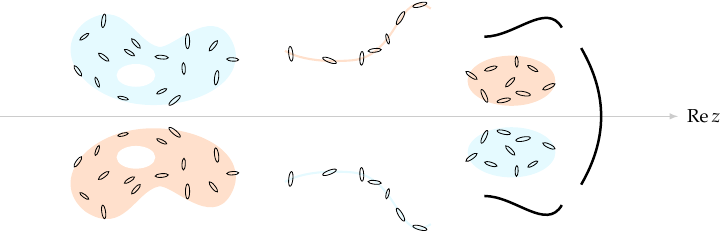}
	\caption{
	The collection of $n$ shrinking loops $\hat{\gamma}^{(n)}$ enclosing each of the $n$ shrinking bands $\gamma^{(n)}$ defining the spectrum of the sequence of finite-gap potentials. These loops together with $\Gamma_G$ (thick black contour) are the contour defining RHP~\ref{rhp:O.n}.
	}
	\label{fig:microloops}
\end{figure}

The new unknown $\vect{O}^{(n)}$ solves the following problem:
\begin{RHP}\label{rhp:O.n}
	Find a $2\times 2$ matrix-valued function $\vect{O}^{(n)}(\, \cdot \,; x, \vect{t})$ such that for each fixed $(x,\vect{t} ) \in \R \times \R^\mathcal{K}$:
	\begin{enumerate}[align=left, leftmargin=*]
	\item[ \textbf{Analyticity}:] $\vect{O}^{(n)}(z;x,\vect{t})$ is analytic for $z \in \C \setminus \left( \hat \gamma^{(n)} \cup \Gamma_G \right)$. 
	\item[ \textbf{Normalization}:] $\vect{O}^{(n)}(z;x,\vect{t}) = \vect{I} + \bigo{z^{-1}}$ as $z \to \infty$. 
	\item[ \textbf{Endpoint growth}:] At any endpoint $p$ of $\Gamma_G$ the matrix $\vect{O}^{(n)}$ admits $\tfrac{1}{4}$-root singularities:
	\[
		\vect{O}^{(n)}(z;x,\vect{t}) = \bigo{ |z-p|^{-1/4} }, \qquad \text{ as } z \to p \in \partial \Gamma_G.
	\]
	\item[ \textbf{Jump Condition}:] $\vect{O}^{(n)}$ has continuous non-tangential boundary values $\vect{O}^{(n)}_\pm(z;x,t)$ as $z \to \Gamma_G \cup \hat \gamma^{(n)}$ satisfying the jump relation
	\begin{equation}
	\begin{gathered}
		\vect{O}^{(n)}_+(z;x,\vect{t}) =\vect{O}^{(n)}_-(z;x,\vect{t})  \vect{V}^{(n)}_O(z;x,\vect{t}), \\ 
		\vect{V}^{(n)}_O(z;x,\vect{t}) = \begin{cases}
			(-\ii \sigtwo) \eee^{2 \ii (\varphi(z;x,\vect{t} ) + \Lambda^{(n)}(z) )\sigthree} & z \in \Gamma_G, \\
			\vect{\psi}^{(n)}_k(z;x,\vect{t})^{-1} & z \in \hat \gamma^{(n)}_{k}, \quad k = 1,\dots,n.
		\end{cases}
	\end{gathered}
	\end{equation}
\end{enumerate}
\end{RHP}

Consider the jump $\vect{V}^{(n)}_O(z;x, \vect{t})$ restricted to any of the shrinking loops $\hat \gamma^{(n)}_{k}$, $k=1,\dots,n$. For $z \in \hat \gamma^{(n)}_k$, it follows from \eqref{one.cut.expand} that $\vect{V}^{(n)}_O(z;x, \vect{t})$  admits the convergent expansion 
\begin{multline}\label{VO.1} 
	\vect{V}^{(n)}_O(z;x,\vect{t}) \restrict{z \in \hat \gamma^{(n)}_{k}}\quad =  \vect{\psi}_{k}^{(n)}(z;x,\vect{t})^{-1} = \\
	\eee^{-\ii \varphi(z;x,\vect{t})\sigthree}
	\begin{bmatrix} 
		1+\sum_{\ell=1}^\infty \eta_{2\ell} \left(\frac{\delta_{k}^{(n)}}{z-\zeta^{(n)}_{k} }\right)^{2\ell}  & 
		-\ii \eee^{-\ii \Lambda_{k}^{(n)}} \sum_{\ell=0}^\infty \eta_{2\ell+1}   \left(\frac{\delta^{(n)}_{k}}{z-\zeta^{(n)}_{j,k}} \right)^{2\ell+1}   \\[1em]
		\ii \eee^{\ii \Lambda_{k}^{(n)}} \sum_{\ell=0}^\infty \eta_{2\ell+1}  \left(\frac{\delta^{(n)}_{k}}{z-\zeta^{(n)}_{k}} \right)^{2\ell+1}   &
		1+\sum_{\ell=1}^\infty \eta_{2\ell} \left(\frac{\delta_{k}^{(n)}}{z-\zeta^{(n)}_{k} }\right)^{2\ell} 
	\end{bmatrix}
	\eee^{\ii \varphi(z;x,\vect{t})\sigthree},
\end{multline}
where the convergence follows from the lower bound in \eqref{loop.dist}.

Assumption~\ref{assume.shrinkrate} (cf.  \eqref{band.gap.ratio}) and \eqref{loop.dist} then imply that on any of the shrinking loops
\begin{align}\label{VO.expand}
	\vect{V}^{(n)}_O(z;x,\vect{t}) \restrict{z \in \hat \gamma^{(n)}_{k}}\ =  \mathbf{I} + 
	\begin{bmatrix} 
		\bigo{\varsigma(n)^2} & 
		\ii \frac{\delta^{(n)}_{k} \eee^{-\ii \Lambda^{(n)}_{k}} \eee^{-2\ii \varphi(z;x,\vect{t})}  }{2(z-\zeta^{(n)}_{k} )}  \left[1 + \bigo{\varsigma(n)^2} \right] 
		\\[1em]
		-\ii \frac{\delta^{(n)}_{k} \eee^{\ii \Lambda^{(n)}_{k}} \eee^{2\ii \varphi(z;x,\vect{t})}  }{2(z-\zeta^{(n)}_{k} )}  \left[1 + \bigo{\varsigma(n)^2} \right] &
		\bigo{\varsigma(n)^2}.
	\end{bmatrix}.
\end{align}

\subsection{Introducing a $(n/2)$-soliton model problem for each finite $n$}
The observation that as each band shrinks to a point, the spectral curve looks increasingly like a set of discrete eigenvalues holds true at the geometric level, but to see solitons emerge it must also hold on the level of the jump matrices. 
The diagonal entires in \eqref{VO.1} are asymptotically near 1 as $n \to \infty$, but the behavior of the off-diagonal entries depend on the asymptotic behavior of the exponential terms $\eee^{\ii \Lambda_{k}^{(n)}}$ as $n \to \infty$.
One implication of condition \eqref{Lambda.cond} in Assumption~\ref{data.assumptions} is that on each loop $ \hat \gamma^{(n)}_{k}$, one of the two off-diagonal entires in \eqref{VO.1} is always negligible. 
Formally setting the negligible entry to zero, the leading order terms in the remaining triangular jump are equivalent, up to a local contour deformation, to a problem having a simple pole at the band center $\zeta^{(n)}_k$. We make this idea precise below. 

For each $n\in 2\N$, recall that $\mathcal{Z}^{(n)}_\mathrm{low}$ denotes the set of midpoints for those shrinking band $\gamma^{(n)}_k$ which accumulate on the set $\Alow$ while $\mathcal{I}{(n)}$ denotes the index set of those $k$'s for which $\zeta^{(n)}_k \in \mathcal{Z}^{(n)}_\mathrm{low}$ (\cf~\eqref{Zn.def}). 
It follows from the Schwarz symmetry of the function $\Lambda$ and contour $\gamma$ that the midpoints in $\overline{\mathcal{Z}^{(n)}_\mathrm{low}}$ accumulate on $\mathcal{A}^{\mathrm{up}}$ as $n \to \infty$. 
Define
\begin{equation}\label{I.components}
	\mathcal{I}^{(n)}_\mathrm{up} := \left\{ k \,:\, 1\leq k \leq n, \ \zeta^{(n)}_k \in \overline{\mathcal{Z}^{(n)}}  \right\} 
	= \{1, \dots,  n \} \setminus \mathcal{I}^{(n)}_\mathrm{low} .
\end{equation}
For $k \in \mathcal{I}^{(n)}_\mathrm{low}$, \eqref{VO.expand} is equivalent to 
\begin{subequations}\label{VO.factor.lower} 
\begin{gather}
	\vect{V}^{(n)}_O(z;x,\vect{t}) \restrict{z \in \hat \gamma^{(n)}_{k}}\ 
	=\quad \eee^{-\ii \varphi(z;x,\vect{t})\sigthree} \vect{ \breve \Psi}_{k}^{(n)}(z)^{-1}
	\begin{bmatrix}
	1 & 0 \\ -\ii \frac{\delta^{(n)}_{k}\eee^{\ii \Lambda_{k}^{(n)}} }{2(z-\zeta^{(n)}_{k} )} & 1 
	\end{bmatrix}
	\eee^{\ii \varphi(z;x,\vect{t})\sigthree}  
\shortintertext{where} 
	 \vect{ \breve \Psi}_{k}^{(n)}(z)^{-1} = 
	 \begin{bmatrix} 
		1+\sum\limits_{\ell=1}^\infty (\eta_{2\ell} - \eta_1 \eta_{2\ell-1}) 
		\left(\frac{\delta_{k}^{(n)}}{z-\zeta^{(n)}_{k} }\right)^{2\ell}  & 
		-\ii  \eee^{-\ii \Lambda_{k}^{(n)}} \sum\limits_{\ell=0}^\infty \eta_{2\ell+1} 
		\left(\frac{\delta^{(n)}_{k}}{z-\zeta^{(n)}_{k}} \right)^{2\ell+1}   \\[1em]
		\ii \eee^{\ii \Lambda_{k}^{(n)}}  \sum\limits_{\ell=1}^\infty (\eta_{2\ell+1} - \eta_1 \eta_{2\ell}) 
		\left(\frac{\delta^{(n)}_{k}}{z-\zeta^{(n)}_{k}} \right)^{2\ell+1}   &
		1+\sum\limits_{\ell=1}^\infty \eta_{2\ell} \left(\frac{\delta_{k}^{(n)}}{z-\zeta^{(n)}_{k} }\right)^{2\ell} 
	\end{bmatrix}
\end{gather}
\end{subequations}

For $k \in \mathcal{I}^{(n)}_\mathrm{up}$, \eqref{VO.expand} is equivalent to 
\begin{subequations}\label{VO.factor.upper} 
\begin{gather}
	\vect{V}^{(n)}_O(z;x,\vect{t}) \restrict{z \in \hat \gamma^{(n)}_{k}}\ 
	= \quad \eee^{-\ii \varphi(z;x,\vect{t})\sigthree} \vect{ \widehat \Psi}_{k}^{(n)}(z)^{-1}
	\begin{bmatrix}
	1 & \ii \frac{\delta^{(n)}_{k} \eee^{-\ii \Lambda_{k}^{(n)}} }{2(z-\zeta^{(n)}_{k} )} \\ 0 & 1 
	\end{bmatrix}
	\eee^{\ii \varphi(z;x,\vect{t})\sigthree}  
\shortintertext{where}
	 \vect{ \widehat \Psi}_{k}^{(n)}(z)^{-1} = 
	 \begin{bmatrix} 
		1+\sum\limits_{\ell=1}^\infty \eta_{2\ell} \left(\frac{\delta_{k}^{(n)}}{z-\zeta^{(n)}_{k} }\right)^{2\ell}  & 
		-\ii \eee^{-\ii \Lambda_{k}^{(n)}} \sum\limits_{\ell=1}^\infty  (\eta_{2\ell+1} -\eta_1 \eta_{2\ell})  
		\left(\frac{\delta^{(n)}_{j,k}}{z-\zeta^{(n)}_{k}} \right)^{2\ell+1}   \\[1em]
		\ii \eee^{\ii \Lambda_{k}^{(n)}} \sum\limits_{\ell=0}^\infty \eta_{2\ell+1} 
		\left(\frac{\delta^{(n)}_{k}}{z-\zeta^{(n)}_{k}} \right)^{2\ell+1}   &
		1+\sum\limits_{\ell=1}^\infty (\eta_{2\ell} - \eta_1 \eta_{2\ell-1}) 
		\left(\frac{\delta_{k}^{(n)}}{z-\zeta^{(n)}_{k} }\right)^{2\ell} 
	\end{bmatrix}
\end{gather}
\end{subequations}

\begin{prop} \label{prop:one.cut.error}
If conditions \ref{assume.shrinkrate}-\ref{assume.phases} in Assumption~\ref{data.assumptions} are satisfied, then the expressions $\vect{ \breve \Psi}_{k}^{(n)}(z)^{-1}$ and $ \vect{ \widehat \Psi}_{k}^{(n)}(z)^{-1}$ defined by \eqref{VO.factor.lower}-\eqref{VO.factor.upper} are near-identity matrices satisfying
\begin{equation}
\begin{aligned}
	&\vect{ \breve \Psi}_{k}^{(n)}(z)^{-1} - \vect{I}  =  
	\begin{bmatrix} 
	   \bigo{ \varsigma(n)^2} & \bigo{ \varsigma(n)^2 n r_n  } \\[0.5em] 
	   \bigo{ \frac{\varsigma(n)^2}{n r_n} } & \bigo{\varsigma(n)^2 } 
	 \end{bmatrix},
	 && &z \in \hat \gamma^{(n)}_{k}, \ k \in \mathcal{I}^{(n)}_\mathrm{low}, \\
	&\vect{ \widehat \Psi}_{k}^{(n)}(z)^{-1} - \vect{I}  =  
	\begin{bmatrix} 
	   \bigo{ \varsigma(n)^2} & \bigo{ \frac{\varsigma(n)^2}{n r_n} } \\[0.5em] 
	   \bigo{ \varsigma(n)^2 n r_n  } & \bigo{\varsigma(n)^2 } 
	 \end{bmatrix},
	&& &z \in \hat \gamma^{(n)}_{k}, \ k \in \mathcal{I}^{(n)}_\mathrm{up}.
\end{aligned}
\end{equation}
\end{prop}
\begin{proof}
	Consider $k \in \mathcal{I}^{(n)}_\mathrm{low}$. Then the infinite sums in formula \eqref{VO.factor.lower} for $\vect{ \breve \Psi}_{k}^{(n)}(z)$ are absolutely convergent for $z \in \hat \gamma^{(n)}_k$.
	The diagonal entry estimates follow immediately from \eqref{band.gap.ratio} and \eqref{loop.dist}. 
	For the off-diagonal entries \eqref{A+} holds. Since $\beta$ is a bounded nonzero function, we have $\left| \delta^{(n)}_k \eee^{\ii \Lambda^{(n)}_k} \right| = \bigo{n^{-1}}$ so that  
\begin{gather*}
	\left| \vect{ [\breve \Psi}_{k}^{(n)}(z)^{-1}]_{21} \right| = \bigo{ \frac{\delta^{(n)}_k \eee^{\ii \Lambda_k^{(n)}}}{z-\zeta_k} \varsigma(n)^{2}} =
	\bigo{ \frac{ \varsigma(n)^{2}}{n r_n}}, \\
	\left| \vect{ [\breve \Psi}_{k}^{(n)}(z)^{-1}]_{12} \right| = \bigo{ \frac{z-\zeta_k}{\delta^{(n)}_k \eee^{\ii \Lambda_k^{(n)}}} \varsigma(n)^{2}} =
	\bigo{ n r_n \varsigma(n)^{2}}. \\
\end{gather*}
For $k \in \mathcal{I}^{(n)}_\mathrm{up}$ \eqref{A-} holds, and the corresponding estimates for $\vect{ \breve \Psi}_{k}^{(n)}(z)$ are obtained similarly.
\end{proof}

These observations motivate the following model problem in which we simply replace the factors $\vect{ \breve \Psi}_{j,k}^{(n)}$ and $\vect{ \widehat \Psi}_{j,k}^{(n)}$ with identity on each shrinking loop:
\begin{RHP}\label{rhp:Pn}
	Find a $2\times 2$ matrix valued function $\vect{P}^{(n)}(z; x,\vect{t})$ with the following properties
	\begin{enumerate}[align=left, leftmargin=*]
	\item[ \textbf{Analyticity}:] $\vect{P}^{(n)}(z;x,\vect{t})$ is analytic for $z \in \C \setminus \left( \hat \gamma^{(n)} \cup \Gamma_G \right)$. 
	\item[ \textbf{Normalization}:] $\vect{P}^{(n)}(z;x,\vect{t}) = \vect{I} + \bigo{z^{-1}}$ as $z \to \infty$. 
	\item[ \textbf{Endpoint growth}:] At any endpoint $p$ of $\Gamma_G$ the matrix $\vect{P}^{(n)}$ admits $\tfrac{1}{4}$-root singularities:
	\[
		\vect{P}^{(n)}(z;x,\vect{t}) = \bigo{ |z-p|^{-1/4} }, \qquad \text{ as } z \to p \in \partial \Gamma_G.
	\]
	\item[ \textbf{Jump Condition}:] $\vect{P}^{(n)}$ has continuous non-tangential boundary values $\vect{P}^{(n)}_\pm(z;x,t)$ as $z \to \Gamma_G \cup \hat \gamma^{(n)}$ satisfying the jump relation
	\begin{equation}\label{Pn.jump}
	\begin{gathered}
		\vect{P}^{(n)}_+(z;x,\vect{t}) =\vect{P}^{(n)}_-(z;x,\vect{t})  \vect{V}_P^{(n)}(z;x,\vect{t}) \\ 
		\vect{V}_P^{(n)}(z;x,\vect{t}) = \begin{cases}
			(-\ii \sigtwo) \eee^{2\ii (\varphi(z;x,\vect{t}) + \Lambda^{(n)}(z)) \sigthree} & z \in \Gamma_G \\
			\begin{bmatrix}
			   1 & 0 \\ -\ii \frac{\delta^{(n)}_{k}\eee^{\ii \Lambda_{k}^{(n)}}}{2(z-\zeta^{(n)}_{k} )}  \eee^{2\ii \varphi(z;x,\vect{t})}  & 1 
			\end{bmatrix} 
			& z \in \hat \gamma^{(n)}_{k}, \quad k \in  \mathcal{I}^{(n)}_\mathrm{low} \\
			\begin{bmatrix}
			  1 &  \ii \frac{\delta^{(n)}_{k} \eee^{-\ii \Lambda_{k}^{(n)}}  }{2( z-\zeta^{(n)}_{k} ) } \eee^{-2\ii \varphi(z;x,\vect{t})}  \\ 0 & 1 
			\end{bmatrix}
			& z \in \hat \gamma^{(n)}_{k}, \quad k \in  \mathcal{I}^{(n)}_\mathrm{up}
		\end{cases}
	\end{gathered}
	\end{equation}
\end{enumerate}
\end{RHP}

The model problem we have constructed in RHP~\ref{rhp:Pn} is equivalent up to a transformation on a compact set in the complex $z$ plane to RHP~\ref{rhp:fg-gen}
describing the evolution of generalized $n/2$-soliton potential on a genus $G$ finite-gap background. This potential is encoded in the scattering data 
\begin{gather}\label{data.gensol.n}
	\mathcal{D}^{(n)}_\mathrm{sol} = \left( \Gamma_G, \Lambda^{(n)} ; \mathcal{Z}^{(n)}_\mathrm{low}, c^{(n)} \right)
\end{gather}
where the norming constant function $c^{(n)}$ is given by  \eqref{Zn.def}.
Specifically, the transformation
\begin{equation}
\mathbf{S}^{(n)}(z;x,\vect{t}) = \begin{cases}
			\mathbf{P}^{(n)}(z;x,t)
			\begin{bmatrix}
			   1 & 0 \\ \ii \frac{\delta^{(n)}_{k}\eee^{\ii \Lambda_{k}^{(n)}}}{2(z-\zeta^{(n)}_{k} )}  \eee^{2\ii \varphi(z;x,\vect{t})}  & 1 
			\end{bmatrix} 
			& z \in \inside(\hat \gamma^{(n)}_{k}), \quad k \in  \mathcal{I}^{(n)}_\mathrm{low} \\
			\mathbf{P}^{(n)}(z;x,t) 
			\begin{bmatrix}
			  1 & - \ii \frac{\delta^{(n)}_{k} \eee^{-\ii \Lambda_{k}^{(n)}}  }{2( z-\zeta^{(n)}_{k} ) } \eee^{-2\ii \varphi(z;x,\vect{t})}  \\ 0 & 1 
			\end{bmatrix}
			& z \in \inside(\hat \gamma^{(n)}_{k}), \quad k \in  \mathcal{I}^{(n)}_\mathrm{up} \\
			\mathbf{P}^{(n)}(z;x,t) & \text{elsewhere}
		\end{cases}
\end{equation}
has the effect of contracting the jump contours along each loop $\hat \gamma_k^{(n)}$ to the point $\zeta_k^{(n)}$, so that the new unknown $\mathbf{S}^{(n)}$ has a simple pole at each $\zeta_k^{(n)}$. The new function $\mathbf{S}^{(n)}(z;x,\vect{t})$ satisfies the generalized soliton problem, RHP~\ref{rhp:fg-gen}, with scattering data $\mathcal{D}^{(n)}_\mathrm{sol}$ given by \eqref{data.gensol.n}. Since $\mathbf{P}^{(n)}(z;x,t) = \mathbf{S}^{(n)}(z;x,t)$ for all sufficiently large $z$, it follows that 
\begin{equation}\label{Pn.recover}
	2\ii\lim_{z\to \infty} z \vect{P}^{(n)}_{12}(z;x,\vect{t}) = \psi(x,\vect{t} \,\vert\, \mathcal{D}^{(n)}_\mathrm{sol}),
\end{equation}
where $\psi(x,\vect{t} \,\vert\, \mathcal{D}^{(n)}_\mathrm{sol})$ is the generalized soliton solution of the NLS hierarchy described by RHP~\ref{rhp:fg-gen}. %
\subsubsection{Accuracy of the soliton model problem for finite $n$} %
Let $\vect{E}^{(n)}(z;x,\vect{t})$ denote the finite $n$ error problem, defined as the ratio between the solutions of the exact and model problems:
\begin{equation}\label{En.def}
	\vect{E}^{(n)}(z;x,\vect{t}) := \vect{O}^{(n)}(z;,x,\vect{t}) \vect{P}^{(n)}(z;,x,\vect{t})^{-1}.
\end{equation}
Then $\vect{E}^{(n)}$ satisfies the following problem

\begin{RHP}\label{rhp:En}
	Find a $2\times 2$ matrix valued function $\vect{E}^{(n)}( z; x,\vect{t})$ with the following properties
	\begin{enumerate}[align=left, leftmargin=*]
	\item[ \textbf{Analyticity}:] $\vect{E}^{(n)}(z;x,\vect{t})$ is analytic for $z \in \C \setminus  \hat \gamma^{(n)} $. 
	\item[ \textbf{Normalization}:] $\vect{E}^{(n)}(z;x,\vect{t}) = \vect{I} + \bigo{z^{-1}}$ as $z \to \infty$. 
	
\item[ \textbf{Jump Condition}:] $\vect{E}^{(n)}$ has continuous non-tangential boundary values $\vect{E}^{(n)}_\pm(z;x,\vect{t})$ as $z \to  \hat \gamma^{(n)}$ satisfying the jump relation
	\begin{equation}\label{En.jump}
		\begin{gathered}
		\vect{E}^{(n)}_+(z;x,\vect{t}) =\vect{E}^{(n)}_-(z;x,\vect{t}) \vect{V}_E^{(n)}(z;x,\vect{t}),   \\
		\vect{V}^{(n)}_E(z;x,\vect{t}) = \vect{P}^{(n)}_-(z;x,\vect{t}) \eee^{-\ii \varphi(z;,x,\vect{t}) \sigthree}   \vect{v}^{(n)}_E(z)  \eee^{\ii \varphi(z;,x,\vect{t}) \sigthree} \vect{P}^{(n)}_-(z;x,\vect{t})^{-1}, \\ 
		\vect{v}^{(n)}_E(z;x,\vect{t}) = \begin{cases}
			\vect{\breve{\Psi}}^{(n)}_{k}(z)^{-1}   & z \in \gamma^{(n)}_{j,k}, \quad k \in \mathcal{I}^{(n)}_\mathrm{low}, \\
			\vect{ \widehat{\Psi}}^{(n)}_{k}(z)^{-1}   & z \in \gamma^{(n)}_{j,k},  \quad k \in \mathcal{I}^{(n)}_\mathrm{up}.
		\end{cases}
		\end{gathered}
	\end{equation}
\end{enumerate}
\end{RHP}

Our goal is to show that the error $\vect{E}^{(n)}(z;x,\vect{t})$ is near identity by proving that it satisfies a small-norm Riemann-Hilbert problem. From 
Prop.~\ref{prop:one.cut.error} we have
\begin{equation}\label{VEn.est.1}
	\left| \vect{V}^{(n)}_E(z;x,\vect{t}) - \vect{I} \right| = \| \vect{P}_-^{(n)}( \cdot;x,\vect{t}) \|^2_{L^{\infty}(\hat \gamma^{(n)})} \eee^{ 2| \Im \varphi(z;x,\vect{t}) | } \, \varsigma (n)^2 \left( n r_n + \frac{1}{n r_n} \right),
	\quad z \in \hat \gamma^{(n)}.
\end{equation}
Here we've used that any solution of RHP~\ref{rhp:Pn} has $\det \vect{P}^{(n)} = 1$ to equate $\| \vect{P}^{(n)}(, \cdot\, ;x,\vect{t})^{-1} \| = \| \vect{P}^{(n)}( \,\cdot \, ;x,\vect{t}) \|$. It follows that if we can show that the model problem $\vect{P}^{(n)}(z;x,\vect{t})$  is bounded, or at least does not grow too quickly in $n$, then $\vect{E}^{(n)}(z;x,\vect{t})$ will satisfy a small-norm Riemann-Hilbert problem. We can then show the solution $\vect{E}^{(n)}$ exists and expand it asymptotically for $n$ large.

\section{From finite model to infinite model}
In the previous section we described an algebraic construction mapping the RHPs for a sequence of finite-gap potentials to a related sequence of RHPs corresponding to generalized soliton potentials. 
The accuracy of this model relies on knowing the rate at which the solution, $\mathbf{P}^{(n)}(z;x,t)$, of the generalized soliton model RHP~\ref{rhp:Pn} grows as the number of solitons $(n/2)$ goes to infinity. 
Direct estimates of the solution $\mathbf{P}^{(n)}(z;x,t)$ are difficult due to its increasing complexity as the number of solitons grows. 
In this section we show how both of these problems are solved by introducing a continuous model to approximate the generalized soliton problem as $n \to \infty$. 
Under the conditions assumed in Assumption~\ref{data.assumptions} we show that in the large $n$ limit, the sequence of generalized solitons is well approximated by an $n$-independent model (\cf\ RHP~\ref{rhp:Pinfty}) described by a RHP of the  primitive potential class described in RHP~\ref{rhp:accumulated}. 
The accuracy of this approximation is taken up in the following section.

\begin{figure}[t]
\label{fig:superloops}
\begin{overpic}[width=.7\textwidth]{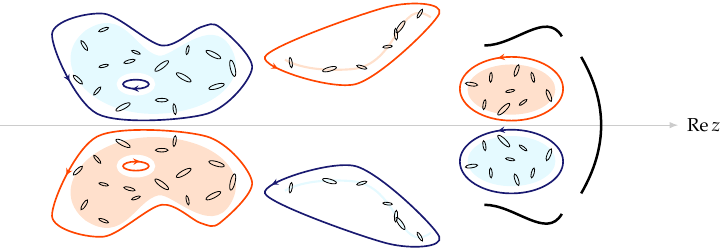}
 \put(1,30){\color{MidnightBlue}{\scalemath{.8}{\widehat{\Gamma}^{(\infty)}}}}
 \put(1,1){\color{OrangeRed}{\scalemath{.8}{\overline{\widehat{\Gamma}^{(\infty)}}}}}
 \put(42,12){\color{MidnightBlue}{\scalemath{.8}{\widehat{\Gamma}^{(\infty)}}}}
 \put(42,19){\color{OrangeRed}{\scalemath{.8}{\overline{\widehat{\Gamma}^{(\infty)}}}}}
 \put(59,8){\color{MidnightBlue}{\scalemath{.8}{\widehat{\Gamma}^{(\infty)}}}}
 \put(59,25){\color{OrangeRed}{\scalemath{.8}{\overline{\widehat{\Gamma}^{(\infty)}}}}}
\end{overpic}
\caption{An example of ``super-loops"  enclosing the accumulations domains (orange and blue regions) on which the bands of the finite-gap potentials accumulate as $n \to \infty$. Orientation is such that the accumulation domains $\mathcal{A}$ always lie in the interior of the super-loops. For non-simply connected regions, like leftmost regions above, the super-loop enclosing this region has multiple components. The small black ``micro-loops" depict $\hat{\gamma}^{(n)}$; the collection of shrinking loops enclosing each bands of the finite-gap potentials. 
 }
\end{figure}
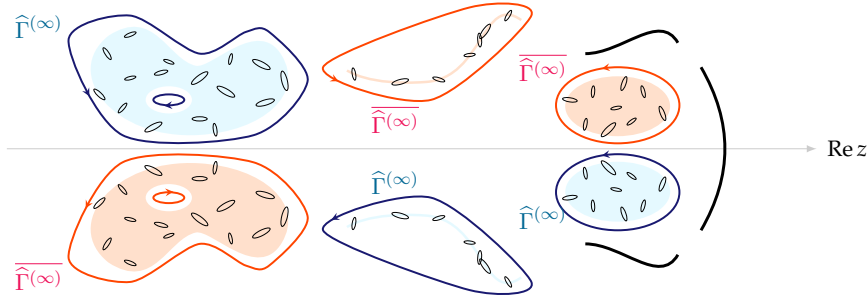
\subsection{Interpolating multi-solitons along super loops} %
Thus far, the geometry of the accumulation set $\mathcal{A}$ has played little to no role in our analysis. Here we make use of the conditions in Assumption~\ref{data.assumptions}.
Specifically, $\mathcal{A}$ can be expressed as the union of two disjoint Schwarz symmetric compact sets $\Alow$ and $\Aup$ such that \eqref{Lambda.cond} holds. 
The sets $\Alow$ and $\Aup$ consists of finitely many disjoint, compact, connected components with piecewise smooth boundaries. 
These components can be both one and two dimensional.  
For simplicity of the initial presentation, we will assume that $\mathcal{A}$ lies a positive distance from the real line (where $\Alow$ and $\Aup$ would necessarily come into contact) and from $\Gamma_G$, the fixed contour describing the background potential.  We will discuss later how to remove this assumption and allow $\partial \mathcal{A}$ to make finitely many, non-tangential, intersections with $\R \cup \Gamma_G$. 

As $\mathcal{A}$ consists of finitely many non-intersecting compact sets, we can introduce a finite collection of ``super-loops'' such that each component of the collection encloses a single connected component of the accumulation set $\mathcal{A}$ in its interior, see Figure~\ref{fig:superloops}. Denote by $\widehat \Gamma^{(\infty)}$ the set of simple nonintersecting closed loops which enclose $\Alow$. By symmetry $\overline{\widehat\Gamma^{(\infty)}}$ encloses $\Aup$. 
Because we assume that $\dist(\mathcal{A}, \Gamma_G \cup \R)>0$ we can arrange that $\widehat\Gamma^{(\infty)}$ and $\overline{\widehat\Gamma^{(\infty)}}$ are disjoint and that $\dist( \widehat \Gamma^{(\infty)}, \mathcal{A}) > 0$.

To extract the leading order asymptotic behavior from our generalized ($n/2$) soliton models $\vect{P}^{(n)}$ as $n \to \infty$, our first step is to introduce a transformation deforming all of the small loops $\hat \gamma_k^{(n)}$ onto the two sets of super-loops $\widehat\Gamma^{(\infty)}$ and $\overline{\widehat\Gamma^{(\infty)}}$.
Make the change of variables 
\begin{equation}\label{Qn.def}
	\begin{gathered}
	\vect{Q}^{(n)}(z;x,\vect{t}) := \vect{P}^{(n)}(z;x,\vect{t}) \vect{T}^{(n)}_{QP}(z;x,\vect{t})^{-1}, \\ 
	\vect{T}^{(n)}_{QP}(z;x,\vect{t}):= \begin{dcases}
		\tril{ -\frac{\ii}{2} \eee^{2\ii \varphi(z;x,\vect{t})} \sum \limits_{\mathclap{\ell \in \mathcal{I}^{(n)}_+}} 
		   \frac{\delta^{(n)}_{\ell}\eee^{\ii \Lambda_{\ell}^{(n)}}  }{z-\zeta^{(n)}_{\ell}} }
		& z \in \inside \left(\widehat \Gamma^{(\infty)} \right)\setminus \inside \left(\hat \gamma^{(n)}\right), \\
		\tril{ -\frac{\ii}{2} \eee^{2\ii \varphi(z;x,\vect{t})} \sum\limits_{\mathclap{\substack{\ell \in \mathcal{I}^{(n)}_+\\ \ell \neq k}}} 
		   \frac{\delta^{(n)}_{\ell}\eee^{\ii \Lambda_{\ell}^{(n)}}  }{z-\zeta^{(n)}_{\ell}} }
		& z \in \inside \left(  \hat \gamma^{(n)}_{k} \right), \quad k \in \mathcal{I}^{(n)}_\mathrm{low}, \\
		 \triu{ \frac{\ii}{2} \eee^{-2\ii \varphi(z;x,\vect{t})} \sum\limits_{\mathclap{\ell \in \mathcal{I}^{(n)}_-}} 
		    \frac{\delta^{(n)}_{\ell}\eee^{-\ii \Lambda_{\ell}^{(n)}}  }{z-\zeta^{(n)}_{\ell}} }
		& z \in \inside \left(\overline{\widehat \Gamma^{(\infty)}} \right)\setminus \inside \left(\hat \gamma^{(n)}\right), \\
		\triu{ \frac{\ii}{2} \eee^{-2\ii \varphi(z;x,\vect{t})} \sum\limits_{\mathclap{\substack{\ell \in \mathcal{I}^{(n)}_- \\ \ell \neq k}}} 
		    \frac{\delta^{(n)}_{\ell}\eee^{-\ii \Lambda_{\ell}^{(n)}}  }{z-\zeta^{(n)}_{\ell}} }
		& z \in \inside \left(  \hat \gamma^{(n)}_{k} \right), \quad k \in \mathcal{I}^{(n)}_\mathrm{up}, \\
		\ \vect{I} & \text{elsewhere}.
	\end{dcases} 
	\end{gathered}
\end{equation}

Notice that $\vect{Q}^{(n)}$ is piecewise analytic; inside each shrinking loop, $\hat \gamma_{k}^{(n)}$, the summation $\vect{T}_{WP}^{(n)}$ skips the simple pole at $\zeta_{k}^{(n)}$ enclosed by that loop. Moreover, the transformation removes the jumps from each of the shrinking loops. 
For example, let $k \in \mathcal{I}^{(n)}_\mathrm{low}$, then for $z \in \hat \gamma^{(n)}_{k}$ we have: 
\begin{multline}
	\vect{Q}_-^{(n)}(z;x,\vect{t})^{-1}\vect{Q}_+^{(n)}(z;x,\vect{t}) = \\
	\tril{ -\frac{\ii}{2} \eee^{2\ii \varphi(z;x,\vect{t})} \sum\limits_{\mathclap{{\ell \in \mathcal{I}^{(n)}_+}}} \frac{\delta^{(n)}_{\ell}\eee^{\ii \Lambda_{\ell}^{(n)}}  }{z-\zeta^{(n)}_{\ell}} }
	\tril{ \frac{\ii}{2} \eee^{2\ii \varphi(z;x,\vect{t})} \frac{\delta^{(n)}_{k}\eee^{\ii \Lambda_{ k}^{(n)}}  }{z-\zeta^{(n)}_{k}} }
	\tril{ \frac{\ii}{2} \eee^{2\ii \varphi(z;x,\vect{t})} \sum\limits_{\mathclap{\substack{\ell \in \mathcal{I}^{(n)}_+ \\ \ell \neq k}}} \frac{\delta^{(n)}_{\ell}\eee^{\ii \Lambda_{\ell}^{(n)}}  }{z-\zeta^{(n)}_{\ell}} } = \vect{I}.	
\end{multline}
The calculation when $k \in \mathcal{I}^{(n)}_\mathrm{up}$ is similar. The resulting unknown $\vect{Q}^{(n)}$ is analytic except for jumps along the fixed contour  $\Gamma_G$ and the superloops $\widehat \Gamma^{(\infty)} \cup \overline{\widehat \Gamma^{(\infty)}}$ 

\begin{RHP}\label{rhp:Qn}
	Find a $2\times 2$ matrix valued function $\vect{Q}^{(n)}( z; x,\vect{t})$ with the following properties
	\begin{enumerate}[align=left, leftmargin=*]
	\item[ \textbf{Analyticity}:] $\vect{Q}^{(n)}(z;x,\vect{t})$ is analytic for $z \in \C \setminus \left( \Gamma_G \cup \widehat \Gamma^{(\infty)} \cup \overline{\widehat \Gamma^{(\infty)}} \right)$. 
	\item[ \textbf{Normalization}:] $\vect{Q}^{(n)}(z;x,\vect{t}) = \vect{I} + \bigo{z^{-1}}$ as $z \to \infty$. 
	\item[ \textbf{Endpoint growth}:] At any endpoint $p$ of $\Gamma_G$ the matrix $\vect{Q}^{(n)}$ admits $\tfrac{1}{4}$-root singularities:
	\[
		\vect{Q}^{(n)}(z;x,\vect{t}) = \bigo{ |z-p|^{-1/4} }, \qquad \text{ as } z \to p \in \partial \Gamma_G.
	\]
	\item[ \textbf{Jump Condition}:] $\vect{Q}^{(n)}$ has continuous non-tangential boundary values $\vect{Q}^{(n)}_\pm(z;x,\vect{t})$ as $z \to \Gamma_G \cup\widehat \Gamma^{(\infty)} \cup \overline{\widehat \Gamma^{(\infty)}}$ satisfying the jump relation
	\begin{equation}\label{Qn.jump}
	\begin{gathered}
		\vect{Q}^{(n)}_+(z;x,\vect{t}) =\vect{Q}^{(n)}_-(z;x,\vect{t})  \vect{V}^{(n)}_Q(z;x,\vect{t}), \\ 
		\vect{V}^{(n)}_Q(z;x,\vect{t}) = \begin{dcases}
			(-\ii \sigtwo) \eee^{2\ii (\varphi(z;x,\vect{t}) + \Lambda^{(n)}(z) ) \sigthree} & z \in \Gamma_G, \\
			\tril{ -\frac{\ii}{2} \eee^{2\ii \varphi(z;x,\vect{t})} \sum\limits_{\ell \in \mathcal{I}^{(n)}_+} \frac{\delta^{(n)}_{\ell}\eee^{\ii \Lambda_{\ell}^{(n)}}  }{z-\zeta^{(n)}_{\ell}} }
			& z \in \widehat\Gamma^{(\infty)},  \\
			\triu{ \frac{\ii}{2} \eee^{-2\ii \varphi(z;x,\vect{t})} \sum\limits_{\ell \in \mathcal{I}^{(n)}_-} \frac{\delta^{(n)}_{\ell}\eee^{-\ii \Lambda_{\ell}^{(n)}}  }{z-\zeta^{(n)}_{\ell}} }
			& z \in \overline{\widehat \Gamma^{(\infty)}}.
			\end{dcases}
	\end{gathered}
	\end{equation}
\end{enumerate}
\end{RHP}

\subsection{Passing to the continuum limit along super-loops}
Now that the jump contours have been pushed onto the set of superloops $\widehat \Gamma^{(\infty)} \cup \overline{\widehat \Gamma^{(\infty)}}$, the factors $(z- \zeta_k^{(n)})^{-1}$ in the jump matrix $\vect{V}^{(n)}_Q$ are uniformly bounded. 
Then conditions~\ref{assume.measure} and \ref{assume.phases} in Assumption~\ref{data.assumptions} suggest that in the large $n$ limit the summations in \eqref{Qn.jump} can be replaced by integrals. 
Let $\vect{Q}^{(\infty)}$ be the solution of the following model problem:
\begin{RHP}\label{rhp:Qinfty}
Find a $2\times 2$ matrix valued function $\vect{Q}^{(\infty)}( z ; x,\vect{t})$ with the following properties
	\begin{enumerate}[align=left, leftmargin=*]
	\item[ \textbf{Analyticity}:] $\vect{Q}^{(\infty)}(z;x,\vect{t})$ is analytic for $z \in \C \setminus \left( \widehat \Gamma^{(\infty)} \cup \Gamma_G \right)$. 
	\item[ \textbf{Normalization}:] $c= \vect{I} + \bigo{z^{-1}}$ as $z \to \infty$. 
	\item[ \textbf{Endpoint growth}:] At any endpoint $p$ of $\Gamma_G$ the matrix $\vect{Q}^{(\infty)}$ admits $\tfrac{1}{4}$-root singularities:
	\[
		\vect{Q}^{(\infty)}(z;x,\vect{t}) = \bigo{ |z-p|^{-1/4} }, \qquad \text{ as } z \to p \in \partial \Gamma_G.
	\]
	\item[ \textbf{Jump Condition}:] $\vect{Q}^{(\infty)}$ has continuous non-tangential boundary values $\vect{Q}^{(\infty)}_\pm(z;x,\vect{t})$ as $z \to \Gamma_G \cup \widehat \Gamma^{(\infty)}$ satisfying the jump relation
	\begin{equation}\label{Qinfty.jump}
	\begin{gathered}
		\vect{Q}^{(\infty)}_+(z;x,\vect{t}) =\vect{Q}^{(\infty)}_-(z;x,\vect{t})  \vect{V}^{(\infty)}_Q(z;x,\vect{t}), \\ 
		\vect{V}^{(\infty)}_Q(z;x,\vect{t}) = \begin{dcases}
			(-\ii \sigtwo) \eee^{2\ii (\varphi(z;x,\vect{t}) + \Lambda^{(\infty)}(z)) \sigthree} & z \in \Gamma_G, \\
			\tril{ \eee^{2\ii \varphi(z;x,\vect{t})} \int\limits_{\Alow} \frac{\beta(\zeta)}{\zeta-z}  \mathrm{d}\mu(\zeta) }
			& z \in \widehat{\Gamma}^{(\infty)}, \\
			\triu{  -\eee^{-2\ii \varphi(z;x,\vect{t})} \int\limits_{\Aup} \frac{\overline{\beta(\bar \zeta)}}{\zeta-z}  \mathrm{d}\mu(\zeta) } 
			& z \in \overline{\widehat \Gamma^{(\infty)}}.
			\end{dcases}
	\end{gathered}
	\end{equation}
\end{enumerate}
\end{RHP}

In the passing to the limit, the matrix $\mathbf{Q}^{(\infty)}$ is no longer directly associated with the $N$-soliton RHP for the NLS hierarchy. The existence of solutions to RHP~\ref{rhp:Qinfty} must be established independently. This is achieved in the following propositions. 

\begin{prop}\label{prop:Qinfty.exists}
For each $(x,t) \in \R^2$ the solution $\vect{Q}^{(\infty)}(z;x,\vect{t})$ of RHP~\ref{rhp:Qinfty} exists and is unique.
\end{prop}

\begin{proof}
Observe that the contour $\Gamma_Q:= \widehat\Gamma^{(\infty)} \cup \Gamma_G$ is Schwarz reflection invariant ($\overline{\Gamma}_Q = - \Gamma_Q$ in the sense of path algebra) and that the jump matrix satisfies the symmetry $\vect{V}^{(\infty)}_Q(\bar z;x,\vect{t})^\dagger = \vect{V}^{(\infty)}_Q(z;x,\vect{t})^{-1}$.
The existence of a solution then follows from Zhou's vanishing lemma\footnote{In \cite{Zhou89} the contour is assumed to be Schwarz symmetric where here we are working with an anti-Schwarz symmetric contour. Reversing the orientation of the contour replaces the jump matrix with its inverse, which is why we have $\vect{V}^{(\infty)}_Q(\bar z;x,\vect{t})^\dagger = \vect{V}^{(\infty)}_Q(z;x,\vect{t})^{-1}$ instead of $\vect{V}^{(\infty)}_Q(\bar z;x,\vect{t})^\dagger = \vect{V}^{(\infty)}_Q(z;x,\vect{t})$ as appears in \cite{Zhou89}.} \cite{Zhou89}.  
Uniqueness follows from Liouville's theorem by observing that if $\vect{Q}^{(\infty)}_1$ and $\vect{Q}^{\infty}_2$ are two solutions of RHP~\ref{rhp:Qinfty} then $\vect{F}(z;x,\vect{t}) = \vect{Q}^{(\infty)}_1(z;x,\vect{t}) \vect{Q}^{(\infty)}_2(z;x,\vect{t})^{-1}$ is an entire, bounded function with $\lim\limits_{z \to \infty} \vect{F}(z;x,\vect{t}) = \vect{I}$. 
\end{proof}

\subsection{Characterization of infinite model via Primitive Potentials}

In this section we make a simplification of our model $\vect{Q}^{(\infty)}$ which reduces it to the form of RHP~\ref{rhp:accumulated}, which more directly relates to scattering data.
To do so we first examine the jump matrices \eqref{Qinfty.jump}. Here, it's useful to express the probability measure $\dd \mu$, supported on the accumulation set $\mathcal{A}$, in the form \eqref{eig.measure.density} of a complex linear or area measure with density $\rho(z)$. Following the notation of \eqref{eig.measure.density} let $\mathcal{A}_1$ and $\mathcal{A}_2$ denote the one and two dimensional components of $\mathcal{A}$ respectively and define $\Alow_k := \mathcal{A}_k \cap \Alow$, $k=1,2$. 
Then the non-trivial elements  in the jump \eqref{Qinfty.jump} can be expressed in terms of the function\footnote{We lightly abuse notation here, by writing $\beta(z) = \beta(z,\bar z)$ and $\rho(z) = \rho(z, \bar z)$ to stress the planar nature of the integral over the two-dimensional set $\Alow_2$. We hope this does not cause undue confusion.}
\[
	F(z) = \int\limits_{\Alow} \frac{\beta(\zeta)}{\zeta-z} \dd \mu(\zeta) 
	= \frac{1}{2\pi \ii} \int\limits_{\Alow_1 } \beta(\zeta)\rho(\zeta) \frac{\dd \zeta}{\zeta-z}  
	+ \frac{1}{2\pi \ii} \int\limits_{\Alow_2 } \beta(\zeta,\bar \zeta)\rho(\zeta,\bar \zeta) \frac{\dd \zeta \wedge \dd\bar\zeta}{\zeta-z}
\]	
which we can recognize as 1 and 2-dimensional Cauchy transforms on the sets $\Alow_1$ and $\Alow_2$ respectively, of the function
\begin{equation}\label{R.def}
	R(z) := \begin{cases}
		\rho(z) \beta(z),  &z \in \Alow_1 \\
		\rho(z,\bar z) \beta(z, \bar z),  &z \in \Alow_2
	\end{cases}
\end{equation}
The function $F(z)$ is analytic in $\C \setminus \Alow$ and, using the properties of the Cauchy operator, satisfies
\begin{equation}\label{Fj.jumps}
	\begin{aligned}
		&F_{+}(z) - F_{-}(z) = R(z)  && &z \in \Alow_1, \ \\
		&\dbar F(z) = R(z)  && &z \in \Alow_2, 
	\end{aligned}	
\end{equation} 
These properties motivate a transformation of our model problem. Define
\begin{equation}\label{P.infty.def}
	\vect{P}^{(\infty)}(z;x,\vect{t}) := \begin{dcases}
		\vect{Q}^{(\infty)}(z;x,\vect{t}) \tril{ -\eee^{2\ii \varphi(z;x,\vect{t})} \int\limits_{\mathcal{A}^{\mathrm{low}}} \frac{\beta(\zeta)}{\zeta-z}  \mathrm{d}\mu(\zeta) }
			& z \in \inside \left( \widehat\Gamma^{(\infty)} \right)   \\
		\vect{Q}^{(\infty)}(z;x,\vect{t}) \triu{  \eee^{-2\ii \varphi(z;x,\vect{t})} \int\limits_{\mathcal{A}^{\mathrm{up}}} \frac{\beta(\zeta)}{\zeta-z}  \mathrm{d}\mu(\zeta) } 
			& z \in \inside \left( \overline{\widehat \Gamma^{(\infty)}} \right)  \\
		\vect{Q}^{(\infty)}(z;x,\vect{t})  & \text{elsewhere}
	\end{dcases}
\end{equation}
The effect of the transformation defining the new unknown $\vect{P}^{(\infty)}$ is to remove the jumps on the contours $\widehat \Gamma^{(\infty)} \cup \overline{\widehat \Gamma^{(\infty)}}$ which were introduced, somewhat artificially, to enclose the accumulation set $\mathcal{A}$ where the microscopic bands of the finite model problem $\vect{P}^{(n)}$ accumulate as $n \to \infty$. The transformation \eqref{P.infty.def} deforms the jumps back onto the accumulation set $\mathcal{A}$. 

The transformation \eqref{P.infty.def} and the analytic properties in \eqref{Fj.jumps} imply that $\vect{P}^{(\infty)}$ satisfies the following problem:
\begin{dbar.rhp}\label{rhp:Pinfty} 
Find a $2\times2$ matrix-valued function $\vect{P}^{(\infty)}(z ;x,\vect{t})$ with the following properties:
\begin{enumerate}[align=left, leftmargin=*]
	\item[\textbf{Continuity}:] $\vect{P}^{(\infty)}$ is a continuous function of $z$ in $\C \setminus \left( \mathcal{A}_1 \cup \Gamma_G \right)$.
	\item[\textbf{Normalization}:] $\vect{P}^{(\infty)}(z;x,\vect{t}) = \vect{I} + \bigo{z^{-1}}$ as $z \to \infty$. 
	\item[\textbf{Analyticity}:]  $\vect{P}^{(\infty)}$ is analytic in $\C \setminus (\Gamma_G \cup \mathcal{A}$).
	\item[\textbf{Endpoint growth}:] At any endpoint $p$ of $\Gamma_G$ the matrix $\vect{P}^{(\infty)}$ admits $\tfrac{1}{4}$-root singularities:
	\[
		\vect{P}^{(\infty)}(z;x,t) = \bigo{ |z-p|^{-1/4} }, \qquad \text{ as } z \to p \in \partial \Gamma_G.
	\]
	\item[\textbf{Jump Conditions}:] $\vect{P}^{(\infty)}$ has continuous non-tangential boundary values $\vect{P}^{(\infty)}_\pm(z;x,\vect{t})$ as $z \to  \Gamma_G \cup \mathcal{A}_1$ satisfying the jump relation
	\begin{equation}
	\begin{gathered}
		\vect{P}^{(\infty)}_+(z;x,\vect{t}) = \vect{P}^{(\infty)}_-(z;x,\vect{t}) \vect{V}_P^{(\infty)}(z;x,\vect{t}) \\  
		\vect{V}_P^{(\infty)}(z;x,\vect{t}) = \begin{dcases}
			(-\ii \sigtwo) \eee^{2\ii ( \varphi(z;x,\vect{t}) + \Lambda^{(\infty)}(z) )\sigthree} & z \in \Gamma_G \\
			\tril{ R(z) \eee^{2\ii \varphi(z;x,\vect{t})}} & z \in \Alow_1 \\
			\triu{ -\overline{R(\bar z)} \eee^{-2\ii \varphi(z;x,\vect{t})}} &  z \in \overline{\Alow_1}
		\end{dcases}
	\end{gathered}
	\end{equation}
	\item[\textbf{Non-analyticity:}] $\vect{P}^{(\infty)}$ is not analytic for $z \in \mathcal{A}_2$; it is locally a (weak) solution of the partial differential equation 
	\begin{equation}
	\begin{gathered}
		\dbar \vect{P}^{(\infty)}(z;x,\vect{t}) = \vect{P}^{(\infty)}(z;x,\vect{t}) \vect{W}(z;x,\vect{t}) \\
		\vect{W}(z;x,\vect{t}) = \begin{dcases}
			\tril[0]{ R(z) \eee^{2\ii \varphi(z;x,\vect{t})}} & z \in \Alow_2 \\
			\triu[0]{ -\overline{R(\bar z) }\eee^{-2\ii \varphi(z;x,\vect{t})}} & z \in \overline{\Alow_2}
		\end{dcases}
	\end{gathered}
	\end{equation}
\end{enumerate}
\end{dbar.rhp}

Comparing the Problems we see that $\vect{P}^{\infty}$ is the solution of RHP~\ref{rhp:accumulated} with scattering data $\mathcal{D}_{\mathrm{prim}} = \left( \Gamma_G, \Lambda^{(\infty)}; \Alow, R \right)$. 

\subsection{Error analysis for the continuous model problem}

For each finite $n$, the error introduced by approximating $\vect{Q}^{(n)}$ by the continuous model $\vect{Q}^{(\infty)}$ is given by 
\begin{equation}\label{E.infty.def}
	\vect{E}^{(\infty)}(z;x,\vect{t}) := \vect{Q}^{(n)}(z;x,\vect{t})  \vect{Q}^{(\infty)}(z;x,\vect{t})^{-1}. 
\end{equation}
This function solves its own Riemann-Hilbert problem:

\begin{RHP}\label{rhp:Einfty}
Find a $2\times 2$ matrix valued function $\vect{E}^{(\infty)}(\, \cdot \,; x,\vect{t})$ with the following properties
	\begin{enumerate}[align=left, leftmargin=*]
	\item[ \textbf{Analyticity}:] $\vect{E}^{(\infty)}(z;x,\vect{t})$ is analytic for $z \in \C \setminus \Gamma^{(\infty)}_E$, \qquad 
	$\Gamma_E^{(\infty)} :=  \widehat \Gamma^{(\infty)} \cup \overline{\widehat \Gamma^{(\infty)}} $. 
	\item[ \textbf{Normalization}:] $\vect{E}^{(\infty)}(z;x,\vect{t}) = \vect{I} + \bigo{z^{-1}}$ as $z \to \infty$. 
	\item[ \textbf{Jump Condition}:] $\vect{E}^{(\infty)}$ has continuous non-tangential boundary values $\vect{E}^{(\infty)}_\pm(z;x,\vect{t})$ as $z \to   \Gamma_E^{(\infty)}$ satisfying the jump relation
	\begin{equation}\label{E.infty.jump}
	\begin{gathered}
		\vect{E}^{(\infty)}_+(z;x,\vect{t}) =\vect{E}^{(\infty)}_-(z;x,\vect{t})  \vect{V}^{(\infty)}_E(z;x,\vect{t}), \\ 
		\vect{V}^{(\infty)}_E(z;x,\vect{t}) = \vect{Q}^{(\infty)}_-(z;x,\vect{t}) \vect{V}_Q^{(n)}(z;x,\vect{t}) \vect{V}_Q^{(\infty)}(z;x,\vect{t})^{-1} \vect{Q}^{(\infty)}_-(z;x,\vect{t})^{-1}. 
	\end{gathered}
	\end{equation}
\end{enumerate}
\end{RHP}

In order to show that $\vect{E}^{(\infty)}$ is near identity, we need the following estimate. 
\begin{prop}\label{prop:integral.conv}
Given a sequence of finite-gap potentials with scattering data $(\Gamma^{(n)}, \Lambda^{(n)})$ which satisfy the conditions in Assumption~\ref{data.assumptions}, it follows that, uniformly for $z \in \Gamma_E^{(\infty)}$, 
\begin{equation}
	\begin{aligned}
		&\left| -\frac{\ii}{2}\sum_{k\in\mathcal{I}^{(n)}_+ } \frac{ \delta_k^{(n)} \eee^{\ii \Lambda_k^{(n)}} }{z-\zeta^{(n)}_k} 
		  - \int_{\Alow} \frac{\beta(\zeta)}{\zeta-z} \dd \mu(\zeta) \right| 
		  =  \bigo{ f(n)}, \\
		&\left| \frac{\ii}{2}\sum_{k\in\mathcal{I}^{(n)}_- } \frac{ \delta_k^{(n)} \eee^{-\ii \Lambda_k^{(n)}} }{z-\zeta^{(n)}_k} 
		  + \int_{\overline{\Alow}} \frac{\overline{\beta(\bar\zeta)}}{\zeta-z} \dd \mu(\zeta) \right| 
		  = \bigo{  f(n) }.
	\end{aligned}	  
\end{equation}
where 
\begin{equation}\label{error.rate}
	f(n) := \begin{dcases}
		\max \{ n^{-q}, n^{-p}, \scripty{r}_n^{\, \alpha} \} & \mathcal{A} \cap (\Gamma_G \cup \R) = \emptyset \\
		\max \{ n^{-q}, n^{-p}, \scripty{r}_n^{\, \alpha},  \scripty{r}_n^{\frac{d}{1+d}} \} & \mathcal{A} \cap (\Gamma_G \cup \R) \neq \emptyset \\
	\end{dcases}
\end{equation}
Here, $\scripty{r}_n$, defined by \eqref{r def}, measures the rate at which shrinking bands accumulate; $p>0$ is given in \eqref{point.measure.limit}; $q>0$ is given in \eqref{Lambda.cond}; $\alpha \in (0,1]$ is the H\"{o}lder exponent of function $\beta$; 
and $d \in \{1,2\}$ is the smallest dimension of any connected component of $\mathcal{A}$ which intersects $\R \cap \Gamma_G$. 
\end{prop}

\begin{proof}[Proof in the case that $\mathcal{A} \cap( \Gamma_G \cup \R) = \emptyset$]
	First note that since $\mathcal{A}$ is compact and $\Gamma_G \cup \R$ is closed, the condition $\mathcal{A} \cap( \Gamma_G \cup \R) = \emptyset$ implies that $\dist(\mathcal{A}, \Gamma_G \cup \R)>0$. This, and the fact that the shrinking bands $\gamma_k^{(n)}$ approach $\mathcal{A}$ uniformly as $n \to \infty$ imply that the super loops $\widehat \Gamma^{(\infty)} \cup \overline{\widehat \Gamma^{(\infty)}}$ can be chosen such that $\dist(\zeta, z) > c > 0$ for any $z \in \widehat \Gamma^{(\infty)} \cup \overline{\widehat \Gamma^{(\infty)}}$ and $\zeta \in \mathcal{A} \cup \gamma^{(n)}$.

It suffices to prove the first estimate as the second follows from symmetry. 	
For each $k \in \mathcal{I}^{(n)}_\mathrm{low}$ \eqref{A+} holds so  
	\begin{equation}\label{prop.pf.1}
		-\frac{\ii}{2}\sum_{k\in\mathcal{I}^{(n)}_+ } \frac{ \delta_k^{(n)} \eee^{\ii \Lambda_k^{(n)}} }{z-\zeta^{(n)}_k} = 
		\frac{1}{n} \sum_{k\in\mathcal{I}^{(n)}_+ } \frac{ \beta(\zeta^{(n)}_k) + \bigo{n^{-q}}}{\zeta^{(n)}_k-z} = 
		\frac{1}{n} \sum_{k\in\mathcal{I}^{(n)}_+ } \left(\frac{ \beta(\zeta^{(n)}_k) }{\zeta^{(n)}_k-z}  \right)+ \bigo{n^{-q}}
	\end{equation}
	since $z \in \widehat{\Gamma}^{(\infty)}$ and $\zeta_k^{(n)} \in \mathcal{Z}_\mathrm{low}^{(n)}$ implies $|\zeta_k^{(n)} - z| > c> 0$ as discussed above. 
	
Then, using condition~4 in Assumption~\ref{data.assumptions} we introduce the measure $\mu$ and partition $\mathcal{A} = \bigcup_{k=1}^n \mathfrak{a}_k^{(n)}$ of $\mathcal{A}$ satisfying \eqref{point.measure.limit}. Using these partition, we have	
	\begin{equation}\label{prop.pf.2}
	\frac{1}{n} \sum_{\mathclap{k\in\mathcal{I}^{(n)}_+ }} \ \frac{ \beta(\zeta^{(n)}_k) }{\zeta^{(n)}_k-z}  
	= 
	\frac{1}{n} \sum_{\mathclap{k\in\mathcal{I}^{(n)}_+ }}  \frac{ \beta(\zeta^{(n)}_k)
	    ( n \mu(\mathfrak{a}_k^{(n)})+ \bigo{n^{-p}} )    }{\zeta^{(n)}_k-z}   
	= 
	\sum_{\mathclap{k\in\mathcal{I}^{(n)}_+ }} \left(  \int_{\mathfrak{a}_k^{(n)}}  \frac{ \beta(\zeta^{(n)}_k) }{\zeta^{(n)}_k-z}  \dd\mu(\zeta)  \right) + \bigo{n^{-p}}.
	\end{equation}
	
Finally, from condition \ref{assume.phases} of Assumption~\ref{data.assumptions}, the function $\beta$ is $\alpha$-H\"{o}lder continuous\footnote{Also, for $z$ bounded away from $\mathcal{A}$ the function $|\zeta-z|$ is Lipschitz} for some $\alpha \in (0,1]$, so
	\begin{equation}\label{prop.pf.3}
		\begin{aligned}
		\sum_{\mathclap{k\in\mathcal{I}^{(n)}_+ }} \ 
		 \int\limits_{ \mathfrak{a}_k^{(n) }}  \frac{ \beta(\zeta^{(n)}_k) }{\zeta^{(n)}_k-z}  \dd\mu(\zeta)  		
		 &=
		\int\limits_{\Alow} \frac{ \beta(\zeta)}{\zeta-z} \dd\mu(\zeta) - 
		\sum_{k \in \mathcal{I}^{(n)}_+} \int\limits_{\mathfrak{a}_k^{(n)}} \left( \frac{ \beta(\zeta)}{\zeta - z} - \frac{ \beta(\zeta^{(n)}_k)}{\zeta^{(n)}_k - z} \right) \dd \mu(\zeta) 
		\leq \scripty{r}_n^{\alpha} 
		\end{aligned}
	\end{equation}
where we have used the second condition in \eqref{point.measure.limit} to bound the difference $|\zeta - \zeta_k^{(n)}|$ in each component of the partition. 
Combining the estimates in \eqref{prop.pf.1} - \eqref{prop.pf.3} gives the result of the Proposition. 
\end{proof}

\subsubsection{Allowing the accumulation domain $\mathcal{A}$ to have non-tangential contact points}
\begin{figure}[ht]
\hspace*{\stretch{1}}
\begin{overpic}[height=.3\textwidth]{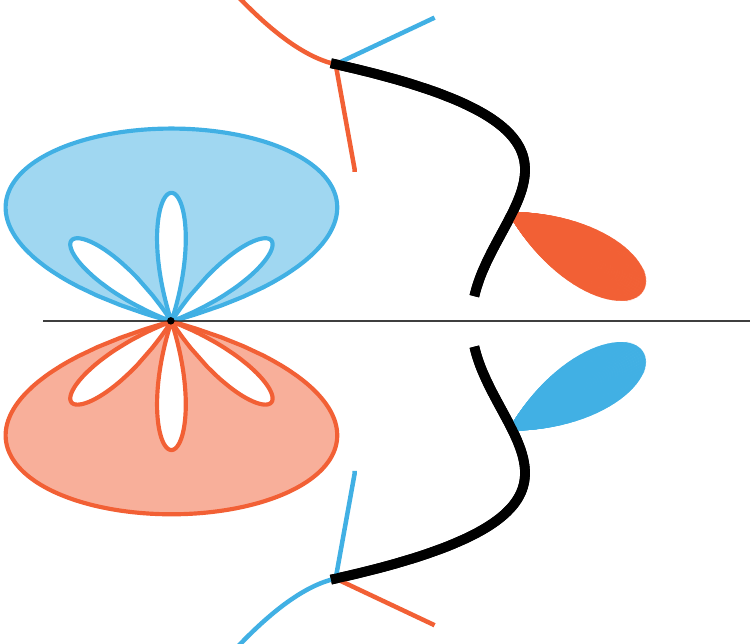}
\put(48,-10){$(a)$}
\end{overpic}
\hspace*{\stretch{1}}
\begin{overpic}[height=.3\textwidth]{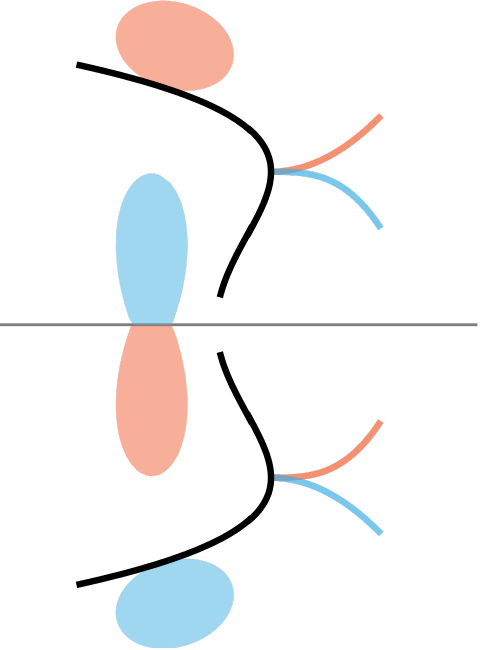}
\put(37,-10){$(b)$}
\end{overpic}
\hspace*{\stretch{1}}
\vspace*{10pt}
\caption{Example of allowable (panel (a)) and non-allowable contact (panel (b)) of the accumulation domain $\mathcal{A}$ (colored contours) with the fixed spectral curve $\Gamma_G$ (black curve)  and the real axis. Multiple touching `events' can occur at isolated points provided it satisfies the non-tangency conditions set out in conditions \ref{assume.measure} and \ref{assume.touching} of Assumption~\ref{data.assumptions}. Tangential contact or contact at non-isolated points are not allowed. 
}
\label{fig:touching}
\end{figure}
In the above argument we have assumed that compact accumulation domain $\mathcal{A}$ is disconnected from both the real axis and fixed spectral curve $\Gamma_G$ describing the background potential. This allowed us to introduce the (collection of) ``super-loops'' $\widehat{\Gamma}^{(\infty)}$ enclosing $\Alow$ and $\overline{\widehat{\Gamma}^{(\infty)}}$ enclosing $\Aup$ to lie at a fixed distance from $\mathcal{A}$ as $n \to \infty$. As a result, the denominators of both the summands and integrals in Proposition~\ref{prop:integral.conv} is bounded away from zero and the function $(\zeta - z)^{-1} \beta(\zeta)$ inherits the $\alpha$-H\"older property of $\beta(\zeta)$ for any fixed $z \in \widehat{\Gamma}^{(\infty)}$. 

Here we briefly described how to modify the arguments from the previous section when $\partial \mathcal{A}$ makes finitely many non-tangential points of contact with $\Gamma_G \cup \R$. The super-loops $\widehat{\Gamma}^{(\infty)}$ are forced to touch the accumulation set $\mathcal{A}$ at each of the points of contact. The non-tangential contact condition at each of the finitely many points of contact (c.f. conditions \ref{assume.measure} and \ref{assume.touching} in Assumption~\ref{data.assumptions}) means that 
one can introduce a disjoint collection $\mathcal{U}$ of open disks centered at each point of contact in $\partial\mathcal{A} \cap (\Gamma_G \cup \R)$ such that: 
\begin{enumerate}[label=(\arabic*)]
	\item There exists an $\eps_0>0$ such that $\dist \lp \widehat \Gamma^{(\infty)}, \ \mathcal{A} \setminus \mathcal{U} \rp \geq \eps_0 $.
	\item There exists a $\phi_0 \in (0,\pi/2)$ such that inside each disk of $\mathcal{U}$ every component of the contour $\widehat\Gamma^{(\infty)}$ is separated from both $\mathcal{A}$ and, for sufficiently large $n$, all $\zeta_k^{(n)} \in \mathcal{U}$ by a sector of interior angle $\phi_0$ with vertex at the point of contact.
\end{enumerate}

See Figure~\ref{fig:touching} for examples of allowed and disallowed touching. Once a contour $\widehat{\Gamma}^{(\infty)}$ is fixed which satisfies these extra conditions, the interpolating transformation is still given by \eqref{Qn.def}, and RHP~\ref{rhp:Qn} still describes the transformed unknown function $\mathbf{Q}^{(n)}(z;x,\vect{t})$. Likewise, we still want to approximate $\mathbf{Q}^{(n)}(z;x,\vect{t})$ by the solution, $\vect{Q}^{(\infty)}(z;x,\vect{t})$, of the model problem, RHP~\ref{rhp:Qinfty}. The 
existence of the solution to the model problem for all $(x,t) \in \R^2$ guaranteed by Proposition~\ref{prop:Qinfty.exists} still applied because the existence relies on the symmetry of the jumps across the real axis which is not effected by allowing the contour $\widehat{\Gamma}^{(\infty)}$ to non-tangentially touch the real axis or the fixed spectral curve $\Gamma_G$. In fact, the only step which requires extra work is the accuracy estimate in Prop.~\ref{prop:integral.conv}.

\begin{lemma}\label{lem:sum.estimate}
For $z \in \widehat \Gamma^{(\infty)}$ and $\zeta^{(n)}_k$ satisfying the conditions of Assumptions~\ref{data.assumptions} we have
\[
	\frac{1}{n} \sum_{k=1}^n \frac{1}{|\zeta_k^{(n)} - z|}  = \bigo{1}, \qquad n \to \infty.
\]
\end{lemma}
\begin{proof}
This result is trivial if $z$ is bounded away from $\mathcal{A}$ (and so from all the $\zeta_k^{(n)}$'s). So let $z_0 \in \mathcal{A} \cap (\R \cup \Gamma_G)$, fix $\eps_0>0$ sufficiently small and consider $z \in \widehat{\Gamma}^{(\infty)} \cap B_{\eps_0}(z_0)$. Then, using the second equation in \eqref{point.measure.limit}, we have  
\begin{equation}\label{4.11}
	\left| \frac{1}{n} \sum_{k=1}^n \frac{1}{\zeta_k^{(n)} - z} \right| 
	\leq 
	\frac{C}{\sin(\phi_0)}\sum_{k=1}^n \frac{  \mu( \mathfrak{a}_k )}{ |\zeta_k^{(n)} - z_0|}.
\end{equation}
Using the representation for $\mu$ in \eqref{eig.measure.density.prob} and the vanishing condition \eqref{measure.vanish} it follows that
\begin{equation}
	\mu(\mathfrak{a}^{(n)}_k) \leq C \lambda(\mathfrak{a}_k^{(n)}) \sup_{\xi \in \mathfrak{a}^{(n)}_k} |\xi - z_0| 
	\leq 
	C \lambda(\mathfrak{a}_k^{(n)}) ( | \zeta_k^{(n)} - z_0| + \scripty{r}_n) ,
\end{equation}
where $C>0$ is some constant, $\lambda$ is Lebesgue measure on $\mathcal{A}$, and we've used the second condition in \eqref{point.measure.limit} in the last step. Plugging this into \eqref{4.11}  using \eqref{gamma separation} gives
\begin{equation}
	 \frac{1}{n} \sum_{k=1}^n \frac{1}{|\zeta_k^{(n)} - z|} 
	\leq  
	\frac{C}{\sin \phi_0} \sum_{k=1}^n \left(1 + \frac{ \scripty{r}_n}{|\zeta_k^{(n)} - z_0|}  \right) \lambda( \mathfrak{a}_k^{(n)}) \leq 
	\frac{C}{\sin \phi_0}  \sum_{k=1}^n  \lambda( \mathfrak{a}_k^{(n)}) = \frac{C}{\sin \phi_0} \lambda( \mathcal{A}).
\end{equation}
\end{proof}

\begin{proof}[Proof of Proposition~\ref{prop:integral.conv} for $\mathcal{A}$ making non-tangential contact with $\R \cup \Gamma_G$]
We will only prove the first estimate in Prop.~\ref{prop:integral.conv} as the second follows from symmetry. Note that even when $\mathcal{A} \cap (\R \cup \Gamma_G) \neq \emptyset$, the earlier proof following the statement of Proposition~\ref{prop:integral.conv} suffices to establish convergence for all $z \in \widehat{\Gamma}^{(\infty)}$ such that $\dist(z, \mathcal{A} )$ is bounded away from zero.  
So, it remains to establish convergence in a set of open balls of fixed radius centered at each point in $\mathcal{A} \cap \left( \R \cup \Gamma_G \right)$. 
Fix a point $z_0 \in \mathcal{A} \cap \left( \R \cup \Gamma_G \right)$ and consider $z \in \widehat{\Gamma}^{(\infty)} \cap B_{\eps_0}(z_0)$. 
Then the first two steps estimates in the proof \eqref{prop.pf.1}-\eqref{prop.pf.2} are the same, except now Lemma~\ref{lem:sum.estimate} is used to bound the sum of errors. 
This gives,
\begin{equation}\label{prop.pf.4}
	\int\limits_{\Alow} \frac{ \beta(\zeta)}{\zeta-z} \dd\mu(\zeta) 
	- \frac{\ii}{2}\sum_{k\in\mathcal{I}^{(n)}_+ } \frac{ \delta_k^{(n)} \eee^{\ii \Lambda_k^{(n)}} }{z-\zeta^{(n)}_k} 
	 = \mathfrak{E}^{(n)} + \bigo{n^{-\flat}},
\end{equation}
where $\flat = \max\{p,q\}$ and
\begin{equation}\label{prof.pf.5}
	\mathfrak{E}^{(n)}:=  \sum_{k \in \mathcal{I}^{(n)}_+} \int\limits_{\mathfrak{a}_k^{(n)}} \left( \frac{ \beta(\zeta)}{\zeta - z} 
	  - \frac{ \beta(\zeta^{(n)}_k)}{\zeta^{(n)}_k - z} \right) \dd \mu(\zeta) .
\end{equation}

Control of the remaining integral terms follows from the fact that the density of the measure $\dd \mu$ is assumed to linearly vanish at each in $\mathcal{A} \cap (\R \cup \Gamma_G)$ (cf. \eqref{eig.measure.density.prob},\eqref{measure.vanish} ). For $z_0 \in \mathcal{A} \cap (\R \cup \Gamma_G)$ and $z \in \widehat{\Gamma}^{(\infty)} \cap B_{\eps_0}(z_0)$ we have 
\begin{align}
	\left| \mathfrak{E}^{(n)} \right| \leq 
	   \sum_{\mathclap{k \in \mathcal{I}^{(n)}_+}} \ \int\limits_{\mathfrak{a}_k^{(n)}} 
	   \left( \left| \frac{ \zeta - z_0}{\zeta-z} \right| \left| \beta(\zeta^{(n)}_k) - \beta(\zeta) \right| + 
	   \left| \beta(\zeta^{(n)}_k) \right| \left| \frac{(\zeta-z_0)(\zeta - \zeta^{(n)}_k)}{(\zeta - z)(\zeta^{(n)}_k-z)} \right| \right)
	   \frac{ \rho(\zeta)}{\left| \zeta -z_0 \right|} \dd \lambda(\zeta),
\end{align}
where $\dd \lambda$ is Lebesgue measure on $\mathcal{A}$. The condition that in a neighborhood of $z_0$ the points $\zeta \in \mathcal{A}$ and $\zeta^{(n)}_k \in \gamma^{(n)}$ are contained in a region bounded to both sides by sectors of angle $\phi_0 \in (0, \pi/2)$ with vertex at $z_0$ separating the region from $z \in \widehat\Gamma^{(\infty)}$ implies that
\begin{equation}
	|\zeta - z| \geq C | \zeta - z_0| \sin(\phi_0)
	\quad \text{and} \quad 
	|\zeta^{(n)}_k - z| \geq C | \zeta^{(n)}_k - z_0| \sin(\phi_0), \quad k=1,\dots, n.
\end{equation}
If follows that 
\begin{equation}
	\left| \mathfrak{E}^{(n)} \right| \leq 
	\frac{C}{\sin(\phi_0)}  \underbrace{ \sum_{\mathclap{k \in \mathcal{I}^{(n)}_+}} \ \int\limits_{\mathfrak{a}_k^{(n)}} 
	   \left| \zeta^{(n)}_k - \zeta \right|^\alpha\, \dd \lambda(\zeta) }_{:= \mathfrak{E}_1}
	+ \frac{C}{\sin(\phi_0)^2}  \underbrace{ \sum_{\mathclap{k \in \mathcal{I}^{(n)}_+}} \ \int\limits_{\mathfrak{a}_k^{(n)}} 
	   \frac{ \left| \zeta_k^{(n)} - \zeta \right| }{  \left| \zeta^{(n)}_k - z_0 \right|} d \lambda(\zeta) }_{:= \mathfrak{E}_2}
\end{equation}
where we've used the H\"older continuity of $\beta$ in the first integral and boundedness of $\beta$ in the second. 
Now, using the second condition in \eqref{point.measure.limit} we have
\begin{equation}
	\mathfrak{E}_1 
	\leq 
	\sum_{k \in \mathcal{I}^{(n)}_+} 
	  \sup_{\xi \in \mathfrak{a}_k^{(n)}} | \zeta^{(n)}_k - \xi|^\alpha \lambda( \mathfrak{a}^{(n)}_k) 
	\leq
	C \scripty{r}_n^{\,\alpha} \sum_{\mathclap{k \in \mathcal{I}^{(n)}_+}} \lambda( \mathfrak{a}^{(n)}_k) 
	\leq  \scripty{r}_n^\alpha \lambda( \mathcal{A} ) = \bigo{\scripty{r}_n^{\,\alpha}}.
\end{equation}
For $\mathfrak{E}_2$ let $\delta_n$ be a sufficiently small $n$-dependent parameter to be chosen later; let $B_{\delta_n}$  be the ball of radius $\delta_n$ centered at $z_0$. Then
\begin{equation}
	\mathfrak{E}_2  = 
	\sum_{\mathclap{k \in \mathcal{I}^{(n)}_+}} \ \int\limits_{\mathfrak{a}_k^{(n)} \cap B_{ \delta_n} }
	   \frac{ \left| \zeta_k^{(n)} - \zeta \right| }{  \left| \zeta^{(n)}_k - z_0 \right|} d \lambda(\zeta) 
	+ \sum_{\mathclap{k \in \mathcal{I}^{(n)}_+}} \ \int\limits_{\mathfrak{a}_k^{(n)} \setminus B_{ \delta_n } }
	   \frac{ \left| \zeta_k^{(n)} - \zeta \right| }{  \left| \zeta^{(n)}_k - z_0 \right|} d \lambda(\zeta)    
\end{equation}
In the first integral we observe that the condition \eqref{gamma separation} and \eqref{point.measure.limit} imply the integrand in the first integral above is bounded. If follows that 
\begin{equation}
	\mathfrak{E}_2  = C \lambda( B_{\delta_n} ) + \frac{\scripty{r}_n}{\delta_n} \lambda(\mathcal{A}).
\end{equation}
Since the Lebesgue measure $\lambda( B_{\delta_n} ) = \bigo{ \delta_n^d }$, where $d \in \{1,2\}$ is the smallest dimension of any connected component of $\mathcal{A}$ intersecting $z_0$. We minimize the error in $\mathfrak{E}_2$ by choosing $\delta_n$ such that the error from each term is of the same order which occurs when $\delta_n = \left( \scripty{r}_n \right)^{\frac{1}{1+d}}$ which results in 
\begin{equation}
	\mathfrak{E}_2  = \bigo{ {\scripty{r}_n}^{\frac{d}{d+1}} }.
\end{equation}
Combing the above estimates gives the stated rate of convergence in Proposition~\ref{prop:integral.conv} when $\mathcal{A} \cap (\Gamma_G \cup \R) \neq \emptyset$. 
\end{proof}

These estimates are enough to establish bounds on $\vect{E}^{(\infty)}$.
\begin{prop}\label{prop:Einfty}
	For all sufficiently large $n$ and $(x,\vect{t})$ in any compact set bounded independently of $n$, the solution $\vect{E}^{(\infty)}(z;x,\vect{t})$ of RHP~\ref{rhp:Einfty} exists and satisfies
	\begin{equation}
		\vect{E}^{(\infty)}(z;x,\vect{t}) = \vect{I} + \bigo{ f(n) }.
	\end{equation}
	Moreover, for large $z$
	\begin{equation}
		\vect{E}^{(\infty)}(z;x,\vect{t}) = \vect{I} + \frac{\vect{E}_1^{(\infty)}(x,\vect{t}) }{z} + \bigo{z^{-2}},  \qquad \left| \vect{E}_1^{(\infty)}(x,\vect{t})  \right| = \bigo{ f(n)},
	\end{equation}  
	where $f(n)$ is described in Proposition~\ref{prop:integral.conv}.

Additionally, if the solution $\vect{Q}^{(\infty)}(z;x,\vect{t})$ of RHP~\ref{rhp:Qinfty} is of exponential order in $x$ and $\vect{t}$, \textit{i.e.}, if
\begin{equation}\label{prop.Q.growth.bound}
	\left\| \vect{Q}^{(\infty)}(\,\cdot\,; x,\vect{t}) \right\|_{L^\infty(\Gamma_E^{(\infty)})} < K \eee^{ b (|x| + |\vect{t}|)},  \quad \text{ for some } K, b > 0,
\end{equation}
then for any $s \in (0, 1)$, there exists an $H>0$ such that the above estimates remain true as $n \to \infty$ with $|x| + |\vect{t}| < H \left[ s \log \left( \frac{1}{f(n)} \right) \right] $ 
except that the error estimates are replaced by the slower decay rate $\bigo{ f(n)^{1-s} }$. 
\end{prop}

\begin{proof}
For $z \in \Gamma_G$, it follows from \eqref{Lambda.G.limit} that 
\[
	 \vect{V}^{(n)}_Q(z;x;\vect{t}) \vect{V}^{(\infty)}_Q(z;x;\vect{t})^{-1} = \vect{I} 
	\qquad z \in \Gamma_G
\]
so that $\vect{E}^{(\infty)}$ is analytic for $z \in \Gamma_G$.
Proposition~\ref{prop:integral.conv} and \eqref{Qn.jump}-\eqref{Qinfty.jump} establish that 
\[
	\left| \vect{V}^{(n)}_Q(z;x;\vect{t}) \vect{V}^{(\infty)}_Q(z;x;\vect{t})^{-1} - \vect{I} \right| = \bigo{ f(n)  \eee^{2h(x,\vect{t})} }
	\qquad z \in \widehat \Gamma^{(\infty)} \cup \overline{\widehat \Gamma^{(\infty)}} 
\]
where 
\begin{equation}\label{h.def.Qinfty}
	h(x,\vect{t}) =  \sup_{\mathclap{z \in  \Gamma_E}} \, \Im \varphi(z;x,\vect{t}) 
	\leq Z_\Gamma |x| + \sum_{k=0}^\mathcal{K} | t_k| Z_\Gamma^k ,  
	\quad Z_\Gamma = \max_{z \in \Gamma_E } |z|.
\end{equation}
Noting that the solution $\vect{Q}^{(\infty)}(z;x,\vect{t})$ of model RHP~\ref{rhp:Qinfty} exists and is independent of $n$,
by inserting the above estimate into \eqref{E.infty.jump} one finds
\begin{equation}\label{V.infty.est}
	\left\| \vect{V}^{(\infty)}_E(\,\cdot\,;x,\vect{t}) - \vect{I} \right\|_{L^\infty(\Gamma_E)} \leq C \| \vect{Q}^{(\infty)} (\,\cdot\,;x,\vect{t}) \|^2_{L^\infty(\Gamma_E)} f(n) \eee^{2h(x,\vect{t})}.
\end{equation}

This establishes RHP~\ref{rhp:Einfty} as a small-norm Riemann Hilbert problem. The existence and bounds on the size of the solution then follow from standard estimates. We sketch the details for completeness. The solution of RHP~\ref{rhp:Einfty} can be represented in the form 
\begin{equation}\label{E.infty.est}
 	\vect{E}^{(\infty)}(z;x,\vect{t}) = \vect{I} + \frac{1}{2\pi \ii} \int_{\Gamma_E} \frac{ (\vect{I}+ \vect{\mu}(\zeta;x,\vect{t}) ) \left( \vect{V}_E^{(\infty)}(\zeta;x,\vect{t}) - \vect{I} \right)}{\zeta-z} \, \dd \zeta
\end{equation}
provided that $\mu$ solve the singular integral equation $(1 - \mathcal{C}_E) \vect{\mu} = \mathcal{C}_E \vect{I}$. Here $\mathcal{C}_E$ is the singular integral operator
\begin{equation}\label{E1.infty.est}
	\mathcal{C}_E \vect {f} = \mathcal{C}_- \left[\vect {f} (\vect{V}_E^{(\infty)} - \vect{I}) \right]  := \lim_{z \to{\Gamma_E}_-}\frac{1}{2\pi \ii} \int_{\Gamma_E} \frac{ \vect{f}(\zeta) ( \vect{V}_E^{(\infty)}(\zeta) - \vect{I})}{\zeta - z} \, \dd \zeta
\end{equation}
and $\mathcal{C}_-$ is the Cauchy projection operator. The minus subscript  ${\Gamma_E}_-$ in the limit indicates that $z$ approaches $\Gamma_E$ from the right with respect to the orientation of $\Gamma_E$. It's well known that $\mathcal{C}_-$ is a bounded $L^2$ operator. So, using \eqref{V.infty.est}, 
\[
\begin{aligned}
	\| \mathcal{C}_E \|_{L^2(\widehat{\Gamma}^{(\infty)}) } 
	&= \bigo{ \| \vect{V}_E^{(\infty)}(\,\cdot\,;x,\vect{t}) - \vect{I} \|_{L^\infty(\Gamma_E ) }} \\ 
	&= \bigo{ f(n) \| \vect{Q}^{(\infty)} (\,\cdot\,;x,\vect{t}) \|^2_{L^\infty(\Gamma_E)} \eee^{2 h(x,\vect{t})} }.
\end{aligned}	
\]
If $(x, \vect{t})$ lies in a compact subset of $\R \times \R^{\mathcal{K}}$, then since the final two factors are bounded independent of $n$ it follows that $\| \mathcal{C}_E \|_{L^2(\Gamma_E) } = \bigo{ f(n) }$. 
If, in addition, we have the growth bound in \eqref{prop.Q.growth.bound}, then it follows that 
$\| \mathcal{C}_E \|_{L^2(\Gamma_E) } = \bigo{ f(n) \eee^{ C (|x|+ |\vect{t}|}) }$ for some constants $C>0$. Given $s \in (0, \min \{q,\frac{\alpha}{d}\})$, taking $H = C^{-1}$
it trivially follows that whenever $|x| + |t| < H s \log\left( \frac{1}{f(n)} \right)$it follows that 
$\| \mathcal{C}_E \|_{L^2(\Gamma_E) } = \bigo{ f(z)^{1-s} }$. 

These bound establishes the existence of a near identity resolvent operator $(1 - \mathcal{C}_E)^{-1}$, which then gives $\vect{\mu}$ as $\vect{\mu} = (1 - \mathcal{C}_E)^{-1} \mathcal{C}_E \vect{I}$. The estimate in \eqref{E.infty.est} follows immediately. The estimate in \eqref{E1.infty.est} also follows upon expanding the Cauchy integral kernel geometrically for large $z$ which gives
\[
	\vect{E}_1^{(\infty)}(x,\vect{t}) = -\frac{1}{2\pi \ii} \int_{\Gamma_E}  (\vect{I}+ \vect{\mu}(\zeta;x,\vect{t}) )\left( \vect{V}_E^{(\infty)}(\zeta;x,\vect{t}) - \vect{I} \right) \, \dd\zeta.
\]
\end{proof}

\section{Error analysis for the finite model problem and proofs of the main theorems}

We're now prepared to study the solution $\vect{E}^{(n)}\!$ of RHP~\ref{rhp:En}, which measures the error introduced by approximating $\vect{O}^{(n)}(z;x,t)$---the solution of RHP~\ref{rhp:O.n} describing a finite-gap potential with $n$ shrinking bands---by $\vect{P}^{(n)}(z;x,t)$---the solution of RHP~\ref{rhp:Pn} which encodes a generalized $n/2$-soliton solution of the NLS hierarchy. 
Our starting point is the estimate \eqref{VEn.est.1}, 
which bounds $| \vect{V}_E^{(n)}-\vect{I} |$ in terms of the asymptotically small band-gap ratio $\varsigma(n)$ and the size of the solution $\vect{P}^{(n)}$ of the multi-soliton model problem RHP~\ref{rhp:Pn} along the jump contour $\hat \gamma^{(n)}$. We will show $|\vect{V}_E^{(n)}-\vect{I} |$ is asymptotically small by showing that $\left| \vect{P}^{(n)} \right|$ is bounded on $\hat \gamma^{(n)}$.

Equations \eqref{Qn.def}, \eqref{E.infty.def}, and Propositions~\ref{prop:Qinfty.exists} and \ref{prop:Einfty} establish the existence of a solution to RHP~\ref{rhp:Pn} given by 
\begin{equation}\label{Pn.prod}
	\vect{P}^{(n)}(z;x,\vect{t}) = \vect{E}^{(\infty)}(z;x,\vect{t}) \vect{Q}^{(\infty)}(z;x,\vect{t}) \vect{T}_{PQ}^{(n)}(z;x,\vect{t}), \qquad z \in \C \setminus \hat \gamma^{(n)}.
\end{equation}
As both $\vect{E}^{(\infty)}$ and $\vect{Q}^{(\infty)}$ are analytic on $\hat \gamma^{(n)}$ its clear that $\vect{P}^{(n)}$ takes continuous boundary values along $\hat \gamma^{(n)}$ given by taking the appropriate boundary value of the piecewise continuous function $\vect{T}_{PQ}^{(n)}$ as $z \to \hat \gamma^{(n)}$. 

Consider the boundary value $\| \vect{P}_-^{(n)}(\,\cdot\, ;x,\vect{t}) \|_{L^{\infty}(\hat \gamma^{(n)})}$. As the contours $\hat \gamma^{(n)}$ are positively oriented, the minus boundary value of each $\hat \gamma^{(n)}_k$ comes from taking the limit as $z \to \hat \gamma^{(n)}_k$ from outside the loop.
For $z$ on any of the component loops $\hat \gamma^{(n)}_{k}$ in $\hat \gamma^{(n)}$ we have from \eqref{Qn.def}
\begin{multline}\label{TPQ.bound}
	\left| {(\vect{T}_{PQ}^{(n)})}_-(z) \right| 
	\leq C \eee^{2 | \Im \varphi(z;x,\vect{t})|} \left|  \sum\limits_{\ell \in \mathcal{I}^{(n)}_\pm  }
	\frac{\delta^{(n)}_{\ell}\eee^{\pm \ii \Lambda_{\ell}^{(n)}}  }{z-\zeta^{(n)}_{\ell}}  \right| \\
	 \leq  \frac{C}{r_n}  \eee^{2 | \Im \varphi(z;x,\vect{t})|}  \frac{1}{n} \sum\limits_{\ell \in \mathcal{I}^{(n)}_\pm} \left| \beta(\zeta^{(n)}_{\ell})\right|  
	 \leq  \frac{C}{r_n}  \eee^{2 | \Im \varphi(z;x,\vect{t})|}, \qquad z \in \hat \gamma^{(n)}.
\end{multline}
Here, we've used Assumption~\ref{Lambda.cond} and \eqref{loop.dist} to simplify the estimate. 

\begin{prop} \label{prop:En.jump}
Suppose that the ensemble of scattering data $(\Gamma^{(n)},\Lambda^{(n)})$ satisfies the conditions of Assumption~\ref{data.assumptions}. Then in the thermodynamic limit, for all sufficiently large $n$ the jump matrix $\vect{V}_E^{(n)}(z;x,\vect{t})$ defined by \eqref{En.jump} satisfies
\begin{align}
	&\left\| \vect{V}^{(n)}_E(\, \cdot\, ;x,\vect{t}) - \vect{I} \right\|_{L^{\infty}(\hat\gamma^{(n)}) } = \bigo{ \varsigma(n)^2 \left(  \frac{n}{r_n} + \frac{1}{n r_n^3} \right) },
	&&\text{for } |x| + |\vect{t}| < K
\intertext{Moreover, provided that the solution $\vect{Q}^{(\infty)}(z;x,t)$ is of exponential order in $x$ and $t$ for $z$ in compact sets, then 
for each sufficiently small $s>0$, there exists $H > 0$ such that }
	&\left\| \vect{V}^{(n)}_E(, \cdot\, ;x,\vect{t}) - \vect{I} \right\|_{L^{\infty}(\hat\gamma^{(n)}) } = \bigo{ n^s \varsigma(n)^2 \left(  \frac{n}{r_n} + \frac{1}{n r_n^3} \right) },
	&&\text{for } |x| + |\vect{t}| < H s \log(n)
\end{align}
\end{prop}
\begin{proof}
	For $(x,\vect{t})$ in any compact subset of $\R\times\R^\mathcal{K}$, Propositions.~\ref{prop:Qinfty.exists} and \ref{prop:Einfty} guarantee that both $\vect{Q}^{(\infty)}(z;x,\vect{t})$ and $\vect{E}^{\infty}(z;x,\vect{t})$ exists and are bounded (in norm) independent of $x,\vect{t}$ and $n$ for $z \in \hat\gamma^{(n)}$. It then follows from \eqref{Pn.prod}-\eqref{TPQ.bound} that 
$
	\left\| \vect{P}_-^{(n)}(z;x,\vect{t}) \right\|_{L^\infty(\hat \gamma^{(n)} )} = \bigo{ r_n^{-1} }.
$
 Inserting this estimate into \eqref{VEn.est.1} gives the stated bounds for $\left\| \vect{V}^{(n)}_E(\, \cdot\, ;x,\vect{t}) - \vect{I} \right\|_{L^{\infty}(\hat\gamma^{(n)}) } $ when $|x| + |\vect{t}|$ is bounded. 
 	
	If we allow for $(x,\vect{t})$ with $|x| + |\vect{t}| < H \log(n)$ for some $H>0$, then under the additional assumption that $\vect{Q}^{\infty}(z,x,\vect{t})$ is of exponential order in $x$ and $\vect{t}$ we can again choose $H$ sufficiently small to ensure that $\vect{E}^{(\infty)}$ still exists (Prop.~\ref{prop:Einfty}) and moreover
$
	\left\| \vect{P}_-^{(n)}(z;x,\vect{t}) \right\|_{L^\infty(\hat \gamma^{(n)} )} = \bigo{ r_n^{-1} \eee^{C_1|x| + C_2 |\vect{t}| } }.
$
for some constants $C_1,C_2>0$. Plugging this into \eqref{VEn.est.1}, we get
\[
	\left| \vect{V}_E^{(n)}(z;x,\vect{t}) - I \right| = \bigo{\eee^{C_1|x| + C_2 |\vect{t}| } \varsigma(n)^2 \left( \frac{n}{r_n} + \frac{1}{n r_n^3} \right) \eee^{C_1|x| + C_2|\vect{t}|} }.
\]
Then by choose $H$ small enough we ensure the exponential factor in the above is bounded above by $n^s$ and the result follows immediately. 
\end{proof}

The following proposition follows immediately from the above estimate and the theory of small-norm Riemann-Hilbert problems. The proof is very similar to the proof of Prop~\ref{prop:Einfty} and so is omitted. 

\begin{prop}\label{prop.En}
	Under the same assumptions as Proposition~\ref{prop:En.jump} for all sufficiently large $n$ a unique solution $\vect{E}^{(n)}$ of RHP~\ref{rhp:En} exists and satisfies
	\begin{equation}
		\vect{E}^{(n)}(z;x,\vect{t}) = \vect{I} + \bigo{ q(n;x,t) \varsigma(n)^2 \left(  \frac{n}{r_n} + \frac{1}{n r_n^3} \right) }.
	\end{equation}
	Moreover, for large $z$
	\begin{equation}
		\vect{E}^{(n)}(z;x,\vect{t}) = \vect{I} + \frac{\vect{E}_1^{(n)}(x,\vect{t}) }{z} + \bigo{z^{-2}},  \qquad 
		\left| \vect{E}_1^{(n)}(x,\vect{t})  \right| = \bigo{ q(n;x,\vect{t}) \varsigma(n)^2 \left(  \frac{n}{r_n} + \frac{1}{n r_n^3} \right) }
	\end{equation}  
	Here 
	\[
		q(n;x,\vect{t}) = \begin{cases}
			1 & |x| + |\vect{t}| = \bigo{1} \\
			n^{-s} & |x| + |\vect{t}| < H s \log(n)
		\end{cases}
	\]
\end{prop}

\begin{proof}[Proof of Theorem~\ref{thm:fg-sol}]
For each $n \in 2\N$, let $\mathcal{D}_{\mathrm{fg}}^{(n)}$ be a sequence of finite-gap scattering data satisfying Assumption~\ref{data.assumptions}. Let $\vect{M}^{(n)}(z;x,\vect{t})$ be the solution of RHP~\ref{rhp:n-phase} parameterized by $\mathcal{D}_{\mathrm{fg}}^{(n)}$ for $x \in \R$ and $\vect{t} \in \R^\mathcal{K}$ for some fixed $\mathcal{K} \in \N$. Then the exact transformations $\vect{M}^{(n)}(z;x,\vect{t}) \mapsto \vect{O}^{(n)}(z;x,\vect{t}) \mapsto \vect{E}^{(n)}(z;x,\vect{t})$ defined by \eqref{O.def} and \eqref{En.def} gives
\begin{equation}\label{pf.main.thm.1}
	\vect{M}^{(n)}(z;x,\vect{t}) =  \vect{E}^{(n)}(z;x,\vect{t})  \vect{P}^{(n)}(z;x,\vect{t}),  \quad |z| \gg 1,
\end{equation}
where $\vect{P}^{(n)}(z;x,\vect{t})$ is the solution of RHP~\ref{rhp:Pn}. As is explained between \eqref{Pn.jump} and \eqref{Pn.recover},  $\vect{P}^{(n)}(z;x,\vect{t})$ is equal outside a compact set to the solution of the generalized $n/2$-soliton Riemann-Hilbert problem, RHP~\ref{rhp:fg-gen} with data $\mathcal{D}^{(n)}_\mathrm{sol}$ given by \eqref{soliton.ensemble.data}. It follows from the result of Proposition~\ref{prop.En}, that
\begin{equation}\label{pf.main.thm.2}
	\begin{aligned}
	\psi \left(x,\vect{t} \, \vert\, \mathcal{D}_\mathrm{fg}^{(n)} \right) 
	= 2\ii \lim_{z \to \infty} \vect{M}^{(n)}_{12}(z;x,\vect{t}) 
	&= 2\ii \lim_{z \to \infty} \vect{P}^{(n)}_{12}(z;x,\vect{t}) + 2\ii \vect{E}^{(n)}_{12}(x,\vect{t}) 
	\\
	&=\psi \left(x,\vect{t} \, \vert\, \mathcal{D}_\mathrm{sol}^{(n)} \right) + \bigo{ q(n;x,\vect{t}) \varsigma(n)^2 \left(  \frac{n}{r_n} + \frac{1}{n r_n^3} \right) }
	\end{aligned}
\end{equation}
Observing that $q(n;x,\vect{t}) = 1$ for $(x, \vect{t})$ bounded, this completes the proof of Theorem~\ref{thm:fg-sol}. 
\end{proof}

\begin{proof}[Proof of Theorem~\ref{thm:fg-prim} ]
Continuing the argument given above from \eqref{pf.main.thm.1}, the transformations \eqref{Qn.def}, \eqref{E.infty.def}, and \eqref{P.infty.def} imply that in a neighborhood of $\infty$, 
\begin{equation}
	\vect{P}^{(n)}(z;x,\vect{t}) = \vect{E}^{(\infty)}(z;x, \vect{t}) \vect{P}^{(\infty)}(z;x, \vect{t}),  \qquad |z| \gg 1.
\end{equation}
where $P^{(\infty)}(z;x,\vect{t})$ is the solution of the primitive-potential problem RHP~\ref{rhp:accumulated} with scattering data given by \eqref{prim.limit.data}. The existence of a solution to RHP~\ref{rhp:accumulated}  is established by Proposition~\ref{prop:Qinfty.exists} and the explicit transformation $\vect{Q}^{(\infty)} \mapsto \vect{P}^{(\infty)}$ defined by \eqref{P.infty.def}. The result of Theorem~\ref{thm:fg-prim} then follows immediately from the above representation and the estimates of Proposition~\ref{prop:Einfty}.
\end{proof}

\section{Explicit examples of condensate gas realizations}\label{sec:examples}

In this section we give some concrete examples of the method presented here. For simplicity, we consider the zero background case, i.e. when $G=-1$ (equivalently  $\Gamma_G = \emptyset$), so every band of the finite-gap sequence has vanishing arc length in the thermodynamic limit. 

\subsection{Eigenvalues sampled from box data}

For any integer $N \in \N$, let $\psi^{(2N)}(x,0) = 1_{ [- N\pi ,0] }(x)$ be the box-shaped potential of height 1 on the interval $-N\pi  \leq  x \leq 0$. The 
Zakharov-Shabat scattering data for a box-potential is exactly computable. The discrete spectrum corresponding to $\psi^{(2N)}(x,0)$ is  $\mathcal{Z}^{(2N)} \cup \overline{\mathcal{Z}^{(2N)}}$, where  $\mathcal{Z}^{(2N)}= \{ \zeta_k^{(n)} \}_{k=1}^{N}$ and the eigenvalues satisfy:
\begin{equation}
	\left[ \sqrt{z^2+1} - \frac{1}{\pi N}  \arcsin(-i z) \right]  \bigg\vert^\ii_{z=\zeta_{k}^{(2N)} } 
	\equiv \int_{\zeta_{k}^{(2N)} }^\ii  \frac{ -\ii z + (\pi N)^{-1} } {\sqrt{1+z^2}} |\dd z|   
	=  \frac{k}{N}, \qquad -\ii \zeta_{k}^{(2N)} \in (0,1) .  
\end{equation}
Clearly, as $N \to \infty$, these eigenvalues fill out the interval $[0,\ii]$ with asymptotic density 
\be\label{rho-box}
	\dd \mu = \frac{ |z|}{\sqrt{1+z^2}} |dz| =  -\frac{2\pi \ii z}{\sqrt{z^2+1}} \frac{\dd z}{2\pi \ii}.  
\ee
Observe, that this density satisfies the condition \ref{assume.touching} in Assumption~\ref{data.assumptions} as it vanishes linearly at $z=0$. 

The norming constants $c^{(2N)}(\zeta_k^{(2N)})$ associated to each $\zeta_{k}^{(2N)}$ are also exactly computable and given by 
\begin{equation}
	c^{(2N)}(z) =   \frac{ 1 + z^2}{N\pi z+ \ii}. 
\end{equation}
Here we encounter a small issue, as the function $\beta$ needed for our main result, defined by \eqref{beta.thm}, is given by
\begin{equation}
	\beta(z) = \lim_{N \to \infty} 2N c^{(2N)}(z) = \frac{2}{\pi} \frac{1+z^2}{z} 
\end{equation}
which is clearly not a H\"{o}lder continuous function at $z=0$, as required. Nor are the box potentials $\psi^{(2N)}(x,0)$ reflectionless; they have a nonzero, reflection coefficient for $z \in \R$ exhibiting rapid oscillations on the scale of $N^{-1}$. 

To produce potentials satisfying Assumption~\ref{data.assumptions} we introduce a small parameter $\eps>0$ and only allow our modified eigenvalues to come from the interval $[\ii \eps, \ii]$, with renormalized density (cf. \eqref{eig.measure.density.complex})
\begin{equation}
	\widetilde{\varrho} (z) = \frac{-2\pi \ii z}{\sqrt{1-\eps^2}\sqrt{1+z^2}}
\end{equation}
so that the eigenvalues $\widetilde{Z}_\mathrm{low}^{(2N)} = \left\{ \tilde{\zeta}_k^{(2N)} \right\}_{k=1}^N$ satisfy
\begin{equation}\label{ex1.band.centers}
	\int_{\tilde{\zeta}_k^{(2N)}}^\ii  \widetilde{\varrho} (z) \frac{ \dd z}{2 \pi \ii} = \frac{k}{N} 
	\qquad \Rightarrow \qquad
	\tilde{\zeta}_k^{(2N)} = \ii \sqrt{ 1 - \frac{(1-\eps^2) k^2}{N^2}}, \quad k=1,\dots, N
\end{equation}

We define the sequence of scattering data $\mathcal{D}_\mathrm{sol}^{(2N)} = \{ \emptyset; \widetilde{Z}_\mathrm{low}^{(2N)}, c^{(2N)} \}$
and let $\widetilde{\psi}(x, \vect{t}\, \vert\, \mathcal{D}_\mathrm{sol}^{(2N)})$ denote the sequence of reflectionless potentials obtained by solving 
 RHP~\ref{rhp:fg-gen} on the zero background (i.e. $\Gamma_G = \emptyset$) with discrete spectral data $\mathcal{D}_{\mathrm{sol}}^{(2N)}$.

Using the results of Theorem~\ref{thm:fg-sol} we can construct a related sequence of finite-gap potentials. For each $N$ fix bands $\gamma^{(2N)} = \bigcup_{k=1}^{2N} \gamma^{(2N)}_k$ where 
\begin{equation}
	\gamma_{k}^{(2N)} = \begin{dcases}
		[\tilde{\zeta}^{(2N)}_k - \delta^{(2N)}_k, \tilde{\zeta}^{(2N)}_k + \delta^{(2N)}_k], 
		&k=1,\dots, N,  \\
		\gamma^{(2N)}_{k} = -\overline{\gamma^{(2N)}_{k-N} } 
		&k=N+1,\dots, 2N. 
	\end{dcases}
\end{equation}
In this setting, with the band centers $\tilde{\zeta}_k^{(2N)}$ satisfying \eqref{ex1.band.centers} we have $r_{2N} = \min_{k\neq j} | \tilde \zeta_k^{(2N)} - \tilde \zeta_j^{(2N)}| \sim N^{-2}$. So in order to satisfy \eqref{band.gap.ratio.rate} we need $|\delta^{(2N)}_k| = \littleo{N^{-9/2}}$, then we choose a phase function $\Lambda^{(2N)} : \gamma^{(2N)} \to \C$ such that 
\[
	\lim_{N\to \infty} \left[ \ii N \delta^{(2N)}_k e^{\ii \Lambda^{(2N)}_k } - \beta \left( \zeta^{(2N)}_k\right) \right] = 0,
\]
which implies
\be\label{Lam}
\Im \Lambda_k^{(2N)}(z) \sim \log \left( \frac{\pi N | \delta_k^{(2N)} |}{2}  \frac{-\ii \zeta_k^{(2n)}}{1 + \left( \tilde \zeta_k^{(2N)} \right)^2} \right) 
\ee
Thus $\mathcal{D}^{(2N)}_\mathrm{fg} = \left( \gamma^{(2N)}, \Lambda^{(2N)} \right)$ denotes a sequence of finite-gap scattering data constructed from the soliton data. Let $\psi \left(x,\vect{t} \, \vert\, \mathcal{D}^{(2N)}_\mathrm{fg} \right)$ be the sequence of finite-gap potentials obtained from solving RHP~\ref{rhp:n-phase} with data $\mathcal{D}^{(2N)}_\mathrm{fg}$. Then as $N \to \infty$ the sequence of finite-gap potentials $\psi \left(x,\vect{t} \, \vert\, \mathcal{D}^{(2N)}_\mathrm{fg} \right)$ and $N$ soliton potentials $\psi \left(x,\vect{t} \, \vert\, \mathcal{D}^{(2N)}_\mathrm{sol} \right)$ converge to each other as described by Theorem~\ref{thm:fg-sol}.

Now, for any fixed $\eps> 0$, $\beta(z)$ is H\"{o}lder---analytic actually---on the accumulation set $\mathcal{A} = [\ii \eps, \ii]$ of the modified eigenvalues. So Theorem~\ref{thm:fg-prim} applies and both the sequence of finite-gap potentials $\psi \left(x,\vect{t} \, \vert\, \mathcal{D}^{(2N)}_\mathrm{fg} \right)$ and $N$ soliton potentials $\psi \left(x,\vect{t} \, \vert\, \mathcal{D}^{(2N)}_\mathrm{sol} \right)$ converge as $N \to \infty$ to the primitive potential $\widetilde{\psi}(x, \vect{t} \, \vert\, \mathcal{D}_\mathrm{prim})$ where
\begin{equation}
	\mathcal{D}_\mathrm{prim} = \left( \emptyset; [\ii \eps, \ii], R \right), \qquad R(z) = \beta(z) \varrho(z) = -4i\sqrt{ \frac{ 1+ z^2}{1-\eps^2} }
\end{equation}
is the solution of RHP~\ref{rhp:accumulated} with data $\mathcal{D}_\mathrm{prim}$. 
The solutions of this RHP were studied in \cite{GGJMM} in the context of mKdV, note, however,  that at $\vect{t} = (0,0,\dots,0)$ the solution is independent of the particular flow, where it was shown that 
\begin{equation}
	\widetilde{\psi}(x, \vect{0} \, \vert\, \mathcal{D}_\mathrm{prim}) =
	\begin{dcases}
	 \bigo{ \eee^{-c x}}, 
	&x \to + \infty,  \\
	\widetilde{\psi}(x, \vect{0} \, \vert\, \mathcal{D}_\mathrm{prim}) = (1+\eps) \dn\left( (1+\eps)(x-x_0), \tfrac{4\eps}{(1+\eps)^2} \right) + \bigo{x^{-1}}, 
	&x \to -\infty,  \\
	\end{dcases}
\end{equation}
where $x_0$ is an explicit phase shift expressed as a certain integral depending on $R(z)$. As $\eps \downarrow 0$ we point out that 
\[
	\widetilde{\psi}(x, \vect{t} \, \vert\, \mathcal{D}_\mathrm{prim}) \sim 1 + (1-4 \sin^2(x-x_0))\eps + \bigo{\eps^2}, \qquad x \ll -1.
\]
Observe, that this implies $\widetilde{\psi}(x, \vect{0} \, \vert\, \mathcal{D}_\mathrm{prim})$ agrees to leading order with the limiting initial data $1_{\R_-}(x) = \lim_{N \to \infty} 1_{[-N\pi, 0]}(x)$ for all sufficiently large $|x|$ as $\eps \downarrow 0$. .

From the perspective of the thermodynamic limit of finite-gap solutions, the most natural object to match the above example would be a fNLS periodic soliton gas, introduced  in \cite{TW2022}, where the semiclassical limit of the fNLS with a nonnegative one hump periodic potential $q(x)$ was considered. 
Using formal WKB analysis, one can show that the Zakharov-Shabat spectrum of this potential consists of $N$ bands locating on $i\R$ between $\ii \min_x q$ and $\ii \max_x q$ and their complex conjugates, where $1/N$ plays the role of the semiclassical parameter. 
It turns out that, generically, the length of the bands is exponentially  small in $N$.
Here, we choose $q(x)$ by setting $q(x) = 1$ for $\pi N <x <0$, followed by an interval where $q(x) = 0$ for $ 0<x<\mu(N)\pi N$, where $\mu(N)\geq 0$. We then extend $q(x)$ to $\R$ as an $(1+\mu(N)) \pi N$ periodic function.  
The fNLS soliton gas corresponding to this type of potential was considered in \cite[Section 4.4]{TW2022}. There it was shown that the resulting density of bands for the sequence of finite-gap potential coincides with \eqref{rho-box}. Additionally, the bandwidth $|\delta^{(2N)}(z)|$ was computed and shown to satisfy $|\delta^{(2N)}(z)| \sim e^{-\nu_\mu(z)N}$, where
\be
\nu_\mu(z)= \frac{\pi\mu |z|}{2}.
\ee
Substituting this into \eqref{Lam}, we derive the leading behavior of the phase as
\be \label{Lam-as}
\Im\Lambda(z) \sim\ - \frac{\pi\mu(n) |z|n}{2}+\ln N -\ln \frac{\pi}{2}-\ln\left(
|z|-\frac 1{|z|}
\right).
\ee
Expression \eqref{Lam-as} defines  a particular realization (up to the phase of $e^{\ii \Lambda}$)  of the periodic fNLS soliton gas. As it was shown in \cite{TW2022}, for this gas the DOS $u(z)=\frac 1{\pi(1+\mu)}$, the spectral density is $\rho(z)=\frac{|z|}{\pi (1+\mu)\sqrt{1+z^2}}$, and the spectral scaling function (see Section \ref{sec-backgr})
	\be
	\s(z)=\pi \mu\sqrt{1+z^2}.
	\ee
Thus, for this realization, the case of a constant $\mu>0$ corresponds to  a linear  dependence of $ \Im\Lambda$ on $n$ (a proper soliton gas)
 whereas the case $\mu(N)\to 0$, for example, $\mu(N)=\frac 1{\sqrt N}$,
 corresponds to the sublinear dependence of $ \Im\Lambda$ on $n$ (fNLS soliton condensate with $\s=0$).

\subsection{Realizations of an mKdV soliton gas}
Fixing $\vect{t} = (0,0,0,4t,0,\dots)$ so that only $t_3=4t$ varies; then the flow of the hierarchy \eqref{flows} reduces to the standard mKdV equation
\begin{equation}
	\psi_t + 6 \psi^2 \psi_x + \psi_{xxx} = 0,
\end{equation}
provided that we enforce an extra symmetry condition on the scattering data so that the potential $\psi$ must be real-valued. Specifically, the solutions of our Riemann-Hilbert problem must satisfy the symmetry $\vect{M}(-z) = \vect{\sigma}_2 \vect{M}(z) \vect{\sigma}_2$ in addition to the conjugation symmetry $\overline{\vect{M}(\bar z)} = \vect{\sigma}_2 \vect{M}(z) \vect{\sigma}_2$ satisfied by all flows of the focusing hierarchy. 
At the level of scattering data, these symmetries imply certain restriction. The spectrum itself must be symmetric with respect to both the real and imaginary axes. 
For any imaginary eigenvalue $\zeta_j = \ii \kappa_j$, the associated norming constant $c_j \in \ii \R$. For finite-gap potentials (cf. RHP~\ref{rhp:n-phase}), a band contour $\gamma_j \subset \Gamma$ which meets the imaginary axis must either lie exactly along the axis, or cross it perpendicularly. If the band lies along $\ii\R$ then the phase constant $\eee^{\ii \Lambda} \in \ii \R$, while if $\gamma_j$ crosses perpendicularly, then $\eee^{\ii \Lambda} \in \R$. Finally, for a primitive potential with $\mathcal{A} \subset \ii \R$, (the case we consider here) we must have $R(z) \in \ii \R$ for $z \in \mathcal{A}$.  

Let $0< \eta_1 < \eta_2$; fix $\mathcal{A} = \ii[\eta_1,\eta_2] \cup \ii[-\eta_2, -\eta_1]$ as the accumulation set of our example.
Since $\mathcal{A}$ is one-dimensional, we have $\mathcal{A} = \mathcal{A}_1$ and $\mathcal{A}_2 = \emptyset$. 
Let $\mathcal{A}^{\mathrm{low}} = \ii [-\eta_2, -\eta_1]$. Define a positive-valued function $r: \ii [-\eta_2, -\eta_1] \to (0,\infty)$. 
The pair $\mathcal{D}_\mathrm{prim} = (\mathcal{A}^{\mathrm{low}}, i r(z) )$ define a soliton condensate gas through RHP~\ref{rhp:accumulated} with $\Gamma = \mathcal{A}_2 = \emptyset$ where the jump \eqref{accumulated.jump} is given by 
\begin{equation}
	V(z;x,t) = \begin{dcases}
		\triu{ \ii r(\bar z) e^{-2i \varphi(z;x,t) }} & z \in \ii [\eta_1, \eta_2], \\
		\tril{ \ii r(z)  e^{2i \varphi(z;x,t) }} & z \in \ii [-\eta_2, -\eta_1], \\
	\end{dcases}
	\qquad
	\varphi(z;x,t) = x z + 4 t z^3.
\end{equation}
Such potentials were recently studied in \cite{GGJMM}. They describe realizations of a soliton condensate which is exponentially small as $x \to -\infty$ and approaches an elliptic (genus one) finite-gap solution of mKdV with spectral band $\ii[\eta_1, \eta_2]$ and its complex conjugate as $x \to + \infty$.

This realization of a soliton condensate can be realized as the limit of either a soliton or finite-gap ensemble. 
Define a strictly positive density $\varrho: [\eta_1, \eta_2] \to (0,\infty)$ with $\int_{\eta_1}^{\eta_2} \varrho = 1$. 
For each $n \in \N$, define eigenvalues $\{ \zeta_k^{(n)} \}_{k=1}^n \subset \ii[ -\eta_2, -\eta_1]$ according to the rule 
\begin{equation}\label{mkdv.ex.eigs}		
	\int_{\eta_1}^{\ii \zeta_k^{(n)}} \varrho(\eta) d\eta = \frac{j-\tfrac{1}{2}}{n}, \quad k=1,\dots, n.
\end{equation}
These eigenvalues satisfy \eqref{point.measure.limit} with $p=2$, with the measure $\dd\mu$ in \eqref{eig.measure.density} given by $\dd \mu(z) = 2\pi \varrho(\ii z) \dd \lambda(z)$. 
For each $n$ we define the soliton ensemble with scattering data\footnote{We omit the dependance on background data $\Gamma_G$ and $\Lambda$ from the scattering data since since in this example we are working with the zero background.}
\begin{equation}
	\mathcal{D}_{\mathrm{sol}}^{(n)} = \left( \mathcal{Z}^{(n)}, c^{(n)} \right) = \left( \left\{ \zeta_k^{(n)} \right\}_{k=1}^n, c^{(n)}(z) \right) , 
	\quad \text{where} \quad 
	c^{(n)}(z) = \frac{ \ii r(z) }{2\pi n \varrho(\ii z)}.
\end{equation}
and $\zeta_k^{(n)}$ is given by \eqref{mkdv.ex.eigs}.

To construct a finite-gap ensemble we enumerate our set of shrinking bands by $\Gamma^{(n)} = \bigcup_{k=1}^n (\gamma_k^{(n)} \cup \gamma_{-k}^{(n)})$ where 
\begin{equation}
	\gamma_k^{(n)} = [\zeta_k^{(n)} - \delta_k^{(n)}, \zeta_k + \delta_k^{(n)}], \qquad
	\gamma_{-k}^{(n)} = [\bar \zeta_k^{(n)} + \bar \delta_k^{(n)}, \bar \zeta_k - \bar \delta_k^{(n)}], \qquad k = 1,\dots,n.
\end{equation}
and the half-band length $\delta_k$ satisfies 
\begin{equation}
	\delta_k^{(n)} \in \R \cup \ii \R, \qquad | \delta_k^{(n)} | = \littleo{n^{-2}}.
\end{equation}
We then choose the initial phases $\Gamma^{(n)}_k$ on each band to satisfy
\begin{equation}
	\eee^{\ii \Lambda_k^{(n)} } = \frac{ r(\zeta_k^{(n)}) }{ \pi \varrho(i \zeta_k^{(n)} ) } \frac{1}{ n \delta_k^{(n)}}.
\end{equation}

\subsection*{Acknowledgements}
The authors would like to thank the Isaac Newton Institute for Mathematical Sciences, Cambridge, for support and hospitality during the programme "Emergent phenomena in nonlinear dispersive waves", where work on this paper was undertaken. This work was supported by EPSRC grant EP/R014604/1. 
R.J. was supported in part by the National Science Foundation under grant DMS-2307142 and by the Simons Foundation under grant 853620. 
A.T. was supported in part by the National science Foundation grant DMS-2009647 and by a grant from the Simons Foundation.
The authors would also like to thank Marco Bertola for interesting discussions and insightful comments and suggestions.

\appendix
	
\section{Riemann-Hilbert problems for $N$-soliton solutions of the NLS hierarchy with arbitrary normalization}
\label{app:normalization}
The typical characterization of the forward/inverse scattering transform for the NLS hierarchy is in terms of Riemann-Hilbert problem normalized to approach identity as the spectral variable tends to infinity is not uniquely defined. The usual problem(s) one sees in the literature are just one of a family of different RHPs which characterize the evolution of the same initial potential $\psi_0(x)$. Our goal here is to describe this family of problems as it applies to soliton solutions. In the interest of simplicity we restrict ourselves to the class of rapidly decaying initial data ($\psi_0 \in L^1(\R)$ is enough for the discussion here) which is sufficient to grasp the essentials ideas we wish to present.  

We begin with a brief overview of the forward scattering map, We omit many details and proofs which the interested reader can find elsewhere in the literature. The first step in the forward scattering transform is for each $z\in \R$ to construct \textit{Jost solutions} $\vect{w}^\pm(x;z)$ of the Zakharov-Shabat scattering problem \eqref{zs} with $\psi(x,\vec{t}) = \psi(x,0) = \psi_0(x)$. That is, solutions normalized such that $\vect{w}^\pm(x;z) \eee^{\ii z x \sigthree} \to \vect{I}$ as $x \to \pm \infty$. For each $z \in \R$ both solutions $\vect{w}^\pm(x;z)$ exist and are bounded (making the real line the continuous spectrum of the scattering operator \eqref{zs}). Both are also a basis for the solution space of \eqref{zs} and so must be linearly related. One finds that
\begin{gather}
	\vect{w}^-(x;z) = \vect{w}^+(x;z) \vect{S}(z), \qquad z \in \R
\intertext{where, due to a symmetry of \eqref{zs} one finds that}
	\label{scattering.mat}
	\vect{S}(z) = \begin{bmatrix} a(z) & -\overline{b(z)} \\ b(z) & \phantom{-}\overline{a(z)} \end{bmatrix}
	\qquad
	\det\vect{S}(z) = |a(z)|^2 + |b(x)|^2 = 1
\end{gather}
One can further show that the column vectors of $\vect{w}^\pm(x;z)$ extend analytically to solutions of \eqref{zs} for $z$ in particular half-planes; specifically, $\vect{w}_1^-$, $\vect{w}_2^+$, and $a(z)$ extend analytically to solutions for $\Im z > 0$, while $\vect{w}^+_1$, $\vect{w}_2^-$, and $\overline{a(\bar z)}$ extend to $\Im z<0$. The analytic properties of the diagonal entries of the scattering matrix follow from their Wronskian representation
\[
	a(z) = \Wr\left[ \vect{w}^-_1(x;z), \vect{w}^+_2(x;z) \right]
	\qquad
	\overline{a(\bar z)} = \Wr\left[ \vect{w}^-_1(x;z), \vect{w}^+_2(x;z) \right]
\]
Eigenvalues of the scattering operator \eqref{zs} in $\C^+$ are precisely the zeros of $a(z)$ in $\C^+$ and their complex conjugates in $\C^-$, at any such point $\zeta_k$, the two analytic solutions are linearly related for all $x$, that is
\[
	\vect{w}^-_1(x; \zeta_k) = \gamma_k \vect{w}^+_2(x;\zeta_k), \qquad \gamma_k \in \C \setminus\{0\},
\]
and the numbers $\gamma_k$ are called the connection coefficients related to each discrete eigenvalue. Finally, given the set of zeros $\{ \zeta_k \}_{k=1}^N$ of $a(z)$ in $\C^+$ and the scattering function $b: \R \to \C$, one can recover the function $a(z)$ from the trace formula
\begin{equation}\label{a.trace}
	a(z) =\lp \prod_{k=1}^N  \frac{z-\zeta_k}{z - \bar \zeta_k} \rp \exp \lp \frac{1}{2\pi \ii} \int_{-\infty}^{\infty} \log(1- |b(\xi)|^2) \frac{d\xi}{\xi-z} \rp
\end{equation}

The time evolution of the scattering coefficients $a(z)$, $b(z)$ and connection coefficients $\gamma_k$ under the flows of the NLS hierarchy are linear. We have
\begin{align*}
	&a(z,\vect{t}) = a(z,0)
	&&b(z,\vect{t}) = b(z,0) e^{i \varphi(z,x, \vect{t}\,)}
	&\gamma_k(\vect{t}) = \gamma_k(0) e^{i \varphi(\zeta_k, x, \vect{t}\,)}
\end{align*}
where $\vect{t} = (t_1, \dots, t_n)$ is the vector of time-flow variables. To set the inverse problem as a Riemann Hilbert problem one defined a piecewise meromorphic function in $\C \setminus \R$ by defining 
\begin{equation}\label{mrhp.soliton}
	\vect{m}(z;x,\vect{t}) := 
	\begin{dcases}
	\begin{bmatrix} 
		\frac{ \vect{w}^-_1(z;x,\vect{t}) }{ a_R(z)} &  
		\frac{ \vect{w}^+_2(z;x,\vect{t}) }{ a_L(z)} 
	\end{bmatrix} 
	\eee^{\ii \theta(z;x,\vect{t}\,) \vect{\sigma}_3} 
	& \Im z > 0 \\
	\begin{bmatrix} 
		\frac{ \vect{w}^+_1(z;x,\vect{t}) }{ {\overline{a_L(\bar z)}}^{\phantom{1}} }  &  
		\frac{ \vect{w}^-_2(z;x,\vect{t}) }{ \overline{a_R(\bar z)}^{\phantom{1}} } 
	\end{bmatrix} 
	\eee^{\ii \theta(z;x,\vec{t}\,) \vect{\sigma}_3}
	& \Im z < 0 
	\end{dcases}
\end{equation}
where $a_R(z)$ and $a_L(z)$ are any functions analytic in $\C^+$ such that 
\begin{equation}
	a(z) = a_L(z) a_R(z)
\end{equation}
The typical set-up in the literature is to take the trivial factorization with one of $a_L(z)$ and $a_R(z)$ to be identically $a(z)$. If $a_R(z) = a(z)$ (resp. $a_L(z) = a(z)$), then $\vect{m} \to \vect{I}$ as $x \to + \infty$ (resp. $x \to - \infty$).

Soliton solutions of the NLS hierarchy correspond precisely to the potentials $\psi_0(x)$ for which the scattering is reflectionless, i.e. $b(z) \equiv 0$, in which case it follows from \eqref{a.trace} that $a(z)$ is a rational function of $z$. Any partition $\{ \mathcal{Z}_L, \mathcal{Z}_R \}$ of the zeros of $a(z)$ in $\C^+$ defines a factorization
\begin{equation}
\begin{gathered}
	a(z) = \prod_{k=1}^N  \frac{z-\zeta_k}{z - \bar \zeta_k}  = a_L(z) a_R(z), \\
	a_L(z) =\prod_{\zeta_k \in \mathcal{Z}_L}   \frac{z-\zeta_k}{z - \bar \zeta_k},
	\qquad  
	a_L(z) =\prod_{\zeta_k \in \mathcal{Z}_R}   \frac{z- \zeta_k}{z - \bar \zeta_k}.
\end{gathered}
\end{equation}
Given this factorization for a reflectionless potential, the matrix $\vect{m}$ in \eqref{mrhp.soliton} satisfies the following Riemann-Hilbert problem

\begin{RHP} \label{rhp:gen.sol}
Given data
\[
	\mathcal{D}^{\mathrm{sol}}:= \left( \mathcal{Z}_L, \mathcal{Z}_R, c_L,c_R \right)
\]
consisting of two finite disjoint sets $\mathcal{Z}_L,\ \mathcal{Z}_R \subset \C^+$ and two functions $c_L:\mathcal{Z}_L \to \C \setminus\{0\}$, $c_R:\mathcal{Z}_R \to \C \setminus\{0\}$,
find a $2\times 2$ matrix valued function $\vect{m}(\, \cdot\, ;x,\vect{t})$ with the following properties.
\begin{enumerate}[label=\arabic*.]
	\item $\vect{m}(\, \cdot\, ;x,\vect{t})$ is analytic in $\C \setminus (\mathcal{Z} \cup \overline{\mathcal{Z}}), \qquad \mathcal{Z} =\mathcal{Z}_L \cup \mathcal{Z}_R$
	\item $\vect{m}(z ;x,\vect{t}) = \vect{I} + z^{-1} \vect{m}_1(x,\vect{t}) + \bigo{z^{-2}}$ as $z \to \infty$.
	\item $\vect{m}(\,\cdot\,;x,\vect{t)}$ has simple poles at each point $\zeta \in \mathcal{Z}$, and their complex conjugates, satisfying the residue relations:
	\begin{itemize}
		\item For $\zeta \in \mathcal{Z}_R$
		\begin{equation}
			\begin{aligned}
			&\res_{z = \zeta} \vect{m}(z;x,\vect{t}) = \lim_{z \to \zeta}  \vect{m}(z;x,\vect{t}) 
			\tril[0]{c_R(\zeta) \eee^{\ii \varphi(z;x,\vect{t})}} \\
			&\res_{z = \bar \zeta} \vect{m}(z;x,\vect{t}) = \lim_{z \to \bar \zeta}  \vect{m}(z;x,\vect{t}) 
			\triu[0]{-\overline{c_R(\zeta)} \eee^{-\ii \varphi(z;x,\vect{t})}}
			\end{aligned}
		\end{equation}
		\item For $\zeta \in \mathcal{Z}_L$
		\begin{equation}
			\begin{aligned}
			&\res_{z = \zeta} \vect{m}(z;x,\vect{t}) = \lim_{z \to \zeta}  \vect{m}(z;x,\vect{t}) 
			\triu[0]{c_L(\zeta) \eee^{-\ii \varphi(z;x,\vect{t} )}} \\
			&\res_{z = \bar \zeta} \vect{m}(z;x,\vect{t}) = \lim_{z \to \bar \zeta}  \vect{m}(z;x,\vect{t}) 
			\tril[0]{-\overline{c_L(\zeta)} \eee^{\ii \varphi(z;x,\vect{t} )}}
			\end{aligned}
		\end{equation}
	\end{itemize}	 
\end{enumerate}
\end{RHP}
 
In terms of the direct, reflectionless, scattering data, $\{ \zeta_k, \gamma_k\}_{k=1}^N$ the norming constants are given by
\begin{equation}
	\begin{aligned}
	&c_R(\zeta_k) = \gamma_k \frac{ a_L(\zeta_k)}{a_R'(\zeta_k)}
	&& \zeta_k \in \mathcal{Z}_R, 
	\qquad
	&c_L(\zeta_k) = \frac{1}{\gamma_k} \frac{ a_R(\zeta_k)}{a_L'(\zeta_k)}
	&& \zeta_k \in \mathcal{Z}_L, 
	\end{aligned}
\end{equation}
but one can forgo the direct scattering problem and instead say that any admissible soliton data set $\mathcal{D}^{\mathrm{sol}} = (\mathcal{Z}_L, \mathcal{Z}_R, c_L, c_R)$ parameterizes a reflectionless potential $\psi$ which is an $N = \left| \mathcal{Z}_L \cup \mathcal{Z}_R \right|$ soliton solution of the NLS hierarchy.
 
\section{General AKNS Riemann-Hilbert Problems}\label{sec:AKNS}

In this paper we consider several different Riemann-Hilbert problems with different jump contour geometries and jump matrices. Our goal in this section is to show that the solution of all of these problems, produce potentials $\psi(x,\vect{t})$ of the NLS hierarchy. Specifically, consider the following general problem. 

\begin{RHP}\label{rhp:akns}
Given a Schwarz symmetric contour $\Gamma$ and a function $\vect{V}_0: \Gamma \to SL_2(\C)$ such that $\vect{V}_0(z) =\vect{V}_0(\bar z)^\dagger$, for each $(x,\vect{t}) \in \R \times \R^{\mathcal{K}}$ find a $2\times2$ matrix-valued function $\vect{m}$ such that
\begin{enumerate}[align=left, leftmargin=*]
	\item[ \textbf{Analyticity}:] $\vect{m}(z;x,\vect{t})$ is analytic for $z \in \C \setminus \Gamma$. 
	\item[ \textbf{Normalization}:] $\vect{m}(z;x,\vect{t}) = \vect{I} + \vect{m}_1(x,\vect{t}) z^{-1} + \bigo{z^{-2}}$ as $z \to \infty$. 
	\item[ \textbf{Jump Condition}:] $\vect{m}$ has continuous non-tangential boundary values $\vect{m}_\pm(z;x,\vect{t})$ satisfying 
	\begin{gather}
		\vect{m}_+(z;x,\vect{t}) =\vect{m}_-(z;x,\vect{t})  \eee^{-\ii \varphi(z;x,\vect{t}) \sigma_3} \vect{V}_0(z) \eee^{\ii \varphi(z;x,\vect{t}) \sigma_3} 
		\qquad z \in \Gamma \\ 
		\nonumber
		\varphi(z;x,\vect{t}) = x z + \sum_{k=0}^\mathcal{K}  t_k z^k
	\end{gather}
\end{enumerate}	
\end{RHP}

\begin{prop}
Suppose that $\vect{m}(z;x,\vect{t})$ is a solution of RHP~\ref{rhp:akns} which depends smoothly on $(x,\vect{t})$. Then 
\[
	\psi(x,\vect{t}) = 2\ii [\vect{m}_1(x,t)]_{12}
\]
is a solution of the first $\mathcal{K}$ flows of the focusing NLS-hierarchy. 
\end{prop}

\begin{proof}
Given $\vect{m}(z;x,t)$, define $\vect{w}(z;x,\vect{t}) := \vect{m}(z;x,\vect{t} ) \eee^{-\ii \varphi(z; x, \vect{t}) \sigthree}$. It follows from RHP~\ref{rhp:akns} that $\vect{w}(z;x,\vect{t})$ is analytic for $z \in \C \setminus \Gamma$ with jump $\vect{w}_+(z;x,\vect{t}) = \vect{w}_-(z;x,\vect{t}) \vect{V}_0(z)
$ independent of $(x, \vect{t})$. Note that as $\det \vect{V}_0(z) \equiv 1$, $\det \vect{w_+}  - \det \vect{w}_- = 0$, so by Morerra's theorem $\det \vect{w}(z)$ is an entire function of $z$. Since $\vect{w} \to \vect{I}$ as $z \to \infty$ if follows that $\det \vect{w}(z;x, \vect{t}) \equiv 1$.
Consider the matrices in $\mathfrak{sl}_2(\C)$ defined by 
\begin{equation}\label{lax.from.w}
	\begin{aligned}
	&\vect{A}(z;x,t) = \vect{w}_x(z;x,\vect{t}) \vect{w}(z;x,\vect{t})^{-1} \\
	&\vect{B}^{ [k]}(z;x,t) = \vect{w}_{t_k}(z;x,\vect{t}) \vect{w}(z;x,\vect{t})^{-1} , \quad k=0,\dots, \mathcal{K}. 
	\end{aligned}
\end{equation} 
It's easy to show, applying Morerra's theorem again, that $\vect{A}$ and $\vect{B}^{[k]}$ are entire functions of $z$. Writing
\begin{equation}\label{lax.from.m}
	\begin{gathered}
	\vect{A}(z) = \vect{m}_{x}(z) \vect{m}(z)^{-1} - \ii z \vect{m}(z) \sigthree \vect{m}(z)^{-1} \\
	\vect{B}^{[k]}(z) = \vect{m}_{t_k}(z) \vect{m}(z)^{-1} - \ii z^k \vect{m}(z) \sigthree \vect{m}(z)^{-1}, \quad k=1,\dots, \mathcal{K}
	\end{gathered}
\end{equation}
(where we have suppressed the $x$ and $\vect{t}$ arguments for brevity), it follows from the asymptotic behavior of $\vect{m}$ that $\vect{B}^{[k]}$ is a polynomial of degree at most $k$. A similar argument shows $\vect{A}$ is a linear polynomial in $z$. 

The above formula shows that $\vect{w}(z;x,t)$ is a simultaneous solution of the system of differential equations
\[
	\vect{w}_x = \vect{A} \vect{w} 
	\qquad
	\vect{w}_{t_k} = \vect{B}^{[k]} \vect{w} , \ k=0,\dots, \mathcal{K}.
\]	
Differentiating \vect{A} and $\vect{B}^{[k]}$ with respect to a time $t_j$ using \eqref{lax.from.w} gives
\[
	\vect{A}_{t_j}(z) = \vect{w}_{x t_j}(z) \vect{w}(z)^{-1} - \vect{A}(z) \vect{B}^{[j]}(z)
	\qquad
	\vect{B}^{[k]}_{t_j}(z) = \vect{w}_{t_k t_j}(z) \vect{w}(z)^{-1} - \vect{B}^{[k]}(z) \vect{B}^{[j]}(z)
\]
which yields the zero curvature relations
\begin{equation}\label{zero.curve}
	\vect{A}_{t_k} - \vect{B}^{[k]}_{x} + \left[ \vect{A},\ \vect{B}^{[k]} \right] = \vect{0}, \qquad k=1,\dots, \mathcal{K}.
\end{equation}
and 
\[
	\vect{B}^{[k]}_{t_j} - \vect{B}^{[j]}_{t_k} + \left[ \vect{B}^{[k]},\ \vect{B}^{[j]} \right] = \vect{0}, \qquad t_j \neq t_k,
\]
which implies that the time flows all commute $\vect{A}_{t_j t_k} = \vect{A}_{t_k t_j}$. 

Introducing the Laurent expansions of $\vect{m}^{\pm 1}$ about infinity, $\vect{m}(z;x,\vect{t}) = \sum_{j=0}^{\infty} \vect{m}_j(x,\vect{t}) z^{-j}$ and $\vect{m}(z;x,\vect{t})^{-1} = \sum_{j=0}^{\infty} \vect{n}_j(x,\vect{t}) z^{-j}$, the coefficients of the entire functions $\vect{A}$ and $\vect{B}^{[k]}$ can be computed from \eqref{lax.from.m}. 
Write 
\begin{gather}
	\vect{B}^{[k]} = \sum_{\ell=0}^k \vect{b}_{\ell} z^{k-\ell} + \bigo{z^{-1}} =  \sum_{\ell=0}^k \vect{b}_{\ell} z^{k-\ell}  
\shortintertext{then using \eqref{lax.from.m} }
	\vect{b}_\ell = -\ii \sum_{j=0}^\ell \vect{m}_{\ell-j} \sigthree \vect{n}_j  =  \sum_{j=0}^{\ell} \left[\ii \sigthree,\, \vect{m}_{\ell-j} \right] \vect{n}_j. 
	\label{b.ell.1}
\end{gather}
Similarly one can compute $\vect{A}$ 
\begin{equation}
	\vect{A} = -\ii z \sigthree + \left[ \ii \sigthree, \vect{m}_1 \right] 
	+ \sum_{k=1}^\infty z^{-k} 
	\left( 	\vect{b}_{k+1}	+ \sum_{j=0}^{k} \left(\partial_x \vect{m}_{k-j} \right) \vect{n}_j	\right) 
\end{equation}
and since $\vect{A}$ must be polynomial this yields
\begin{equation}
	\vect{A} = \vect{B}^{[1]} = -\ii z \sigthree + \vect{P},  \qquad 
	\vect{P} := \left[ \ii \sigthree, \, \vect{m}_1 \right] 
	= \begin{bmatrix} 0 & q(x,t)  \\ p(x,t) & 0 \end{bmatrix}.
\end{equation}
as well as a second expression for the coefficients of $\vect{b}_\ell$
\begin{equation}
	\vect{b}_\ell =  -\sum_{j=0}^{k} \left(\partial_x \vect{m}_{k-j} \right) \vect{n}_j.
	\label{b.ell.2}
\end{equation}
The zero curvature relations \eqref{zero.curve} then reduces to 
\begin{equation}\label{app.b.last}
	\vect{P}_{t_k} - \vect{B}_{k,x} + \left[\vect{P},\,\vect{B}_k \right] = 0
\end{equation}
Moreover, using the pair of relations \eqref{b.ell.1}, \eqref{b.ell.2} for the coefficients $\vect{b}_\ell$, it can be shown \cite{FNR} that the coefficients of each $\vect{b}_\ell$ can be uniquely expressed as differential polynomials in the potential $\vect{P}$ of the spatial part of the Lax-pair hierarchy. Inserting the expressions for each $\vect{b}_\ell$ into \eqref{app.b.last} yields the hierarchy of flows satisfied by the potential matrix $\vect{P}$. 
\end{proof}
 
\medskip

\printbibliography

\end{document}